\documentclass[a4paper,fleqn,usenatbib]{mnras}

\usepackage{newtxtext,newtxmath}

\usepackage[T1]{fontenc}
\usepackage{ae,aecompl}

\usepackage{graphicx}	% Including figure files
\usepackage{amsmath}	% Advanced maths commands
\usepackage{amssymb}	% Extra maths symbols
\usepackage{xspace}
\usepackage{breqn}
\usepackage{paralist}
\usepackage[noabbrev]{cleveref}
\makeatletter

\newcommand\Autoref[1]{\@first@ref#1,@}
\def\@throw@dot#1.#2@{#1}% discard everything after the dot
\def\@set@refname#1{%    % set \@refname to autoefname+s using \getrefbykeydefault
    \edef\@tmp{\getrefbykeydefault{#1}{anchor}{}}%
    \xdef\@tmp{\expandafter\@throw@dot\@tmp.@}%
    \ltx@IfUndefined{\@tmp autorefnameplural}%
         {\def\@refname{\@nameuse{\@tmp autorefname}s}}%
         {\def\@refname{\@nameuse{\@tmp autorefnameplural}}}%
}
\def\@first@ref#1,#2{%
  \ifx#2@\autoref{#1}\let\@nextref\@gobble% only one ref, revert to normal \autoref
  \else%
    \@set@refname{#1}%  set \@refname to autoref name
    \@refname~\ref{#1}% add autoefname and first reference
    \let\@nextref\@next@ref% push processing to \@next@ref
  \fi%
  \@nextref#2%
}
\def\@next@ref#1,#2{%
   \ifx#2@ and~\ref{#1}\let\@nextref\@gobble% at end: print and+\ref and stop
   \else, \ref{#1}% print  ,+\ref and continue
   \fi%
   \@nextref#2%
}

\makeatother

\defcitealias{Sesar:17}{S17}
\defcitealias{Dehnen:1998}{DB98}
\defcitealias{Portail:17}{P17}

\newcommand{\nrrab}{{\ensuremath{15813}}\xspace}
\newcommand{\nbadplx}{{\ensuremath{82}}\xspace}
\newcommand{\nsample}{{\ensuremath{15651}}\xspace}

\newcommand{\nbigmock}{{\ensuremath{870, 000}}\xspace}

\newcommand{\nrbins}{{\ensuremath{9}}\xspace}
\newcommand{\rmin}{{\ensuremath{1.5\kpc}}\xspace}
\newcommand{\rmax}{{\ensuremath{20\kpc}}\xspace}
\newcommand{\ntheta}{{\ensuremath{5}}\xspace}
\newcommand{\thetabinedges}{{\ensuremath{0\degr,\,25\degr,\,40\degr,\,50\degr,\,60\degr,\,70\degr}\xspace}}

\newcommand{\qrho}{{\ensuremath{q_\rho}\xspace}}
\newcommand{\qrhomeas}{{\ensuremath{1.00 \pm 0.09}\xspace}}

\newcommand{\qpot}{{\ensuremath{q_\Phi}\xspace}}
\newcommand{\qphimeas}{{\ensuremath{1.01 \pm 0.06}\xspace}}

\newcommand{\rhodm}{{\rho_{\rm dm}\xspace}}

\newcommand{\ie}{{\it i.e.}\xspace}
\newcommand{\eg}{{\it e.g.}\xspace}
\newcommand{\cf}{{\it cf.}\xspace}

\newcommand{\kpc}{\ensuremath{\,{\rm kpc}}\xspace}
\newcommand{\pc}{\ensuremath{\,{\rm pc}}\xspace}

\newcommand{\Gyr}{\ensuremath{\,{\rm Gyr}}\xspace}
\newcommand{\kms}{\ensuremath{\,{\rm km}\,{\rm s}^{-1}}\xspace}
\newcommand{\feh}{\ensuremath{{\rm [Fe/H]}}\xspace}

\newcommand{\masyr}{\ensuremath{\,{\rm mas/yr}}\xspace}
\newcommand{\uas}{\ensuremath{\,\mu {\rm as}}\xspace}
\newcommand{\msun}{\ensuremath{{M_\odot}}\xspace}

\newcommand*\azmean[1]{\ensuremath{\langle {#1}\rangle}}
\newcommand*\bigazmean[1]{\ensuremath{\left\langle {#1}\right\rangle}}

\newcommand*\mean[1]{\ensuremath{\overline{#1}}}
\newcommand*\samplemean[1]{\ensuremath{\langle {#1}\rangle}}

\title[Milky Way Force Field and Dark Matter Shape]{The Gravitational Force Field of the Galaxy Measured From the Kinematics of RR Lyrae in Gaia}

\author[C. Wegg et al.]{
Christopher Wegg $^{1,2}$\thanks{E-mail: \href{mailto:chris.wegg@gmail.com}{chris.wegg@gmail.com} (CW)}, Ortwin Gerhard$^{1}$ and Marie Bieth$^{3}$
\\
$^{1}$Max-Planck-Institut f\"ur Extraterrestrische Physik, Giessenbachstrasse, 85748 Garching, Germany\\
$^{2}$Laboratoire Lagrange, Universit\'e C\^ote d'Azur, Observatoire de la C\^ote d'Azur, CNRS, Blvd de l'Observatoire, F-06304 Nice, France\\
$^{3}$TNG Technology Consulting, Betastr. 13a, 85774 Unterf\"{o}hring, Germany
}

\date{Accepted XXX. Received YYY; in original form ZZZ}

\pubyear{2018}

\begin{document}
\label{firstpage}
\pagerange{\pageref{firstpage}--\pageref{lastpage}}
\maketitle 
%%%%%%%%%%%%%%%%%%%%%%%%%%%%%%%%%%%%%%%%%%%%%%%%%%

%%%%%%%%%%%%%%%%%%%% ABSTRACT %%%%%%%%%%%%%%%%%%%%
\begin{abstract}
From a sample of \nsample RR Lyrae with accurate proper motions in Gaia DR2, we measure the azimuthally averaged kinematics of the inner stellar halo between 1.5\kpc and 20\kpc from the Galactic centre. We find that their kinematics are strongly radially anisotropic, and their velocity ellipsoid nearly spherically aligned over this volume. Only in the inner regions $\lesssim 5\kpc$ does the anisotropy significantly fall (but still with $\beta > 0.25$) and the velocity ellipsoid tilt towards cylindrical alignment. In the inner regions, our sample of halo stars rotates at up to $50\kms$, which may reflect the early history of the Milky Way, although there is also significant angular momentum exchange with the Galactic bar at these radii. We subsequently apply the Jeans equations to these kinematic measurements in order to non-parametrically infer the azimuthally averaged gravitational acceleration field over this volume, and by removing the contribution from baryonic matter, measure the contribution from dark matter. We find that the gravitational potential of the dark matter is nearly spherical with average flattening $q_\Phi=\qphimeas$ between 5\kpc and 20\kpc, and by fitting parametric ellipsoidal density profiles to the acceleration field, we measure the flattening of the dark matter halo over these radii to be $q_\rho=\qrhomeas$.
\end{abstract}

%%%%%%%%%%%%%%%%%%%%%%%%%%%%%%%%%%%%%%%%%%%%%%%%%%

%%%%%%%%%%%%%%%%% BODY OF PAPER %%%%%%%%%%%%%%%%%%

\section{Introduction}

Simulations of structure formation in a $\Lambda$CDM universe have been extremely successful in producing many of the observational properties of galaxies across cosmic time. While dark matter only simulations produce dark matter halos with a characteristic profile \citep{nfw} and highly flattened triaxial shapes with flattening $q_\rho \equiv \samplemean{c/a}_\rho \sim 0.5$ \citep[\eg][]{Dubinski:91,Jing:02,Allgood:06,Schneider:12}, this is altered by the uncertain interplay between baryons and dark matter. In particular, dark matter halos are expected to respond to baryonic infall by deviating less from axial symmetry and becoming less flattened \citep{Dubinski:94,Abadi:10}. However, even for the massive, near maximal disk seen in the Milky Way \citep[\eg][]{Bovy:13,Wegg:16} the density typically becomes less flattened by $\Delta q_\rho \sim 0.2-0.3$  \citep[][]{Debattista:08}, corresponding to a typical increase in the flattening of the dark matter potential, $\samplemean{c/a}_\Phi \equiv q_\Phi$, of $\Delta q_\Phi \sim 0.1-0.2$ \citep{Kazantzidis:10,Dai:18}.
 
In external galaxies, we are typically only able to measure dark matter halo properties using samples of galaxies \citep[\eg][]{vanUitert:12,Martinsson:13,Aniyan:16}. In the Milky Way however, we can measure the detailed kinematics of individual stars. We can use these unique measurements to infer the properties of our dark matter halo in much more detail than is possible in external galaxies, and use this as a prototype, a process referred to as near-field cosmology. However, despite the observational advantages of studying the Milky Way's dark matter halo, there is still no consensus on either its shape or profile.

Probes of the shape of the Milky Way's halo include tidal streams, halo kinematics, the flaring of the HI gas disk, and comparison of the local dark matter density with enclosed densities required by the rotation curve \citep[see for example the review by][]{Read:14}. 

The tightest recent constraints on the shape of the halo have arisen from measurements of tidal streams. Initial work focused on the Sagittarius stream suggested the halo to be spherical \citep{Ibata:01} while later models pointed to a oblate halo \citep{Law:10}. The stability of these models was questioned \citep{Debattista:13}, although this problem may be lessened by a halo whose shape changes with radius \citep{VeraCiro:13}.

However, the difficulty of using the complex Sagittarius stream to constrain the halo has led to a recent focus on other colder streams, particularly GD-1 \citep[][]{Grillmair:gd1} and Pal-5 \citep{Odenkirchen:01}. These streams lie $\approx 14\kpc$ and $\approx 18\kpc$ from the Galactic centre, and while the modelling methods vary, the results generally point to a dark matter potential consistent with a  spherical halo. For example, at the location of GD-1, the flattening of the overall potential has been measured to be $q_\Phi\equiv\samplemean{c/a}_\rho=0.87\substack{+0.07\\-0.04}$ by \citet{Koposov:10}, $q_\Phi=0.90\substack{+0.05\\-0.10}$ by \citet{Bowden:15} and $q_\Phi=0.95\pm0.04$  by \citet{Bovy:16}. Similarly at the location of Pal-5 the overall potential was measured to be $q_\Phi=0.95\substack{+0.05\\-0.10}$ by \citet{Kupper:15} and $q_\Phi=0.94\pm0.05$ by \citet{Bovy:16}. Combining these constraints on the potential with baryonic models results in a dark matter halo with axes ratio $q_\rho=1.05\pm0.14$ \citep{Bovy:16}, consistent with spherical, and therefore in tension with the expectations of cosmological $\Lambda$CDM simulations \citep{Dai:18}. This tension is a tantalising prospect because halo shape can, in principle, be a probe of the nature of dark matter and its possible interactions \citep[\eg][]{Peter:13}.

The work here takes a different approach, instead applying Jeans modelling to the kinematics of halo stars under the assumption of dynamical equilibrium. This approach has also been used several times recently to constrain the dark matter halo shape. However, unlike the stream modelling approach where different modelling techniques have produced similar results, the results using halo kinematics are more diverse. For example, \citet{Loebman:14} is the most conceptually similar work to ours. They apply the Jeans equations to SDSS Segue halo star kinematic measurements by \citet{Bond:10}, finding the dark matter to have a flattened potential with $q_\Phi=0.8\pm0.1$ and a corresponding density flattening of $q_\rho=0.4\pm0.1$ within $20\kpc$. However, in contrast, \citet{Bowden:16} favours a highly prolate dark matter potential with flattening $q_\Phi=1.5-2.0$. Typically this Jeans modelling approach requires  parametric models to be fitted, although the size and extent of our sample allows us to largely avoid these parameterisations.

The present state of the art is therefore that stream modelling is providing consistent constraints that the halo is nearly spherical at radii $\approx 14\kpc$ and $\approx 18\kpc$, while, inside this, the shape is highly uncertain. This situation is expected to rapidly change: the recent release of Gaia DR2 has provided measurements of the radial velocities of tens of millions of stars, and accurate astrometry of more than a billion, covering a large fraction of the Galaxy. Here, we take advantage of new accurate Gaia DR2 measurements of proper motions of RR Lyrae in the stellar halo and use them as kinematic tracers in order to measure the properties of the dark matter halo within 20\kpc of the Galactic center. 

Our primary motivation in this work was to constrain the shape of the Milky Way's dark matter halo and its variation with radius. However, we also present results that impact two further important areas. (i) The dark matter density and mass profile inside 20\kpc. This is because the inner parts of our studied volume is a region which is particularly important in understanding whether the dark matter profile has a core as implied by the bulge measurements of \citet{Portail:17}, or a cusp as seen in recent cosmological simulations of Milky Way mass haloes \citep{Grand:17a,Chan:15}. (ii) The kinematics of the Galactic halo which is an extremely interesting topic in its own right. 

The kinematics of the stellar halo are of particular interest  because they provide a probe into the history of this fundamental population of stars (for a review of the stellar halo and its kinematics see section 6.1 of \citealt{Ortwin:araa}, or \citealt{Helmi:08} for a dedicated but older introduction). Comparisons between samples must be made with care: the kinematics of the halo depends on metallicity \citep{Kafle:13,Das:16,Deason:17,Belokurov:18final}, and is therefore sensitive to the sample choice. Here, we study a sample of RR Lyrae without selection with respect to metallicity. The bulk of the halo has $\feh>-2$ and therefore, by using RR Lyrae as a tracer, we largely sample from this, more metal rich part, of the halo. The reader most interested in the kinematics of the stellar halo should concentrate on \autoref{sec:kin}.

The paper proceeds as follows: in \autoref{sec:sample} we construct a sample of RR Lyrae away from the Galactic plane with accurate transverse velocities, in \autoref{sec:kin} we  measure the kinematics of this sample, in \autoref{sec:jeans} we apply the Jeans equations to these kinematics to measure the Galactic acceleration field, and in \autoref{sec:dmmodels} we fit parametric dark matter profiles to these forces. We discuss and place our results in context in \autoref{sec:discussion}, and conclude in \autoref{sec:conc}.

Throughout, we use a distance to the Galactic center of $R_0=8.2\kpc$ \citep{Ortwin:araa}, and an absolute value of the solar velocity $(U,V,W)_\odot = (11.1, 12.24 + 238,7.25)\kms$ \citep{Schonrich:12}. We assess the impact of these assumptions when we assess our systematics. 

\section{A Sample of RR Lyrae With Transverse Velocities}
\label{sec:sample}

We construct a sample with which to trace the dynamics of the halo from the catalogue of RR Lyrae in PanSTARRS1 (PS1) provided by \citet[][hereafter \citetalias{Sesar:17}]{Sesar:17}. \citetalias{Sesar:17} classify stars observed in the PS1 3$\pi$ survey as RR Lyrae using a machine learning approach. In their catalogue, each star has a score (score$_{3,\rm ab}$), where high numbers indicate higher likelihood that the star is a type $ab$ RR Lyrae. We use a threshold of 0.6 which, for the relatively nearby RR Lyrae considered in this work, will provide a sample with greater than  95\% purity and 95\% completeness (table 3 of \citetalias{Sesar:17}). The provided distances to these type $ab$ RR Lyrae are accurate to 3\% \citepalias{Sesar:17}.

\begin{figure}
	\includegraphics[width=\columnwidth]{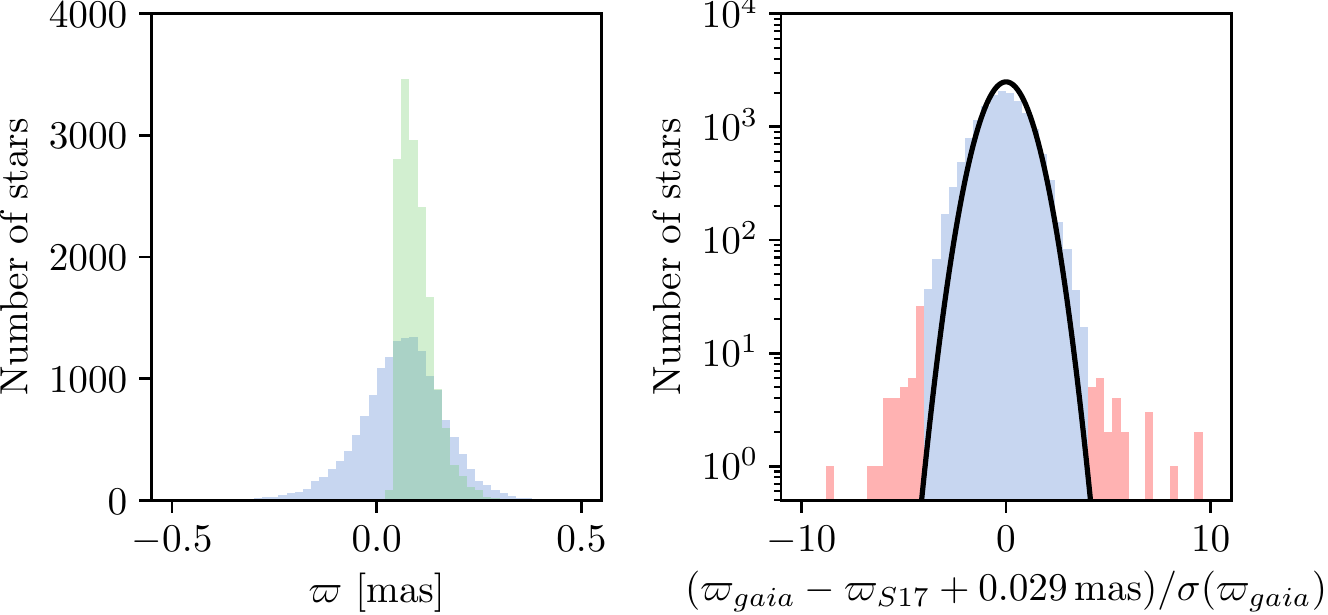}
    \caption{Left panel: The distribution of Gaia DR2 parallax measurements of sample stars (blue) and the \citetalias{Sesar:17} distances inverted to a parallax. Note that because these stars are too faint and distant to have reliable Gaia parallaxes, a significant number have negative parallax.  Right panel: The distribution of residuals between the Gaia DR2 parallax and the \citetalias{Sesar:17} distances \citep[\cf Fig. 8 of][]{Lindegren:18}. If all the residuals were normally distributed we would expect them to follow the black line. We exclude those stars in red with residuals greater than $4\sigma$. In making this cut we assume a parallax zero-point of $-29\uas$ \citep{Lindegren:18}, although the overall sample has a slightly different zero-point of $-38\uas$.}
    \label{fig:plxdifference}
\end{figure}

In our analysis, we consider RR Lyrae with Galactocentric radius between 1.5\kpc and 20\kpc,  which lie more than 20$\degr$ from the Galactic plane in Galactocentric coordinates. However, we make several further cuts to ensure that the sample is clean and complete over a defined volume:
\begin{inparaenum}
\item We only consider RR Lyrae with $|b|>10\deg$ because, closer to the Galactic plane, extinction causes the completeness of the the sample to drop \citepalias{Sesar:17}. 
\item The nominal area of the PS1 survey is ${\rm dec} > -30 \deg$, however, to simplify selection near this boundary, we consider only stars with ${\rm dec} > -29 \deg$.
\item We remove RR Lyrae that in projection lie within a conservative 10 half light radii of a galactic globular cluster. We use the catalogue of \citet{Harris:96} and use 2 arcmin as the half light radius where none has been measured.
\item To remove the Sagittarius dwarf galaxy and stream, we remove all RR Lyrae that lie more than 15\kpc from the Sun and lie within 10 degrees of the plane of the Sagittarius stream as defined by \citet{Majewski:03}. Tests with the Sagittarius stream model of \citet{Law:10} indicate that this should remove more than 95\% of the stream \citep[see also][for an analysis of the Sagittarius stream in the \citetalias{Sesar:17} RR Lyrae]{Hernitschek:17}. 
\end{inparaenum}

The resulting \nrrab RR Lyrae are cross matched with Gaia DR2 which provides astonishingly accurate absolute proper motions. We use a cross match radius of $0.5\arcsec$ and remove cross matches without a measured proper motion, or whose astrometric fit was poor (those with \verb|astrometric_excess_noise_sig| > 10 ).

\begin{figure}
	\includegraphics[width=\columnwidth]{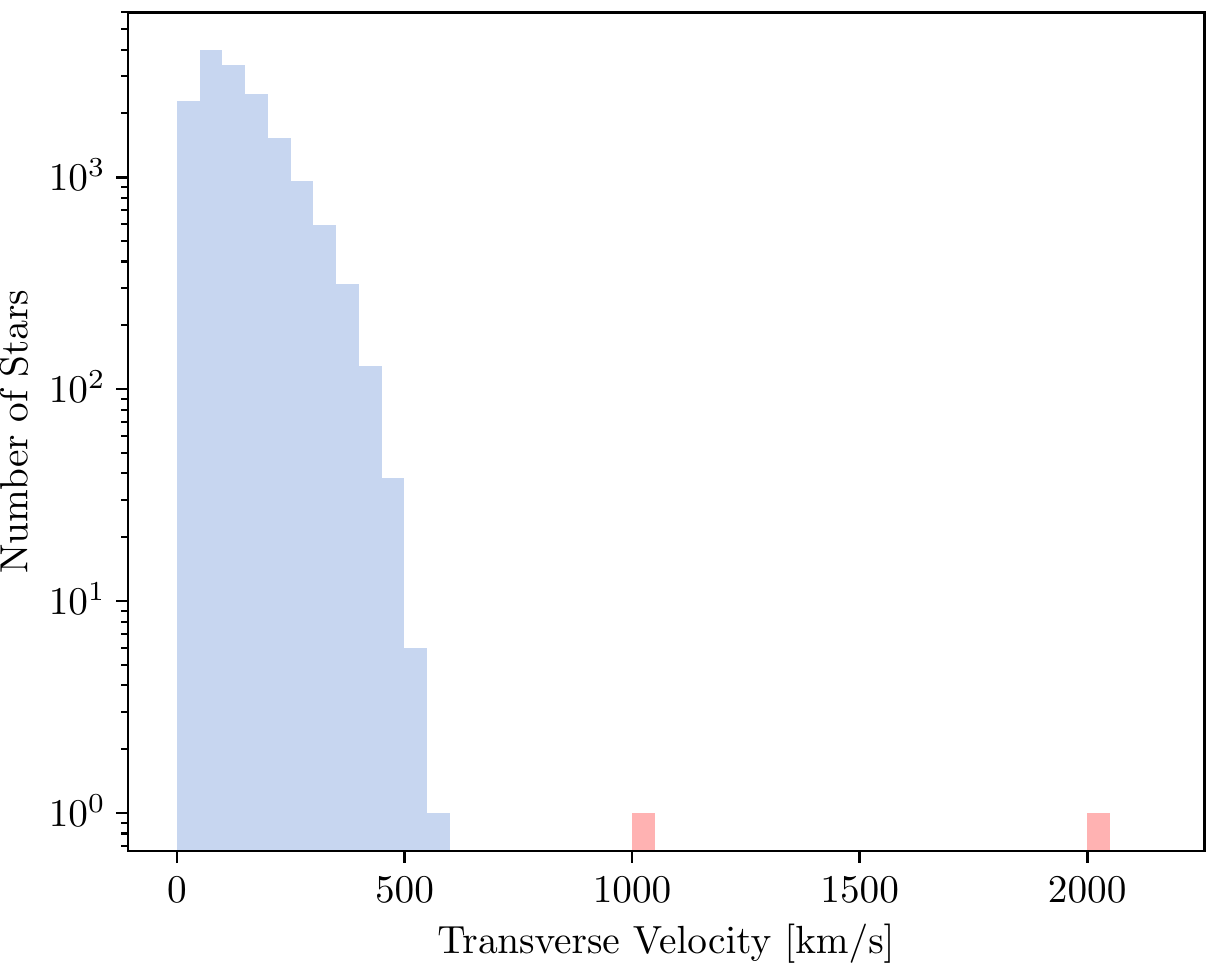}
    \caption{The transverse velocity distribution of the entire sample, the two clear outliers with transverse velocity more than 1000\kms were removed from the sample. }
    \label{fig:transverseveldist}
\end{figure}

\begin{figure}
	\includegraphics[width=\columnwidth]{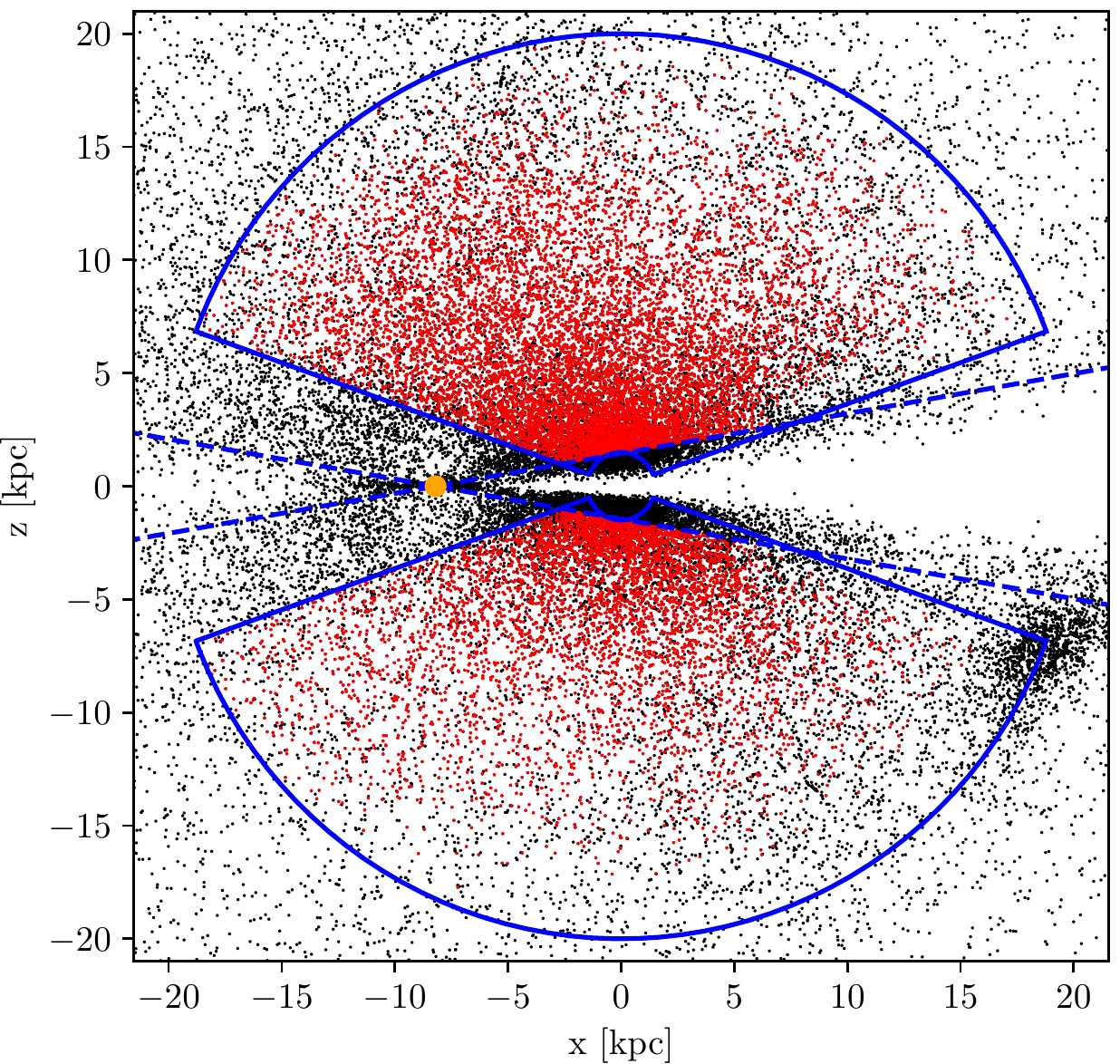}
    \caption{The sample of RR Lyrae considered viewed side on.  In black we plot all likely type $ab$ RR Lyrae from \citetalias{Sesar:17} (score$_{3,\rm ab}>0.6$). In red we show RR Lyrae in our sample after making the cuts described in \autoref{sec:sample}. In particular the volume we consider with galactocentric distance $1.5\kpc \leq r_{\rm gc } < 20\kpc$ and  with Galactocentric angle to the plane $\theta > 20\degr$ is outlined in blue. The Sun is the orange circle, and stars below the dashed blue line with $b<10\degr$ are excluded. Note that the Sagittarius dwarf galaxy at $(x,z) \approx (20,-7)\kpc$, and the Sagittarius are also excised.}
    \label{fig:sample_side_on}
\end{figure}

\begin{figure}
	\includegraphics[width=\columnwidth]{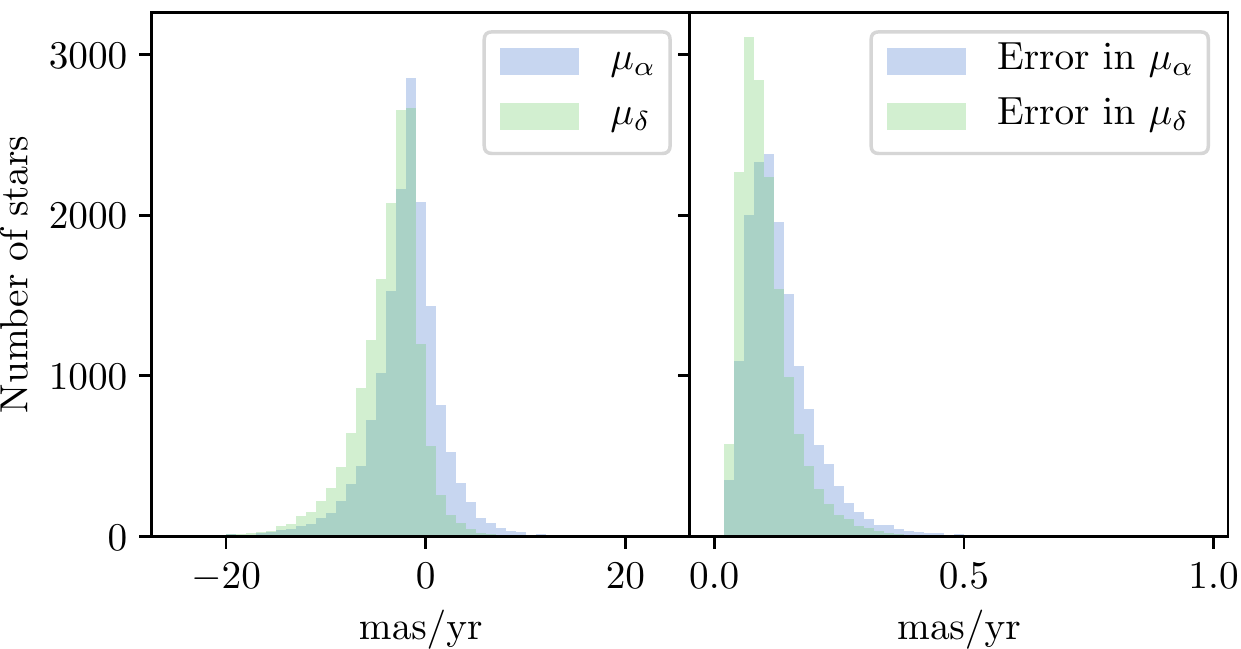}
    \caption{The distribution Gaia DR2 proper motion and proper motion errors of the sample. Note that most of the proper motion errors are less than 0.25 mas/yr and almost all are less than 0.5 mas/yr.}
    \label{fig:pmdists}
\end{figure}

We show in \autoref{fig:plxdifference} the difference in parallax between that measured in Gaia DR2 and the measured \citetalias{Sesar:17} distance modulus converted to parallax. The sample RR Lyrae are too distant and faint to have accurate parallax measurements in Gaia DR2. We therefore use the \citetalias{Sesar:17} RR Lyrae distances throughout our analysis, and their $\approx 3\%$ accuracy was the motivation for using this sample. We do however remove the \nbadplx stars whose Gaia DR2 parallax lies more than $4\sigma$ from that predicted from their distances as measured by \citetalias{Sesar:17}; these are likely to either not be genuine RR Lyrae, or have poor proper motion estimates. We note in passing that our sample has a slightly negative parallax zero point in Gaia DR2 of $-38\uas$. This is similar to the $-29\uas$ found by \citet{Lindegren:18} with the difference likely resulting from the different distribution on the sky of our sources, which are more concentrated towards the Galactic centre than the quasar sample of \citet{Lindegren:18}.

Finally, we remove two RR Lyrae which are clear outliers with respect to their transverse velocity. In \autoref{fig:transverseveldist}, we plot the transverse velocity distribution, and remove the two stars with apparent transverse velocity $>1000\kms$. This very small number of outliers is reassuring: it confirms that nearby contaminants are removed by requiring the Gaia parallax to be consistent with the derived RR Lyrae distance. More distant contaminants would be extremely rare because of the steep density profile of the halo combined with the scarcity of giant stars brighter than the horizontal branch.

Of the original \nrrab RR Lyrae, \nsample remain after the Gaia cross matching. In \autoref{fig:sample_side_on}, we show the distribution of the sample, while in \autoref{fig:pmdists} we show the distribution of the proper motion and proper motion errors. Note that although the Gaia DR2 parallaxes of our sample were not accurate (\autoref{fig:plxdifference}), the proper motions are: the proper motion errors are generally less than 0.25\masyr and almost all smaller than 0.5\masyr. These proper motions are around two orders of magnitude more accurate than Hipparcos. The median error on each individual star of 0.15\masyr corresponds to an error of $14\kms$ at 20\kpc, allowing us to measure kinematics across the entire volume of our sample, provided this error is taken into account.

\begin{figure}
	\includegraphics[width=\columnwidth]{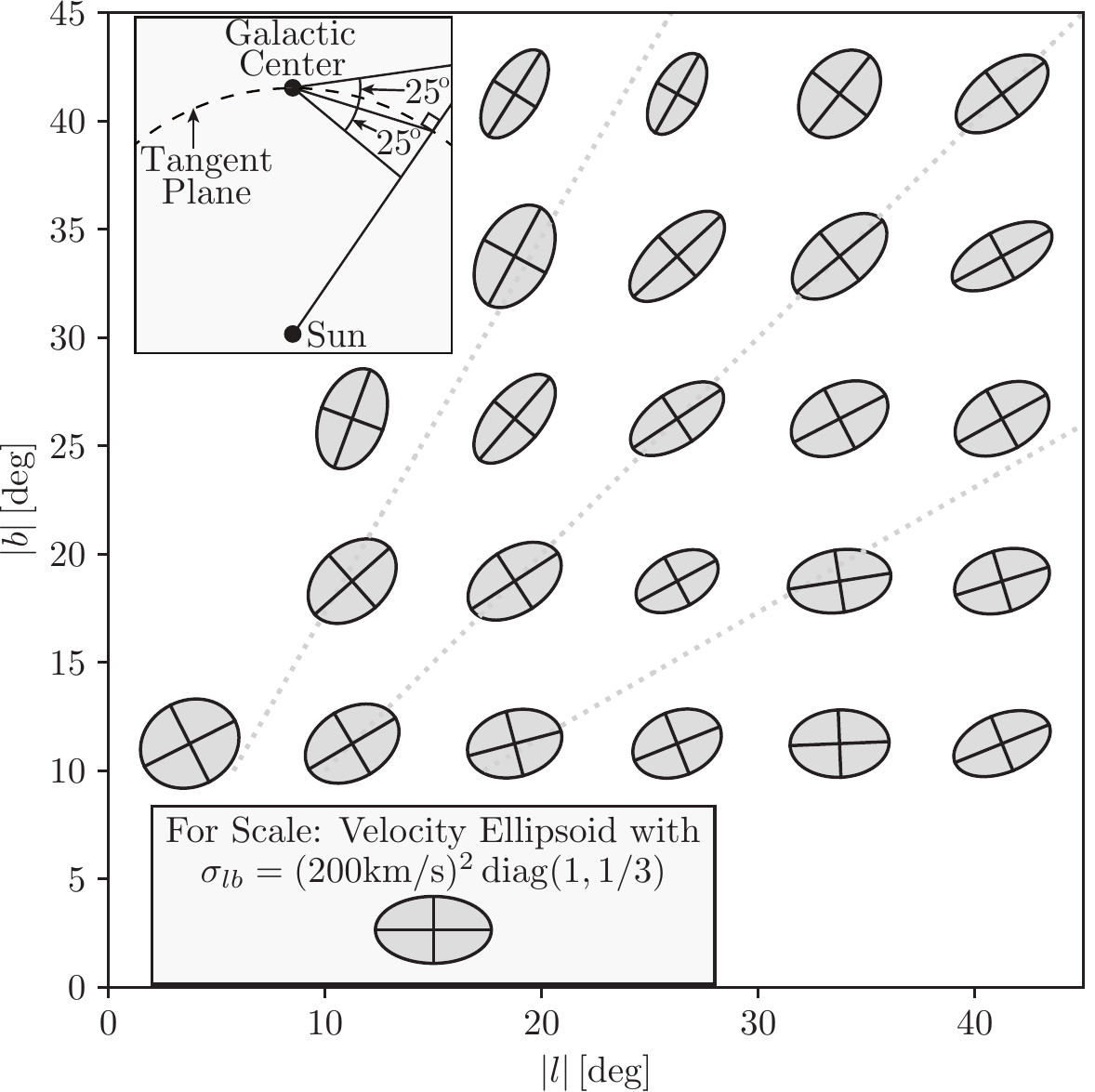}
    \caption{The transverse velocity ellipsoid in the sky plane for stars which are within 25\degr of the tangent plane when projected onto the Galactic plane (we illustrate the stars that would be selected along one sightline by this geometry in the top left inset). This plot is a qualitative indication of the velocity ellipsoid in the meridional plane. The dotted lines are at 30\degr, 45\degr and 60\degr. If the velocity ellipsoid were near spherically aligned we would expect the ellipses to align with these dotted lines, while cylindrical alignment would correspond to near horizontal ellipses. We see that RR Lyrae in the inner halo are strongly radially anisotropic and closer to a spherically aligned velocity ellipsoid that cylindrical. This motivates our choice of spherically sligned coordinates throughout the paper. The bottom inset box shows a velocity ellipsoid with 200\kms semi-major axis for scale. We plot only those bins in $(l,b)$ which contain more than 25 stars.}
    \label{fig:skyplanevelellip}
\end{figure}

As we will see in \autoref{sec:kin}, our selected RR Lyrae in the inner halo are strongly radially anisotropic and have a nearly spherically aligned velocity ellipsoid. To preempt this, and motivate our choices of coordinates and binning, we illustrate the radial anisotropy directly from the data in \autoref{fig:skyplanevelellip}. In making this plot, we selected stars which, when projected onto the Galctic plane, lie within 25\degr of the tangent plane (see figure inset). For small $|l|$ and $|b|$, the transverse velocities of these stars trace the velocities in the meridional plane \ie $(v_l,v_b)\approx(v_{R},v_z)$. In the figure we extend to $|l|$ and $|b|$ values much larger than the strict applicability of this approximation, but we use this figure merely as a clear visual indication directly from the data that the inner halo, as traced by RR Lyrae, has a strongly radially anisotropic nature. On the basis of this figure, we choose to work in spherical coordinates throughout this work, these being more natural for our tracer population than cylindrical coordinates.

\begin{figure}
	\includegraphics[width=\columnwidth]{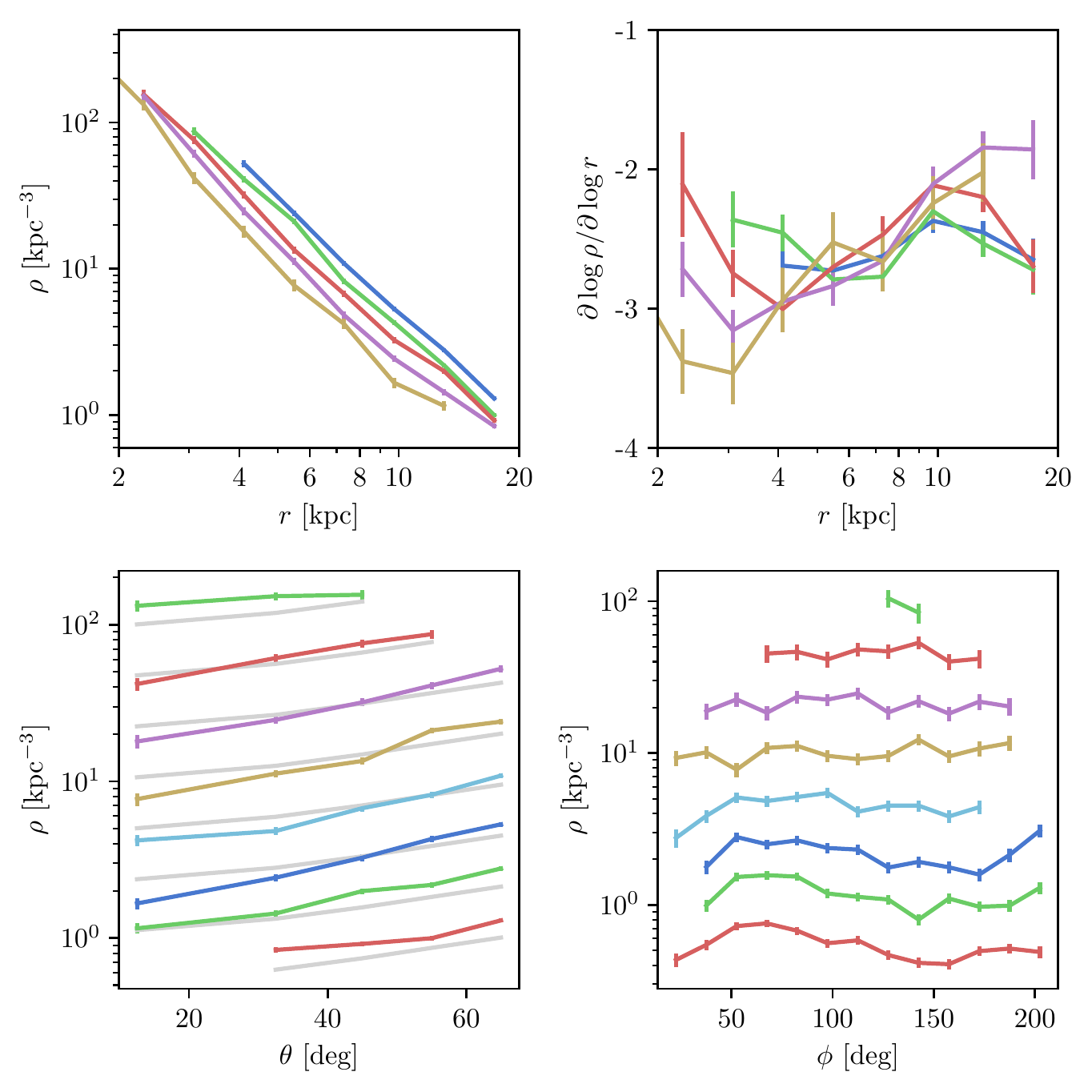}
    \caption{The density of RR Lyrae from \citetalias{Sesar:17} selected as described in \autoref{sec:sample} which appear in our sample. Upper left: The density as a function of radius for each of the 5 bins in $\theta$ (see text for bin details). Because the sample is flattened, the highest density corresponds to the lowest $\theta$ bin. Upper right: The Logarithmic density gradient with line colours corresponding to the same theta bins. Lower left: The density as a function of $\theta$ in each of the logarithmically spaced radial bins (see text for bin details). In light grey we show an ellipsoidal density with $q_{rr}=0.72$ and $\alpha = \partial \log \rho/\partial \log r= -2.6$ for comparison (see \autoref{eq:rho_parametrised}).  Lower right: Density as a function of azimuthal angle, $\phi$, in the same radial bins. $\phi$ is defined so that it is $180\deg$ in the direction of the Sun.}
    \label{fig:density_structure}
\end{figure}

In \autoref{fig:density_structure}, we examine the density of the sample. Throughout this work, we work in spherical bins centred on the Galactic centre where $\theta$ is $90\degr$ in the Galactic plane and $0\degr$ towards the North Galactic Pole. We use \nrbins logarithmically spaced bins between \rmin and \rmax in radius, and bins in $\theta$ with edges at \thetabinedges. These bins extend over all azimuthal angles. To convert from number counts in these bins to densities, we must account for selection effects. Our sample covers only regions of each bin because of the selection cuts (i)-(iv) described above \ie the removal of RR Lyrae with ${\rm dec}<-29\degr$, $|b|<10\degr$, near globular clusters, or in the plane of the Sagittarius stream. One could compute the volume of each bin observed, however because the halo is radially concentrated, doing so would introduce a bias when, for example the inner or outer part of a bin was missing. To account for this, we simulate stars drawn from an ellipsoidal power-law profile:
\begin{equation}
\rho \propto \left[ R^2 + \frac{z^2}{q_{rr}^2} \right]^{\alpha/2}
\propto r^{\alpha} \left[ \sin^2 \theta + \frac{\cos^2 \theta}{q_{rr}^2}  \right]^{\alpha/2} ~.
\label{eq:rho_parametrised}
\end{equation}
where $(R,z)$ are galactocentric cylindrical coordinates, and $(r,\theta)$ galactocentric spherical coordinates. In this estimate of selection fraction, we use gradient $\alpha=\partial \log \rho/\partial \log r= -2.6$ and flattening $q_{rr}=0.72$ which we found best fit the sample overall. \citet{Hernitschek:18} investigated the structure of the \citetalias{Sesar:17} sample beyond 20\kpc and found a similar flattening of the RR Lyrae of $q_{rr}=0.8$ at the 20\kpc inner edge of their sample. We then compute the fraction of simulated stars in each bin which pass the selection cuts described above and use this fraction to correct the number of counts in each bin (these fractions for each bin are plotted later in \autoref{fig:mwkino_phys}). The have checked that our densities are not sensitive to the values used for $\alpha$ and $q_{rr}$, and they only used in this volume correction. Because almost all the selected stars pass our Gaia DR2 selection cuts, regardless of position, we do not simulate these. Instead we consider only the much more important cuts described above as (i)-(iv). 

In what follows, we perform non-parametric modelling of the tracer population of RR Lyrae. We use the parameterisation in \autoref{eq:rho_parametrised} only to compute the observed fraction of each bin. We have found that the results are not sensitive to the details of the parametrisation used in the selection fraction because it is only important that its variation is approximately correct over a bin, and not globally. We also also use this simulation of the selection function to remove poorly sampled bins: Any bin where less than 30\% of the Monte Carlo simulated stars pass the selection cut is removed. This affects the bins which lie at low Galactic latitude due to the $|b|>10\deg$ selection, and one distant bin along the Galactic minor axis which is heavily contaminated by the Sagittarius stream.

We then compute the density using this selection fraction as a correction. In the upper left panel of \autoref{fig:density_structure}, we show the density as a function of galactocentric radius in each $\theta$ bin. It is noteworthy that, as predicted by \citet[][Fig. 1]{PerezVillegas:16}, these densities appear to smoothly connect the RR Lyrae densities measured near the Sun to those measured in the Bulge. In the upper right panel, we show the logarithmic density gradient computed from these densities using a finite difference scheme (see \autoref{sec:jeans} for details). The logarithmic gradient is generally between -3 and -2. In the lower left panel, we show the same density information, but plotted as a function of $\theta$, each line corresponding to a radial bin. We also plot the parameterised density used in estimating the selection fraction. Finally, in the lower right panel, we show the variation of density with azimuthal angle $\phi$ within each of the radial bins. Notice that we have good coverage of azimuthal angle, and that variations with azimuth are relatively small. The variations with azimuth are larger  in our outermost radial bins, and this likely reflects the non-axisymmetric nature of the halo as traced by RR Lyrae in Gaia by \citet{Iorio:18}. We discuss the impact of this in \autoref{sec:nonaxi}.

\section{The Kinematics of the Galactic Halo Traced by RR Lyrae}
\label{sec:kin}

Having previewed the halo kinematics in \autoref{fig:skyplanevelellip}, we now proceed to the more formal analysis of the kinematics of our sample, before applying the Jeans equations to these kinematics in \autoref{sec:jeans}.

For each star in the sample, 5 of the 6 phase-space coordinates are available \ie we have measurements of the 3D position and the transverse velocity of each star, but the radial velocities are unobserved. We consider two methods for recovering the intrinsic kinematics of our sample: 
\begin{inparaenum}
\item a generative method assuming that the velocities are normally distributed, and
\item a method which assumes only that the dispersion tensor is constant with azimuthal angle $\phi$ inspired by \citet[][hereafter \citetalias{Dehnen:1998}]{Dehnen:1998}. 
\end{inparaenum}
The advantages of each method are that (i) is statistically efficient, while (ii) measures the velocity moments while making no assumptions about the form of the velocity distribution.
As we shall show, practically both methods recover the kinematics of observations of mock halos extremely well, and both agree on their reconstruction of the intrinsic kinematics of the sample.

\subsection{The Intrinsic Kinematics Assuming Gaussian Velocities}
\label{sec:gauss}

If we assume that the velocities are Gaussian then, at each point in space, the distribution of velocities is 
\begin{dmath}
f(\mathbf{v})	= \frac{1}{\sqrt{ (2\pi)^3 |\Sigma|}} \exp\left[ -\frac{1}{2} (\mathbf{v}-\mean{\mathbf{v}})^\intercal  \mathbf{\Sigma}^{-1}  (\mathbf{v}-\mean{\mathbf{v}}) \right]
= \frac{1}{\sqrt{ (2\pi)^3 |\Sigma|}} \exp(-Q/2)
\end{dmath}
where $\mathbf{v}$ is the velocity in spherical coordinates $(r,\theta,\phi)$ \ie $\mathbf{v}=(v_r,v_\theta,v_\phi)$, and ${\Sigma}$ is the velocity dispersion tensor in spherical coordinates:
\begin{dmath}
\mathbf{\Sigma}  \hiderel{=}
\begin{bmatrix}
 	\sigma_{rr}^2 & \sigma_{r\theta}^2 & \sigma_{r\phi}^2 \\
 	\sigma_{r\theta}^2 & \sigma_{\theta\theta}^2 & \sigma_{\theta\phi}^2 \\
 	\sigma_{r\phi}^2 & \sigma_{\theta\phi}^2 & \sigma_{\phi\phi}^2 
\end{bmatrix} ~.
\end{dmath}
The off diagonal elements depend on the alignment of the velocity ellipsoid and may be negative. When plotting the velocity dispersion tensor we therefore plot \citepalias[\eg][]{Dehnen:1998}
\begin{equation}
	\sigma_{ij}'={\rm sign} (\sigma_{ij}) \sqrt{| \sigma_{ij}^2 |} ~.
\end{equation}

To find the resultant velocity distribution on the sky, we rotate this coordinate system into cartesian coordinates aligned with $\hat{d},\hat{l},\hat{b}$ \ie into $\mathbf{v'}=(v_d,v_l,v_b)$. Denoting this transformation as $\mathbfss{R}$, then $\mathbf{v'}=\mathbfss{R}  \mathbf{v}$ and the quadratic form $Q$ becomes
\begin{dmath}
	Q=(\mathbf{v'}-\mean{\mathbf{v'}})^\intercal  \mathbfss{R}  \mathbf{\Sigma}^{-1}  \mathbfss{R}^\intercal  (\mathbf{v'}-\mean{\mathbf{v'}})
	 =(\mathbf{v'}-\mean{\mathbf{v'}})^\intercal  \mathbf{\Lambda}^{-1}  (\mathbf{v'}-\mean{\mathbf{v'}})
\end{dmath}
where $\mean{\mathbf{v'}}=\mathbfss{R}  \mean{\mathbf{v}}$ and $\mathbf{\Lambda}=\mathbfss{R}  \mathbf{\Sigma}  \mathbfss{R}^\intercal$. The rotation $\mathbfss{R}$, between spherical coordinates and $(\hat{d},\hat{l},\hat{b})$ is given explicitly in \citet[][eqns. A1-A4]{Ratnatunga:1989} and we do not repeat it here.

Marginalising over the unobserved radial velocity $v_d$ provides the probability distribution of transverse velocities, $\mathbf{v}_\perp=(v_l,v_b)$. This is a two-dimensional multivariate Gaussian:
\begin{equation}
f(\mathbf{v}_\perp) = \frac{1}{2\pi\sqrt{|\mathbf{\Lambda}_\perp|}} \exp\left[ - \frac{1}{2} (\mathbf{v}_\perp - \mean{\mathbf{v}}_\perp)^\intercal  \mathbf{\Lambda}_\perp^{-1}  (\mathbf{v}_\perp - \mean{\mathbf{v}}_\perp) \right] 
\label{eq:skygaussian}
\end{equation}
where $\mathbf{\Lambda}_\perp$ are the components of $\mathbf{\Lambda}$ without the $\hat{d}$ direction \ie
\begin{dmath}
\mathbf{\Lambda}_\perp = 
	\begin{bmatrix}
		\Lambda_{ll} & \Lambda_{lb} \\
		\Lambda_{lb}  & \Lambda_{bb} 
	\end{bmatrix} ~,
\end{dmath}
and $\mean{\mathbf{v}}_\perp = \mathbfss{R}_\perp \mean{\mathbf{v}} $ where $\mathbfss{R}_\perp$ is the rotation matrix $\mathbfss{R}$ without the $\hat{d}$ row \ie it is a $(2\times 3)$ matrix.

Because every star has a different position, the projection of the velocity ellipsoid is different for every star. If we then assume that the velocity ellipsoid is constant in each of our bins in $(r,\theta)$, then this allows us to recover the velocity ellipsoid without measurements of the radial velocities. For example, in bins near the Galactic plane, for stars that are in front or behind the Galactic centre, $v_l$ measures the velocity in the $\hat{\phi}$ direction, while when tangent to the Galactic centre $v_l$ measures the velocity in the $\hat{r}$ direction.

For each star, the likelihood of measuring $\mathbf{v}_{\perp}$ is given by the convolution of \autoref{eq:skygaussian} with the error in transverse velocities. This error, when significant, is dominated by the uncertainty in the Gaia proper motions. Denoting the measurement covariance as $\mathbfss{S}_{\perp}$ then the likelihood of measuring $\mathbf{v}_\perp$ is
\begin{dgroup}
\begin{dmath*}
\mathcal{L}(\mathbf{v}_\perp)	\propto \int {\exp\left[ - \frac{1}{2} (\mathbf{v'}_\perp - \mean{\mathbf{v}}_\perp)^\intercal  \mathbf{\Lambda}_\perp^{-1}  (\mathbf{v'}_\perp - \mean{\mathbf{v}}_\perp) \right]} \times {\exp\left[ - \frac{1}{2} (\mathbf{v}_\perp - \mathbf{v'}_\perp)^\intercal  \mathbfss{S}_\perp^{-1}  (\mathbf{v}_\perp - \mathbf{v'}_\perp) \right] \, d\mathbf{v}'_\perp}
\label{eq:convolgaussian}
\end{dmath*}
\begin{dmath*}
= \frac{1}{2\pi \sqrt{|\mathbfss{C}_\perp|}} \exp\left[ - \frac{1}{2} (\mathbf{v}_\perp - \mean{\mathbf{v}}_\perp)^\intercal  \mathbfss{C}_\perp^{-1}  (\mathbf{v}_\perp - \mean{\mathbf{v}}_\perp) \right]
\end{dmath*}
\end{dgroup}
where $\mathbfss{C}_\perp=\mathbfss{S}_\perp+\mathbf{\Lambda}_\perp$.

In order to estimate the mean velocity, $\mean{\mathbf{v}}$, and dispersion tensor, $\mathbf{\Sigma}$, we consider the total log likelihood of all measurements:
\begin{dgroup*}
\begin{dmath*}
	\log \mathcal{L}=\sum_i \log \mathcal{L}(\mathbf{v}_{\perp i})
\end{dmath*}	
\begin{dmath*}
= - \frac{1}{2} \sum_i (\mathbf{v}_{\perp i} - \mean{\mathbf{v}}_{\perp i})^\intercal  \mathbfss{C}_{\perp i}^{-1}  (\mathbf{v}_{\perp i} - \mean{\mathbf{v}}_{\perp i}) - \frac{1}{2} \log | \mathbfss{C}_{\perp i} | - \log ( 2 \pi )
\end{dmath*}
\end{dgroup*}
where $\mathbf{v}_{\perp i}$ is the measured transverse velocity of star $i$, $\mathbfss{C}_{\perp i}=\mathbfss{S}_{\perp i}+\mathbf{\Lambda}_{\perp i}=\mathbfss{S}_{\perp i}+\mathbfss{R}_{\perp i}\mathbf{\Sigma} \mathbfss{R}_{\perp i}^\intercal$, and $\mean{\mathbf{v}}_{\perp i} = \mathbfss{R}_{\perp i}  \mean{\mathbf{v}}$.

We wish to estimate the kinematic properties of the population \ie $\mean{\mathbf{v}}$ and $\mathbf{\Sigma}$. To do so, we adopt a Bayesian approach and Markov Chain Monte Carlo sample from the posterior distribution generated from this likelihood together with flat priors on both $\mean{\mathbf{v}}$ and $\mathbf{\Sigma}$. In every bin, the number of stars is so large that the results would be insensitive to prior choice. 

\subsection{The Intrinsic Kinematics For non-Gaussian Velocities}
\label{sec:nongauss}

The reader may be concerned by the assumption of Gaussianity in \autoref{sec:gauss}. To alleviate these fears, in this section we derive estimators for the mean velocity and second dispersion tensor which do not depend on the specific form of the velocity distribution. To do so, we generalise the method of \citetalias{Dehnen:1998}. The key assumption that allowed \citetalias{Dehnen:1998} to recover the intrinsic kinematics from transverse velocities was that velocities and positions were uncorrelated \ie that the velocity distribution did not depend on position. This was a good assumption for the solar neighboorhood sample of Hipparcos stars analysed in that work, but here we have Gaia data on stars across the inner Galaxy. We therefore make a different assumption: that positions and kinematics in \emph{spherical Galactocentric coordinates} are uncorrelated throughout each of our individual bins. Because our bins are of limited extent in $r$ and $\theta$ but extend over all azimuthal angles this assumption corresponds to recovering the velocity moments, despite the missing radial velocity, by using the assumption that kinematics are independent of azimuth.

We define the vector $\mathbf{p}$ to be the transverse velocity measurement of our star in spherical Galactocentric coordinates with zero radial velocity:
\begin{equation}
	\boldsymbol{p} \equiv \mathbfss{R}^\intercal  
	\begin{bmatrix}
	0 \\ v_l \\ v_b	
	\end{bmatrix} ~.
\end{equation}
This measurement results from measurements of a star with velocity $\mathbf{v}$ through
\begin{equation}
\boldsymbol{p} = \mathbfss{A} \mathbf{v}
\label{eq:projontop}
\end{equation}
where $\mathbfss{A}$ is the projection matrix which projects velocities onto $v_r=0$. The projection matrix can be derived from linear algebra to be $\mathbfss{A}=\mathbfss{R}_\perp^\intercal \mathbfss{R}_\perp$ but is more frequently expressed in the form used by \citetalias{Dehnen:1998}
\begin{equation}
\mathbfss{A} = \mathbfss{I}_3 - \mathbfss{R}_{\shortparallel}^\intercal \mathbfss{R}_{\shortparallel} \label{eq:matA}
\end{equation}
where $\mathbfss{I}_3$ is the identity matrix, $\mathbfss{R}_\shortparallel$ is the matrix $\mathbfss{R}$, but with only the rows corresponding to $\hat{d}$, and the $\hat{l}$ and $\hat{b}$ rows zeroed. 
Note that although \autoref{eq:matA} mirrors \citetalias{Dehnen:1998} (and \citealt{Schonrich:11,Schonrich:18} which also use the same method), because $\boldsymbol{p}$ is in Galactocentric velocities and not cartesian $(U,V,W)$ velocities, the matrix $\mathbf{R}_{\shortparallel}$, and therefore $\mathbfss{A}$ are thus concretely quite different. Using this coordinate system together with our binning also circumvents the concerns of \citet{McMillan:09} that the \citetalias{Dehnen:1998} method should not be applied to a significant volume of the Galaxy.

The insight of \citetalias{Dehnen:1998} was that, while \autoref{eq:projontop} obviously cannot be inverted, the mean velocity over each bin can be recovered from
\begin{equation}
\samplemean{\boldsymbol{p}} = \samplemean{\mathbfss{A}} \samplemean{\mathbf{v}}
\end{equation}
where we have used the key assumption that positions and velocities are independent. Then trivially
\begin{equation}
\samplemean{\mathbf{v}} = \samplemean{\mathbfss{A}}^{-1} \samplemean{\boldsymbol{p}} ~.
\label{eq:DB98meanv}
\end{equation}
Similarly the dispersion tensor $\boldsymbol{\sigma}$ can be obtained through the inversion of
\begin{equation}
\samplemean{p'_i p'_k} = \sum_{jl} \samplemean{A_{ij} A_{kl}} \sigma_{jl}
\label{eq:DB98dispersion}
\end{equation}
where $\boldsymbol{p}'=\boldsymbol{p} - \mathbfss{A} \samplemean{\mathbf{v}}$. In \autoref{sec:jeans}, we will apply the Jeans equations to the second-moments of the velocity distribution, $\langle v_i v_j \rangle$. We estimate these more directly from the inversion of\footnote{Note that because $\samplemean{p_i p_k}=\samplemean{p_k p_i}$ then from \autoref{eq:DB98moments} $\langle v_j v_l \rangle$ is also symmetric. One can use these symmetries to reduce the number of equations for the computer to solve, as \citetalias{Dehnen:1998} did, but for clarity we leave \autoref{eq:DB98dispersion} and \autoref{eq:DB98moments} unaltered.}
\begin{equation}
\samplemean{p_i p_k} = \sum_{jl} \samplemean{A_{ij} A_{kl}} \langle v_j v_l \rangle ~.
\label{eq:DB98moments}
\end{equation}
Observational errors in transverse velocity do not affect the mean velocity (\autoref{eq:DB98meanv}), but must be accounted for when estimating the dispersion (\autoref{eq:DB98dispersion}) and moments (\autoref{eq:DB98moments}). To do so, we subtract in quadrature the variance caused by these errors (Eqn 18 of \citetalias{Dehnen:1998}). We have tested the accuracy of this correction using mock data.

\begin{figure*}
	\includegraphics[width=0.9\textwidth]{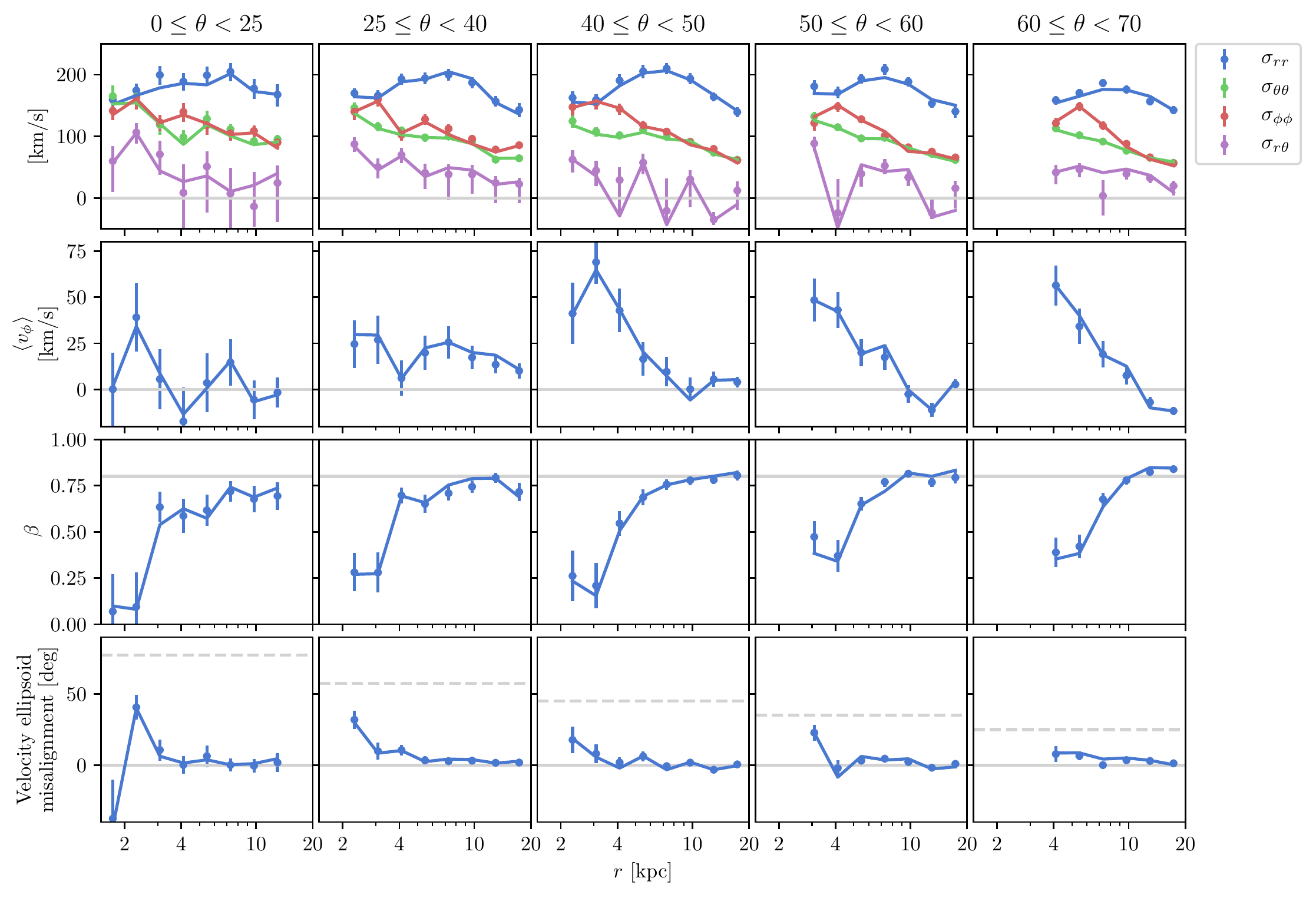}
    \caption{The kinematics of the RR Lyrae sample computed from the 5D phase space coordinates using the methods described in \autoref{sec:kin}. Results assuming a Gaussian velocity distribution  (\autoref{sec:gauss}) are shown as the points with error bars. Results using the \citetalias{Dehnen:1998} method (\autoref{sec:nongauss}) are shown as the line (the errors of this method are similar to those of the Gaussian velocity distribution, being on average 2\% larger). Each column shows the kinematics of a different angular bin. The upper row of figures shows the elements of the velocity dispersion tensor coloured as the legend. The second row shows the rotation with the zero rotation line as light grey. The third row shows the radial anisotropy parameter $\beta\equiv1 - (\sigma_{\theta\theta}^2+\sigma_{\phi\phi}^2)/(2\sigma_{rr})$, we also show $\beta=0.8$ in grey to guide the eye. The lowest row shows the misalignment of the velocity ellipsoid from spherical. Spherical alignment corresponds to the solid grey line and cylindrical alignment to the dashed line.}
    \label{fig:mwkino}
\end{figure*}

\begin{figure*}
	\includegraphics[width=0.85\textwidth]{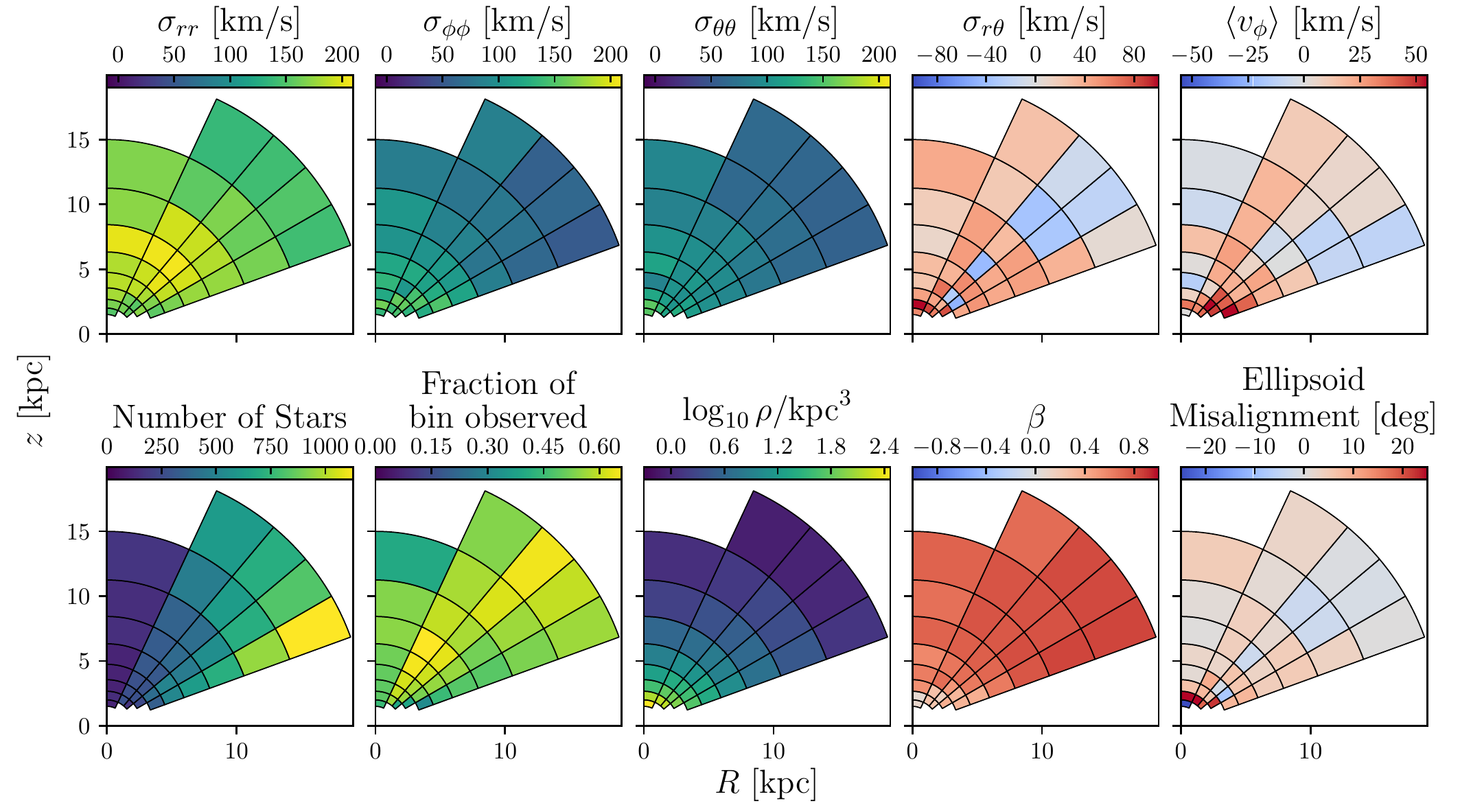}
    \caption{The azimuthally averaged sample kinematics plotted in physical space. The kinematic data is plotted in less intuitive manner, but with statistical errors in \autoref{fig:mwkino}. Upper row: the kinematics \ie the elements of the dispersion tensor $\sigma_{rr}$, $\sigma_{\phi\phi}$, $\sigma_{\theta\theta}$ and $\sigma_{r\theta}$  and the rotation $\azmean{v_\phi}$. Lower row: the number of stars in each bin and the completeness of each bin computed according to \autoref{sec:sample}. These are combined to compute the azimuthally averaged density in the third plot. The final two plots show the radial anisotropy parameter $\beta$ and the misalignment of the velocity ellipsoid from spherical.}
    \label{fig:mwkino_phys}
\end{figure*}

\begin{figure}
\centering
	\includegraphics[width=0.8\columnwidth]{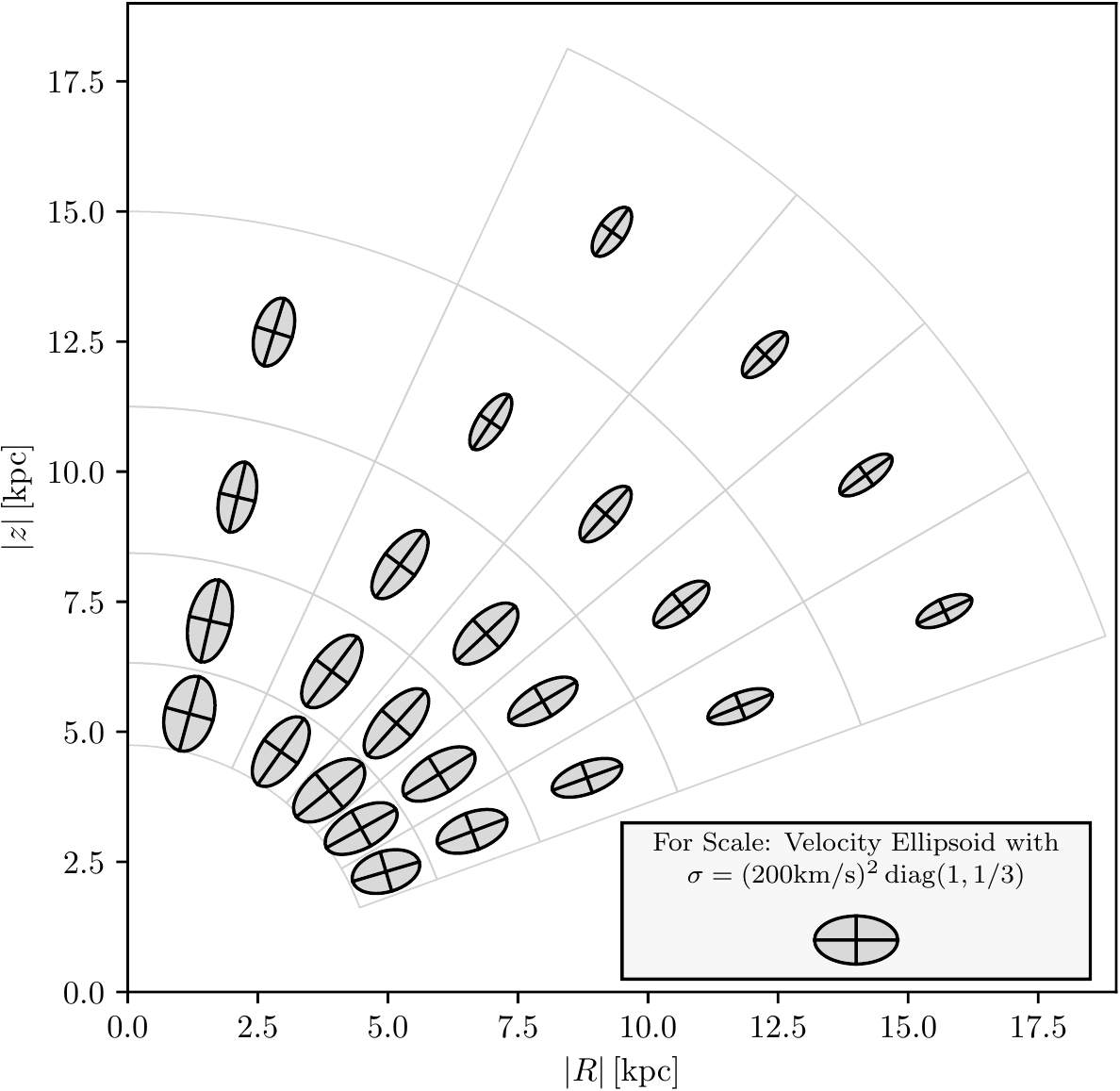}
	\includegraphics[width=0.8\columnwidth]{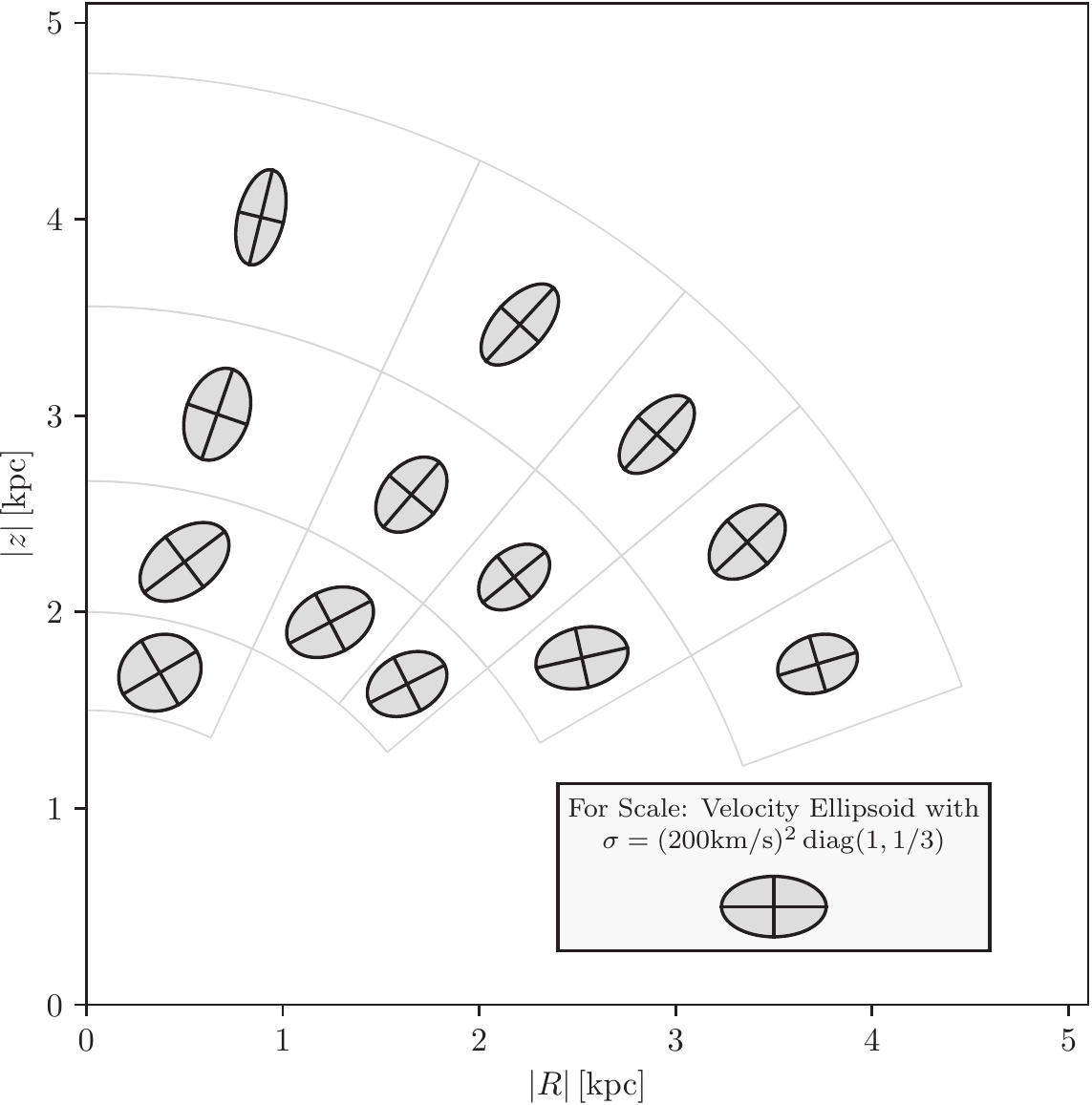}
    \caption{The velocity ellipsoid in the meridonal plane. To maintain roughly equal stars per bin we use a logarithmic binning in radius and therefore split this figure to maintain visibility: the upper panel shows the kinematics between 5\kpc and 20\kpc, and the lower panel shows a zoom in on the region between 1.5\kpc and 5\kpc. The scale of each ellipsoid is shown in the inset.}
    \label{fig:mwkino_vel_ellip}
\end{figure}

\subsection{Testing the Kinematic Reconstruction}

To test the kinematic measurements in \Autoref{sec:gauss,sec:nongauss} (and later our reconstruction of the force field and dark matter halo properties) we have constructed a series of mock halos. These were constructed in the potential of the dynamical models of \citet[][hereafter \citetalias{Portail:17}]{Portail:17}. These made-to-measure models were fitted to a range of data on the bar, bulge and inner Galaxy, while simultaneously matching the rotation curve and stellar surface density near the Sun. They however have no stellar halo, so, to construct our mock halos, we used dark matter halo particles as test particles from which to construct a stellar halo. These particles were then selected with a weighting by energy to have a $\sim r^{-3}$ profile similar to our RR Lyrae sample, and by orbital radial extent to have a similar radial anisotropy. The details of this process are described in \autoref{sec:mockconstruction}, and the \citetalias{Portail:17} models are described in more detail in \autoref{sec:dmmodels} where we use their baryonic part as our fiducial baryonic model.

We have applied the same code and methods on samples generated from these mocks with the results summarised in \autoref{sec:mockkinematics}. To avoid interrupting the flow of the paper, we relegate these tests of our methods to this appendix. Here we draw attention to the comparison of the two kinematic reconstruction methods in \Autoref{sec:gauss,sec:nongauss} on the mock halos. This is shown in \autoref{fig:mockkino} without considering the selection function, and in \autoref{fig:mockkinocut} with the selection function. When the survey is spatially complete, both perform equally well, but the method inspired by \citetalias{Dehnen:1998} performs slightly better when the mocks are folded through the selection function. Both methods give very similar results on the real halo (see \autoref{fig:mwkino}). On the basis of the slightly better recovery of the kinematics of the mock halo by the \citetalias{Dehnen:1998} based method (\autoref{sec:nongauss}), we decided to use this as our fiducial method of reconstructing the intrinsic kinematics from the transverse velocities. 

\subsection{The Measured Kinematics of RR Lyrae in the Inner Halo}

In \autoref{fig:mwkino}, we show the resultant measured kinematics of the sample of halo RR Lyrae. In \Autoref{fig:mwkino_phys,fig:mwkino_vel_ellip}, we show the same data in a more physically informative manner: \autoref{fig:mwkino_phys} shows the kinematics plotted in physical space, while \autoref{fig:mwkino_vel_ellip} shows the measured velocity ellipsoid in the meridional plane.

Several features are noteworthy in these measured kinematics:
\begin{enumerate}
\item The dispersion tensor displays near spherical alignment, tilting towards cylindrical only in the innermost regions. This spherical alignment has been measured in local samples previously \citep{Smith:09tilt,Bond:10,Evans:16,Posti:18ellipsoid}, but here we see that it is close to spherically aligned over the entire range from $4\kpc$ to $20\kpc$. This near spherical alignment does not necessarily mean that the potential must be spherical \citep{Evans:16}, although in many cases it is likely to be \citep{An:16}. As we shall see in \autoref{sec:forcefield}, the potential does appear nearly spherical in the Milky Way. 
\item The RR Lyrae in the halo have a high radial anisotropy of $\beta\approx0.8$. This decreases inside 5\kpc, but remains above $\beta\approx0.25$ even in these inner regions. The $\beta\approx0.8$ measured at solar galactocentric radii is slightly higher than the $\beta=0.7$ measured in the overall halo locally \citep{Smith:09,Bond:10}, but as \citet{Belokurov:18final} show, the local anisotropy of the stellar halo depends strongly on metallicity. Our result of $\beta\approx 0.8$ for RR Lyrae, which are likely to be drawn from the bulk of the halo metallicity distribution at $\feh\gtrsim-2$, agrees with the \citet{Belokurov:18final} measurements at these metalicites. They argue that the extreme anisotropy of these higher metallicity halo stars, which have higher anisotropy than the $\feh<-1.7$ stars in their sample, can be most easily explained by a large fraction of the inner halo forming by the accretion of a massive satellite \citep[see also][]{Deason:18,Helmi:18final}. Here, we see the wider view that the entire inner halo is strongly radial anisotropic. Note that both features (i) and (ii) could be qualitatively anticipated directly from the data in \autoref{fig:skyplanevelellip}. Our kinematics are also in qualitative agreement with the recent 3D kinematic measurements of \citet{Bird:18} who measured $\beta\approx0.85$ inside 20\kpc for metal rich halo stars by combining LAMOST giants with Gaia DR2 proper motions.
\item In the outer regions, beyond 10\kpc, the halo has mild counter rotation of $\sim 10\kms$. This measured outer rotation depends on the assumed velocity of the Sun, but this value is relatively well constrained by the proper motion of Sagittarius A*, assuming that the black hole is at rest with respect to the Galaxy. The mild counter rotation at 20\kpc galactocentric distance is at a similar level to that seen previously in diverse samples (\citealt{Beers:12,Kafle:17,Helmi:18final}, although care must be taken: \citealt{Fermani:13}). It likely results from the halo being built by a limited number of large mergers fragments at these radii \citep{Koppelman:18}, or the accretion of a single large SMC sized object \citep{Helmi:18final,Belokurov:18final}.
\item The halo at solar radii and inside has mild rotation. This rotation could reflect, in part, the accretion history of the inner halo. However, it is interesting that the shape of the rotation profile matches extremely well the rotation of the mock halo (see Fig. \ref{fig:mockkino}). The mock halos were constructed from an initially isotropic dark matter halo, and the stars were selected without reference to rotation direction. Instead, the mock halos acquired their rotation by transfer of angular momentum from the bar \citep[\eg][]{Athanassoula:03}. In the bulge, \citet{PerezVillegas:16} compared the kinematics of the RR Lyrae to barred models, finding that the rotation there could be matched by the spin up of an initially non-rotating population. Here however, the observed rotation is somewhat larger than the rotation in the mock. Whether this difference in level of rotation is a result of differing halo properties between our mock halo and the Milky Way, or whether this reflects the formation history is unclear. In addition if there are a significant fraction of in-situ, thick disk origin, stars in these bins, this would also increase the rotation profile \citep{Haywood:18}. It is worth noting that this rotation has been observed locally previously \citep[\eg][]{Deason:17}, and a similar result in LAMOST K giants was recently found using Gaia DR2 by \citet{Tian:19}. 
\end{enumerate}

\section{Measuring the Galactic Force Field}
\label{sec:jeans}

In this section, we apply the Jeans equations to the measured halo RR Lyrae kinematics in order to measure the acceleration field in the inner halo of the Galaxy.  We first derive discretised versions of the Jeans equations in spherical coordinates (\autoref{sec:jeanseqn}). The novelty of this section is that we derive azimuthally averaged versions of the Jeans equations which do not assume axisymmetry either of the tracer or potential. From the subsequent application of discretised versions of these equations, we measure the azimuthal average of the gravitational acceleration field of the Galaxy in \autoref{sec:forcefield}.
 
\subsection{The Azimuthally Averaged Jeans Equations}
\label{sec:jeanseqn}

We begin from the collisionless Boltzmann equation in spherical coordinates $(r,\theta,\phi)$ where $\phi$ is the azimuthal angle and $\theta$ is the angle from the $z$-axis (BT87, P4-3):
\begin{multline}
	\frac{\partial f}{\partial t} + v_r \frac{\partial f}{\partial r} + \frac{v_\theta}{r} \frac{\partial f}{\partial \theta} + \frac{v_\phi}{r \sin \theta} \frac{\partial f}{\partial \phi} \\
	+ \left( \frac{v_\theta^2 + v_\phi^2}{r} - \frac{\partial \Phi}{\partial r} \right) \frac{\partial f}{\partial v_r} 
	+ \frac{1}{r} \left( v_\phi^2 \cot \theta - v_r v_\theta -\frac{\partial \Phi }{\partial \theta} \right) \frac{\partial f}{\partial {v_\theta}} \\
	-\frac{1}{r} \left[ v_\phi (v_r + v_\theta \cot \theta) + \frac{1}{\sin \theta} \frac{\partial \Phi}{\partial \phi}\right] \frac{\partial f}{\partial v_\phi} = 0
	\label{eq:cbe}
\end{multline}
where $f$ is the distribution function, $\Phi$ is the gravitational potential, and $(v_r,v_\theta,v_\phi)$ are the velocities in the $(\hat{r},\hat{\theta},\hat{\phi})$ directions respectively. Multiplying by $v_r$ and integrating over velocity space gives
\begin{multline}
\frac{\partial (n \mean{v_r}) }{\partial t} 
+ \frac{\partial (n \mean{v_r^2} )}{\partial r} 
+ \frac{1}{r} \frac{\partial (n \mean{v_r v_\theta})}{\partial \theta} 
+ \frac{1}{r \sin \theta} \frac{\partial (n \mean{v_r v_\phi} )}{\partial \phi} \\
+ \frac{n}{r}\left[2\mean{v_r^2}  - \mean{v_\theta^2} - \mean{v_\phi^2} 
+ \mean{v_r v_\theta} \cot \theta\right] 
= -n \frac{\partial \Phi}{\partial r}
\end{multline}
where $n=\int f \, dv_r \, dv_\theta \, dv_\phi$ is the density the tracer and a bar denotes the distribution function weighted mean of the tracer at each point in space \ie $n \mean{x}=\int f x \, dv_r \, dv_\theta \, dv_\phi$.  If we integrate over all azimuthal angles $\phi$ then the first and fourth terms vanish for a system in equilibrium rotating rigidly about the $\hat{z}$ axis. This leaves 
\begin{multline}
\frac{\partial \rho \azmean{ v_r^2 } }{\partial r} 
+ \frac{1}{r} \frac{\partial  \rho  \azmean{v_r v_\theta}}{\partial \theta} 
+ \frac{\rho}{r}\left[2\azmean{v_r^2}  - \azmean{v_\theta^2} - \azmean{v_\phi^2}
 + \azmean{v_r v_\theta} \cot \theta\right] \\
= - \rho \bigazmean{\frac{ \partial \Phi}{\partial r}} \label{eq:jeansr}
\end{multline}
where $\rho$ is the azimuthally averaged density, and $\azmean{}$ denotes the density weighted azimuthal average of a quantity \ie $\rho \equiv \frac{1}{2\pi}\int f \, dv_r \, dv_\theta \, dv_\phi \, d\phi$ and $\azmean{x} \equiv \frac{1}{2\pi\rho} \int f x \, dv_r \, dv_\theta \, dv_\phi \, d\phi$.

\begin{figure*}
	\includegraphics[width=\textwidth]{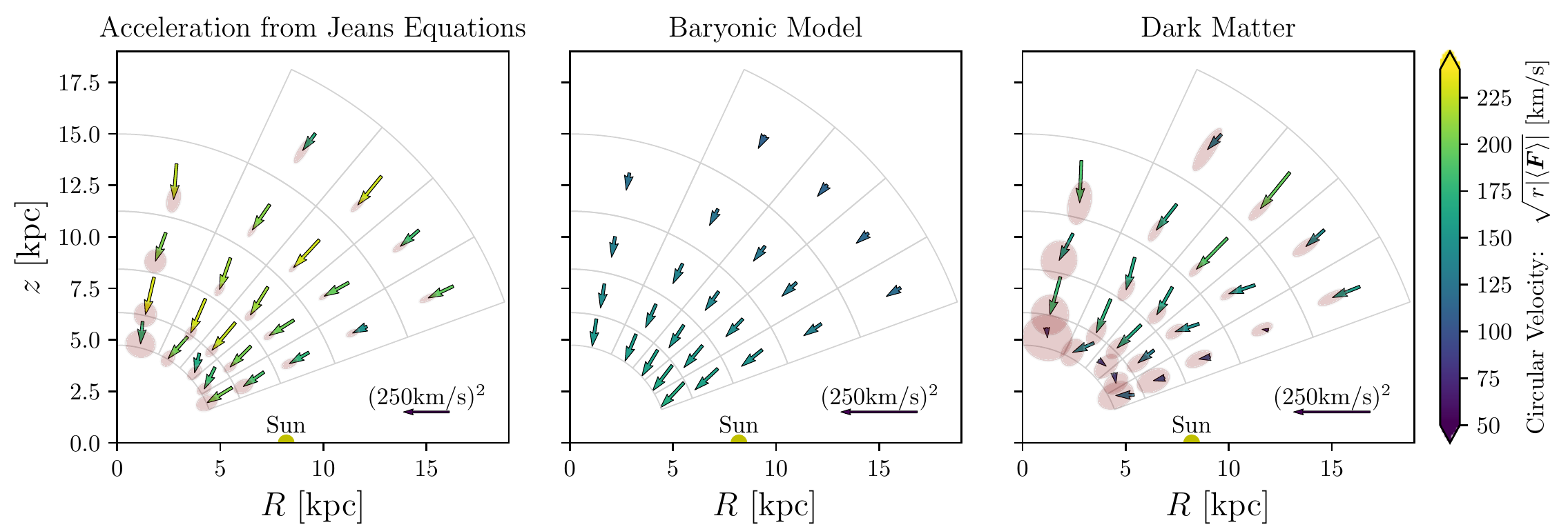}
    \caption{In the left plot, we show the azimuthally averaged gravitational acceleration field measured by applying the discretised Jeans equations (\autoref{eq:discrete_jeans}) to the kinematics of the inner halo measured in \autoref{sec:kin}. To keep the arrows of similar size we plot $r\azmean{\boldsymbol{F}}$ in $({\rm km}/{\rm s})^2$, which is analogous to the square circular velocity used in-plane. Arrows are coloured by the square root of this quantity. Statistical errors are shown by the red ellipses: the arrowheads can lie anywhere within the ellipses, which have semiaxis lengths of $1\sigma$. The inset arrow shows the scale and the Sun is a yellow sphere. The middle panel shows the acceleration field generated by our fiducial baryonic model. The right panel shows the resultant acceleration generated by the dark matter only, once this baryonic contribution is subtracted. We plot only the forces at galactocentric radius larger than 5\kpc; inside this radius, the errors in direction are too large for the accelerations to be usefully individually plotted.}
    \label{fig:jeans_force_field}
\end{figure*}

Performing the same procedure of multiplying by $v_\theta$ integrating over both velocity space and azimuth gives
\begin{multline}
\frac{\partial \rho \azmean{ v_r v_\theta } }{\partial r} 
+ \frac{1}{r} \frac{\partial \rho \azmean{ v_\theta^2}}{\partial \theta} 
+ \frac{\rho}{r}\left[3\azmean{ v_r v_\theta} + (\azmean{v_\theta^2} - \azmean{v_\phi^2}) \cot \theta\right] \\
= -\frac{\rho}{r} \bigazmean{\frac{\partial \Phi}{\partial \theta}} \label{eq:jeanstheta} ~.
\end{multline}
This angular Jeans equation, termed the flattening equation by \citet{Bowden:16}, is important because it allows the direction of the gravitational acceleration, and thereby the flattening of the potential and the dark matter, to be measured.

Note that \Autoref{eq:jeansr,eq:jeanstheta} are the same as the axisymmetric Jeans equations in spherical coordinates \citep[\eg][]{deZeeuw:96}, but do not assume axisymmetry, instead taking the density weighted azimuthal average of quantities. This is because the collisionless Boltzman equation (\autoref{eq:cbe}) is linear in $f$, and so, since the same terms vanish when averaging over azimuth as the axisymmetric case, the same equations result but with the moments and potential replaced by their azimuthal average.

We will use \Autoref{eq:jeansr,eq:jeanstheta} in a discretised form. Before discretising, we rewrite them as
\begin{subequations}
\begin{multline}
\azmean{ v_r^2 } \frac{\partial \log \rho }{\partial \log r} 
+ \frac{\partial \azmean{ v_r^2 } }{\partial \log r} 
+ \frac{\partial \azmean{v_r v_\theta}}{\partial \theta}
+ \azmean{v_r v_\theta} \frac{\partial \log \rho}{\partial \theta} \\
+ \left[2\azmean{v_r^2}  - \azmean{v_\theta^2} - \azmean{v_\phi^2}
 + \azmean{v_r v_\theta} \cot \theta\right] 
= - \bigazmean{\frac{\partial \Phi}{\partial \log r}} \label{eq:jeansr_logr} ~,
\end{multline}
\begin{multline}
\azmean{ v_r v_\theta } \frac{\partial \log \rho }{\partial \log r} 
+ \frac{\partial \azmean{ v_r v_\theta } }{\partial \log r} 
+ \frac{\partial \azmean{v_\theta^2}}{\partial \theta}
+ \azmean{v_\theta^2} \frac{\partial \log \rho}{\partial \theta} \\
+ \left[3\azmean{v_r v_\theta} + (\azmean{v_\theta^2} - \azmean{v_\phi^2}) \cot \theta\right]
= - \bigazmean{\frac{\partial \Phi}{\partial \theta}} \label{eq:jeanstheta_logr} ~.
\end{multline}
\label{eq:jeans_logr}
\end{subequations}
These are more appropriate to use as the basis for the discretised equations because, while $\rho$ changes quickly with $r$ (approximately as $\sim r^{-3}$), the logarithmic gradient, ${\partial \log \rho }/{\partial \log r}$, changes slowly.

We measure the kinematics in bins across $(r,\theta)$ and insert these measurements into the discretised Jeans equations. Radially, we use $n_{\log r}=\nrbins$ bins evenly spaced by $\delta \log r$ in $\log r$ between \rmin and \rmax. In elevation, we use \ntheta bins of $\theta$ with edges at \thetabinedges. These choices were made to have roughly the same number of stars in all bins while minimising the discretisation errors. It is necessary to choose a broad two-dimensional binning in order to obtain a large enough number of stars per bin, and therefore accurate force measurements. In order to reassure the reader that any systematic errors introduced by the discretisation are small, we analyse the mock halos, where the true potential and forces are known, with the same bins in \autoref{sec:mockhalos}. 

Denoting the differentials as $\Delta_{\log r}$ and $\Delta_\theta$, we use second order accurate differences apart from at the endpoints. For the evenly spaced grid in $\log r$ this is
\begin{equation}
	\Delta_{\log r\,i,j} (x) =
	 \begin{cases}
	 	(x_{i+1,j} - x_{i,j})/(\delta \log r)\condition{for $i=0$} \\
	 	 (x_{i,j} - x_{i-1,j})/(\delta \log r) \condition{for $=n_{\log r}-1$} \\
	 	 (x_{i+1,j} - x_{i-1,j})/(2 \delta \log r) \condition{otherwise} 
	 \end{cases}
\end{equation}
where we have labeled the grid cells in $\log r$ by $i$ and $\theta$ by $j$.
For the unevenly spaced grid in $\theta$, we use
\begin{equation}
	\Delta_{\theta\,i,j} (x) =
	 \begin{cases}
	 	(x_{i,j+1} - x_{i,j})/(\theta_{j+1}-\theta_{j})~, & \text{for } j=0 \\
	 	 (x_{i,j} - x_{i,j-1})/(\theta_{j}-\theta_{j-1})~, & \text{for } j=n_\theta-1 \\
	 	 (\alpha x_{i,j+1} + \beta x_{i,j} + \gamma x_{i,j-1})~, & \text{otherwise} 
	 \end{cases}
\end{equation}
where $\theta_{j}$ is the mid-point of the $j$th bin in $\theta$ and $\alpha$, $\beta$ and $\gamma$ are chosen to give second order accuracy in the derivative (see \eg \verb|numpy.gradient| documentation). 

Then, at each $(i,j)$ point on our grid, \autoref{eq:jeans_logr} becomes
\begin{dgroup}
\begin{dmath}
  \azmean{ v_r^2 } \Delta_{\log r} (\log \rho) + \Delta_{\log r} (\azmean{ v_r^2 }) + \azmean{ v_r v_\theta } \Delta_\theta(\log \rho) + \Delta_\theta (\azmean{v_r v_\theta})
  + \left[2\azmean{v_r^2}  - \azmean{v_\theta^2} - \azmean{v_\phi^2}
 + \azmean{v_r v_\theta} \cot \theta \right] 
= r \azmean{F_{r}} \label{eq:discrete_jeansr_logr} ~,
\end{dmath}
\begin{dmath}
 \azmean{ v_r v_\theta }\Delta_{\log r} (\log \rho) + \Delta_{\log r} (\azmean{ v_r v_\theta })
+ \azmean{ v_\theta^2 }\Delta_\theta (\log \rho) + \Delta_\theta (\azmean{v_\theta^2})
+ \left[3\azmean{v_r v_\theta}  + (\azmean{v_\theta^2} - \azmean{v_\phi^2}) \cot \theta \right] 
= r \azmean{F_{\theta}} \label{eq:discrete_jeanstheta_logr} ~.
\end{dmath}
\label{eq:discrete_jeans}
\end{dgroup}
where we have ommited the $(i,j)$ subscripts for clarity, and substituted $\azmean{F_{r}}$ and $\azmean{F_{\theta}}$,  the accelerations in the radial and azimuthal directions respectively, for the derivatives of the potential.

\subsection{The Gravitational Force Field of the Inner 20\,kpc of the Galaxy}
\label{sec:forcefield}

With the discretized Jeans equations (\autoref{eq:discrete_jeans}) and the required measurements of the velocity moments in hand, we proceed to measure $\azmean{F_r}$ and $\azmean{F_\theta}$. These accelerations fully specify the density-weighted azimuthally averaged gravitational acceleration field of the Galaxy: $\azmean{\boldsymbol{F}} = \azmean{F_r} \hat{r} + \azmean{F_\theta} \hat{\theta}$. In the left panel of \autoref{fig:jeans_force_field}, we plot these accelerations as vectors. We plot each arrows length to be proportional to $r \vec{\boldsymbol{F}}$, analogous to the square circular velocity in the galactic plane. Because the circular velocity curve of the Milky Way is fairly flat, this has the advantage that the arrows have roughly equal length.  We computed the statistical errors (plotted in pink in \autoref{fig:jeans_force_field}) using 2 methods: (i) computing the 1-$\sigma$ ellipses of the 10,000 bootstrapped resamples of the data each propagated though the entire acceleration calculation and (ii) using this resampling to estimate the 1-$\sigma$ errors on the kinematics, and propagating these linearly though the computation of the forces. Both (i) and (ii) gave very similar errors and so we used the computationally faster (ii) in making the plot. It is immediately clear that the accelerations appear consistent with being nearly radial throughout the volume probed. In the next sections, we perform a more quantitative analysis of the acceleration field and its implications.

\section{The Properties of the Inner Dark Matter Halo}
\label{sec:dmmodels}

We now proceed to subtract models of the baryonic contribution to the forces in order to measure the properties of the dark matter halo. 

As our fiducial baryonic model, we use a slightly modified version of the baryonic part of the model of \citet[][hereafter \citetalias{Portail:17}]{Portail:17}. This model was constructed by using the made-to-measure method \citep{Syer:96,DeLorenzi:07} to adapt a barred N-body model to fit data on the inner Galaxy. Fitted data consisted of the 3D shape of the bulge measured by \citet{Wegg:13}, combined near-infrared star counts from the VVV, UKIDSS and 2MASS surveys \citep{Wegg:15}, and kinematics from the BRAVA \citep{Kunder:12} and ARGOS \citep{Ness:13IV} surveys. The result is a dynamical model that fits a range of data on the central 5\kpc of the Galaxy, which is where the majority of the stars lie, extremely well.

However, it is also important that the model is accurate outside the central 5\kpc. The \citetalias{Portail:17} model uses a local stellar surface density of $38\msun/\kpc^2$ with an exponential scale length of 2.4\kpc, while for ISM it uses $13\msun/\kpc^2$ with a scale length of 4.8\kpc. The scale heights of these components were set to 300\pc and 130\pc. The difference here to \citetalias{Portail:17} is that, while in that work these disks were truncated at 10\kpc, here we do not truncate them. We test the effect of varying the baryonic model, and in particular the disk scale lengths and surface densities, in \autoref{sec:systematics}.

The resultant accelerations from our fiducial baryonic model are shown as the middle panel of \autoref{fig:jeans_force_field}. In the right panel, we subtract these from the measured accelerations to show the accelerations from the dark matter alone. Here the errors are larger, particularly in the central regions where the force from the baryonic component dominates. It is already clear however that the forces are largely radial, meaning that the dark matter potential must be near spherical.

\subsection{Ellipsoidal Fits to the Dark Matter}
\label{sec:dmellipfits}

To quantify the shape of the dark matter potential, we fit an ellipsoidal potential to the inferred acceleration field provided by the dark matter in each radial bin using the ansatz that the dark matter potential is ellipsoidal:
\begin{equation}
	\Phi_{\rm dm} (m) = \Phi_{\rm dm} \left( \left[R^2 + z^2/q_\Phi^2 \right]^{1/2} \right)  \quad\mbox{and}\quad \frac{\partial \Phi_{\rm dm}}{\partial \log m} = V_c^2
	\label{eq:ellippot}
\end{equation}
where $q_\Phi$ is the flattening of the potential, and $V_c$ is the in-plane circular velocity. Taking the derivative of \autoref{eq:ellippot} with respect to $r$ and $\theta$ provides the accelerations. Concretely for each set of radial bins we fit the parameters $V_c$ and $q_\Phi$ to the accelerations:
\begin{dgroup}
\begin{dmath}
\azmean{F_r}_{\rm dm} = - \frac{\partial \Phi_{\rm dm}}{\partial r} {=} - \frac{V_c^2}{r}
\end{dmath}
\begin{dmath}
\azmean{F_\theta}_{\rm dm} =  - \frac{1}{r} \frac{\partial \Phi_{\rm dm}}{\partial \theta} {=} - \frac{V_c^2 r}{2 m^2} \left( q_\Phi^{-2}-1 \right) \sin 2\theta
\end{dmath}
\label{eq:discrete_jeans}
\end{dgroup}
Because we fit these force measurements at a constant radius $r$, and not-constant ellipsoidal radius $m$, in principle these forces involve a term of order $O( [1-q_\Phi^{-2}] \Phi_{\rm dm}'' )$, which we neglect. In practice, we find near spherical potentials, making this term small. Furthermore, our tests on the mock halo which has $q_\rho\approx 0.8$ (see \autoref{fig:mock_ellipdmhalo}) show that we accurately recover the profiles of $\qpot$ and $V_c$ in this case.

\begin{figure}
	\includegraphics[width=\columnwidth]{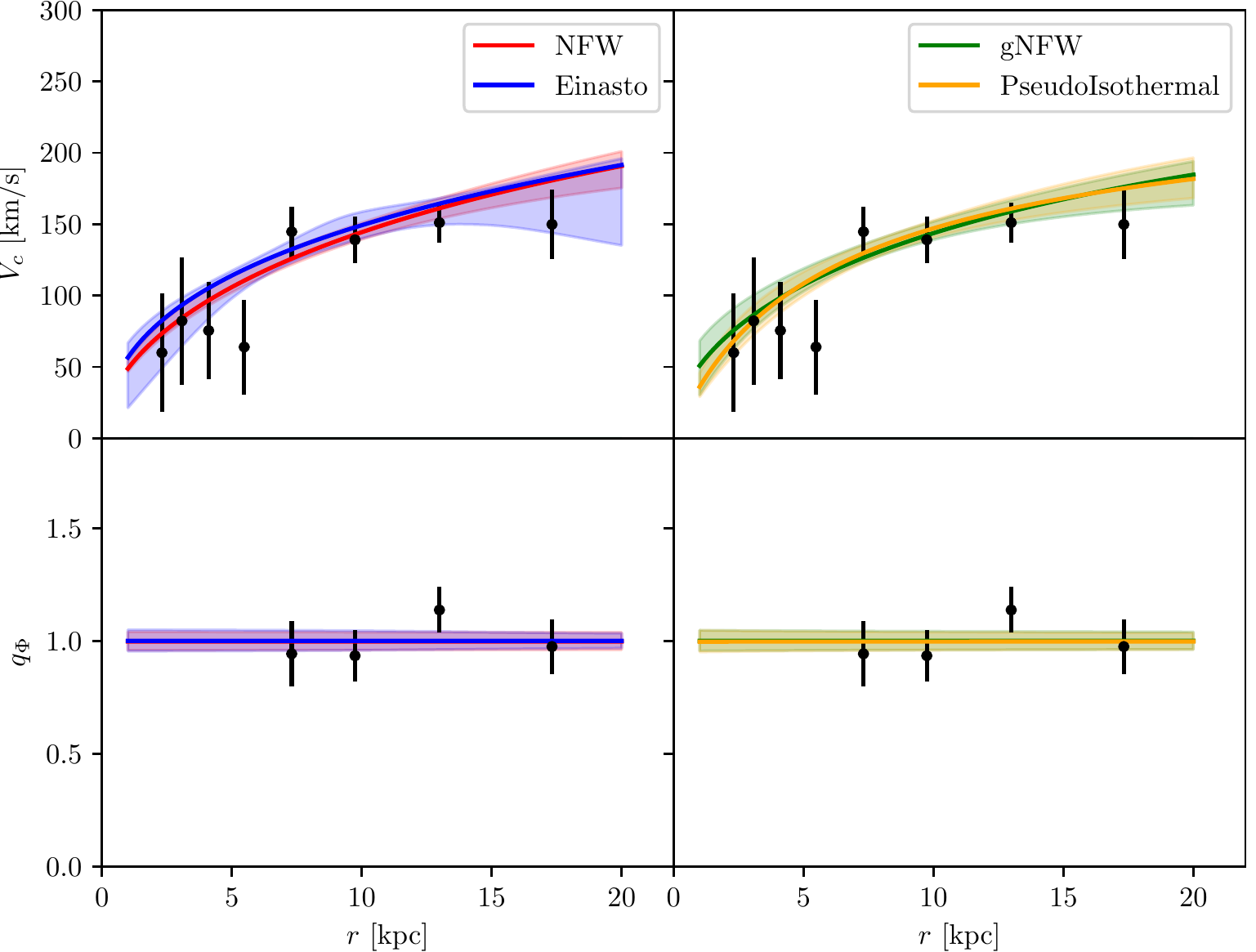}
    \caption{The black data points with errors are the results of fitting an axisymmetric spheroidal potential to each radial bin. In the upper panels, we show the circular velocity, $V_c$, as a function of radius, while in the lower panels we show the flattening of the potential, $q_\Phi$. We also plot the range of dark matter profiles which fits the forces: In the left column we show the NFW (red) and Einasto (blue), and in the right column the gNFW (green) and pseudo-isothermal (orange). For the models at each radial point, we plot the 1-$\sigma$ range of curves as the shaded, and the median as the solid line. In the models the flattening of the iso-potential surface, \qpot, is formally a function of angle, for which we use $\theta=45\degr$. In practice, for the potentials used, $\qpot$ is a very weak function of angle unless $\qrho$ is highly non-spherical.}
    \label{fig:ellipdmhalo}
\end{figure}

We show the result of fitting for $\qpot$ and $V_c$ using the measured forces at each radius in \autoref{fig:ellipdmhalo}. From this figure, we see that, while the circular velocity is farily flat outside the Sun, it drops inside. Meanwhile, from the lower panel we see that the potential is nearly spherical at all radii. Indeed the measurements between 5 and 20\kpc are consistent with a single value of potential flattening of $q_\Phi=\qphimeas$. In order to extract more quantitative overall measurements of the dark matter distribution from the acceleration field, we proceed to fit parametric dark matter density models.

\subsection{Parametric Fits to the Dark Matter}
\label{sec:dmparametricfits}

In this section, we fit parametric dark matter halos to the gravitational acceleration field to measure the properties of the central 20\kpc of the Milky Way's dark matter halo.

\subsubsection{Dark Matter Profiles}

We explore four dark matter parameterisations: NFW, Einasto, pseudo-isothermal and generalised NFW. We treat all as ellipsoidal, writing them as a function of the ellipsoidal radius: $m^2 = R^2 + z^2/q_\rho^2$. For the NFW profile, we use \citep{Navarro:96}
\begin{equation}
	\rhodm \propto \frac{1}{m/r_s (1 + m/r_s)^2} ~.
\end{equation}
The use of this profile is inspired by dark matter only simulations. In these simulations, the halo mass inside the virial radius is correlated to the scale radius $m_s$. However, the dark matter profile, and therefore this relation, are altered by the uncertain interplay between dark and baryonic matter, and therefore we do not use this mass-concentration relation. We also fit a generalised version of this profile where the inner slope is free \citep{Zhao:96gnfw}. This is referred to as the gNFW profile:
\begin{equation}
	\rhodm \propto \frac{1}{(m/r_s)^\gamma (1 + m/r_s)^{3-\gamma}} ~.
\end{equation}

We also fit two other profiles whose central region is less cusped than the NFW profile: an Einasto profile \citep{Einasto:65}
\begin{equation}
	\rhodm \propto \exp \left[ \frac{-2}{\alpha} \left\{ \left( \frac{m}{r_{-2}} \right)^\alpha - 1 \right\} \right] ~,
\end{equation}
and a pseudo-isothermal profile \citep{Sackett:94} 
\begin{equation}
	\rhodm \propto \frac{r_c^2}{r_c^2 + m^2} ~.
\end{equation}

When fitting these profiles, we assume uninformative flat priors in all parameters with the exception of $\gamma$ in the gNFW profile, for which we use a flat prior between -5 and 5,  $\alpha$ in the Einasto profile, for which use a flat prior between 0 and 8. 

\subsubsection{Fitting Process}

For each of the dark matter densities, we compute the accelerations $\azmean{F_r}_{\rm dm}$ and $\azmean{F_\theta}_{\rm dm}$ at the centre of each grid cell due to the dark matter, add these to the baryonic model, and fit these model forces to the measured forces. The process is complicated by the measured forces in each grid cell being correlated. The correlations are introduced by the finite difference approximations used in the discretised Jeans equations (\autoref{eq:discrete_jeans}) which connect measurements at neighbouring points. As a result, neighbouring force measurements are correlated and this must be taken account of during fitting. 

To compute this correlation, we use bootstrap resampling of the data and compute the resultant forces from each resampling. We then estimate the covariance from the bootstrap resampled forces. The force measurements may be written as a vector $\boldsymbol{F}$, of length $N=90$, representing measurements from \ntheta angular bins, \nrbins radial bins, and 2 force directions. Using this notation, we estimate the covariance matrix of the forces $\mathbfss{W}$ from
\begin{equation}
	{W}_{mn} = \frac{1}{N} \sum_{i} (F_{mi}-F_{m}) (F_{ni}-F_n)
\end{equation}
where $F_{mi}$ is the $i$-th resampling of force measurement $F_m$ and we have used $N=10,000$ bootstrap resamplings. 

We assume that the forces have normally distributed errors. This is expected from the central limit theorem because they arise from the kinematic measurements of more than 100 stars in each bin, and it appears to be a good approximation from the bootstrap resampling.
Denoting the $N$ forces predicted from a dark matter profile with parameters $\boldsymbol{X}$ as $\boldsymbol{Y}(\boldsymbol{X})$ then the likelihood of measuring the forces $\boldsymbol{F}$ is  
\begin{equation}
	\mathcal{L} (\boldsymbol{F} | \boldsymbol{X}) = \frac{1}{\sqrt{ (2\pi)^N |\mathbfss{W}|}} \exp\left[ -\frac{1}{2} (\boldsymbol{F}-\boldsymbol{Y}(\boldsymbol{X}))^\intercal  \mathbfss{W}^{-1}  (\boldsymbol{F}-\boldsymbol{Y}(\boldsymbol{X})) \right] ~.
\end{equation}

\begin{figure}
	\includegraphics[width=\columnwidth]{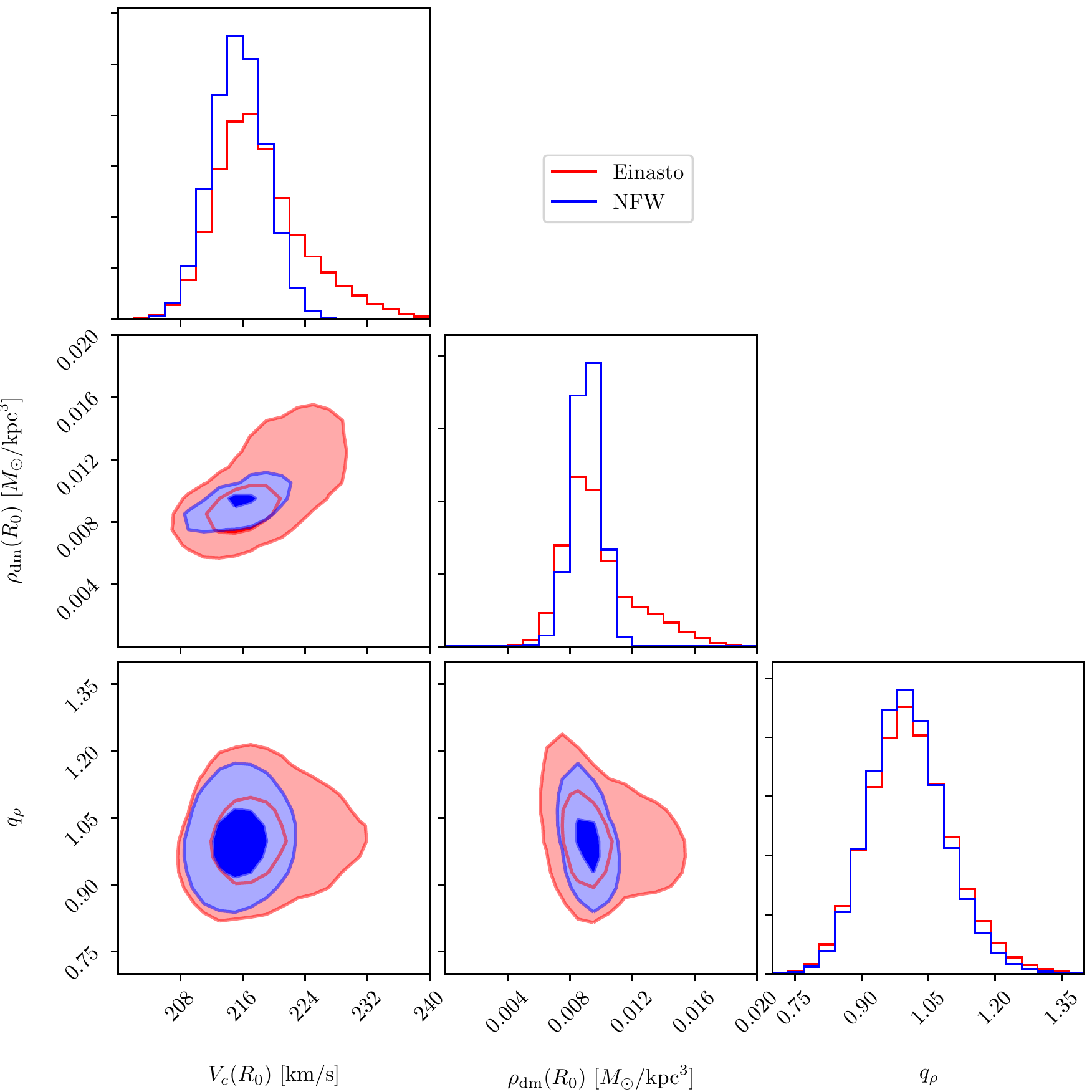}
	\vskip 0.5cm
	\includegraphics[width=\columnwidth]{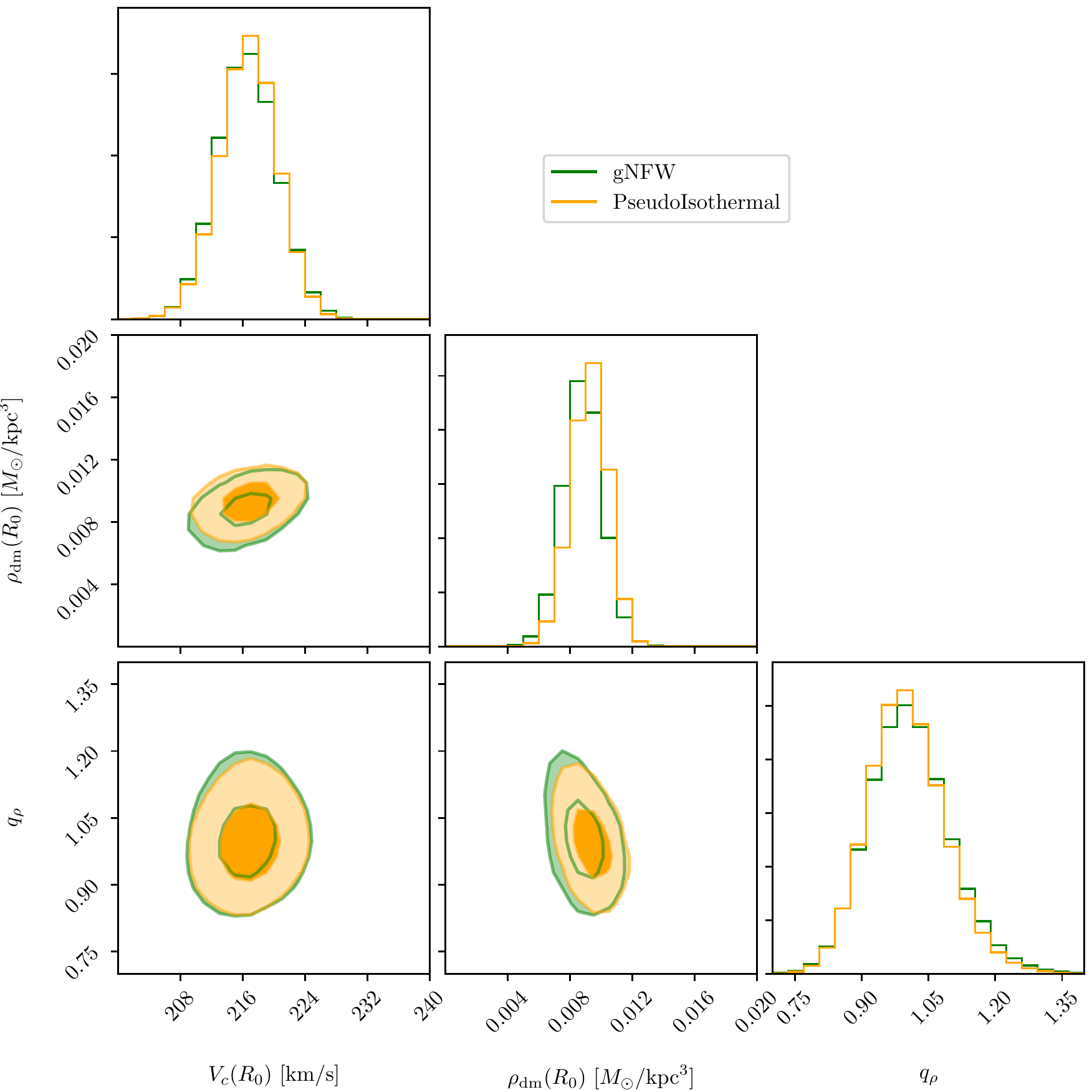}
    \caption{The results of fitting axisymmetric ellipsoidal dark matter density distributions to the forces measured in \autoref{sec:jeans}. Rather than plotting the parameters of the potentials directly, we instead show more physical parameters: (i) the total circular velocity at the Sun including both dark and baryonic matter, $V_c(R_0)
    $, (ii) the dark matter density at the Sun, $\rho_{\rm dm}(R_0)$ and (iii) the dark matter density flattening, $q_\rho$. These parameters 
    \label{fig:dmproperties}}
\end{figure}

We use an MCMC to sample from the posterior distribution of the parameters of our dark matter halos \citep{ForemanMackey:13}. We show the resultant maximum likelihood (or maximum a posteriori probability) parameters of the fits in \autoref{tab:dmparameters}. The parameters of the models are opaque and in some cases highly correlated, making their errors large and their values uninformative. We therefore plot in \autoref{fig:dmproperties} the parameters transformed into more physical quantities: the total circular velocity at the Sun including both dark and baryonic matter, $V_c(R_0)$, the dark matter density at the Sun, $\rho_{\rm dm}(R_0)$ and the dark matter density flattening, \qrho. Note that the Einasto and gNFW profiles have four parameters and so for these profiles there is one additional unplotted nuisance parameter to fully specify the dark matter profile. This extra parameter, which in both cases effectively describes the shape of the density profile, is poorly constrained.  Note that the dark matter density $\rho_{\rm dm}(R_0)$ derives from an extrapolation to the Solar position from our measured accelerations away from the Galactic plane using our ansatz that the dark matter density has an ellipsoidal shape. Comparison between this value, and those measured using more local data therefore represent an interesting test. In addition, our ellipsoidal dark matter parametrizations are fitted to the azimuthal average of the accelerations, and so non-axisymmetries would result in the values $V_c(R_0)$ and $\rho_{\rm dm}(R_0)$ reflecting $\sqrt{ \azmean{ V_c(R_0)^2 }}$ and $\azmean{\rho_{\rm dm}(R_0)}$ respectively.

All the parametric models fit the force field well, having $\chi^2$ values per degree of freedom of $\approx 0.95$. The AIC differs by less than 1 across all 4 models, meaning that we have insufficient information to distinguish between them in our sample. The errors are very similar with the exception of the Einasto profile, where the errors are larger. This is because the Einasto profile has more freedom to change its shape as can be seen from the range of shapes taken by the Einasto profile in \autoref{fig:ellipdmhalo} (and later in \autoref{fig:dmprofiles}). Because of this we conservatively select the Einasto profile as our fiducial and conclude that $\qrho=\qrhomeas$, $\azmean{V_c(R_0)}=(217 \pm 6) \kms$ and $\azmean{\rho_{\rm dm}(R_0)} = (0.0092 \pm 0.0022) M_\odot/{\rm kpc}^3$. From the MCMC samples of all of the profiles we find $\qrho>0.8$ at greater than 99\% significance.

\begin{table*}
\begin{minipage}{\textwidth}
\caption{Parameters of the fitted ellipsoidal dark matter profiles. The fitted parameters can be highly correlated and we therefore give the fitted total circular velocity at the Sun which is better constrained and more physically relevant. We also give for each model the maximum likelihood, this maximum likelihood converted to $\chi^2$ per degree of freedom, and the Akaike information criterion.}
\label{tab:dmparameters}
\centering
\begin{tabular}{lccccccc}
\hline
Profile & $q_\rho$  & $V_c(R_0)$ & $\rho_{\rm dm}(R_0)$ & Best Fitting Parameters & Max $\log \mathcal{L}$ & $\chi^2/{\rm DOF}$ & AIC \\
& & $[\kms]$ & $[\msun/{\rm pc}^3]$ & &&&\\
\hline
NFW & $1.00 \pm 0.08$ & $215.5 \pm 3.5$ & $0.0092 \pm 0.0009$ &$r_s=27\kpc$  & $-782.7$ & $0.90$ & $1571.4$\\
Einasto & $1.00 \pm 0.09$ & $217 \pm 6$ & $0.0093 \pm 0.0022$ &$r_{-2}=8.4\kpc$ $\alpha=1.4$  & $-781.5$ & $0.87$ & $1571.0$\\
gNFW & $1.00 \pm 0.09$ & $217 \pm 4$ & $0.0088 \pm 0.0012$ &$r_s=1.6\kpc$ $\gamma=-3.9$  & $-781.8$ & $0.87$ & $1571.6$\\
PseudoIsothermal & $1.00 \pm 0.08$ & $217 \pm 4$ & $0.0093 \pm 0.0012$ &$r_c=2.7\kpc$  & $-782.4$ & $0.89$ & $1570.9$\\
\hline
\end{tabular}
\end{minipage}
\end{table*}

\subsection{Systematics}
\label{sec:systematics}

In this section, we assess the errors induced by possible systematic errors in the data, and choices made in the analysis. In particular, we examine the effect of possible systematic errors in RR Lyrae distances, and in the the baryonic model. 

To estimate the uncertainties due to the baryonic model, we adjust each component in turn. When adjusting the stellar disk or ISM scale lengths, we fix the density at 4\kpc from the Galactic centre so that, by adjusting these scale lengths, we also adjust the surface density at the Sun. This is by design, so that our range of models encompasses the full range of possible baryonic contributions to the accelerations. In particular our fiducial stellar model has a local stellar surface density $38\msun/{\rm pc}^2$ inside $|z|<1.1\kpc$, chosen to be consistent with the local estimate of $(38\pm4)\msun/{\rm pc}^2$ by \citealt{Bovy:13}. Our shorter and longer scale length models have local stellar densities $32\msun/{\rm pc}^2$ and $44\msun/{\rm pc}^2$ which encompasses this range, and the entire reasonable range of contributions to the accelerations from the stellar disk in general.

Other tests we have performed include: varying the mass in the central 5kpc of the Galaxy by varying the \citetalias{Portail:17} dynamical model used, systematically changing the distance modulus of all the RR Lyrae, altering the distance to the galactic center, and changing the solar velocity. All variation models are summarised in \autoref{tab:systematics}. We find that our fitted parameters are relatively insensitive to any of the tested variations, changing within the formal statistical errors.

\begin{table}
\begin{minipage}{\columnwidth}
\caption{Systematic variations of the fiducial baryonic model and their affect on the fitted parameters. We show only the results of fits to an Einasto profile.}
\label{tab:systematics}
\begin{tabular}{lccc}
\hline
Variation & $V_c(R_0)$ & $\rho_{\rm dm}(R_0)$ & $q_\rho$ \\
& $[\kms]$ & $[M_\odot/{\rm kpc}^3]$ & \\
\hline
Fiducial\footnote{Uses stellar disk with scale length $h_{R,\star}=2.4\,{\rm kpc}$, gas disk with scale length $h_{R,{\rm ism}}=2\times2.4\,{\rm kpc}$, and best fitting model of  \citetalias{Portail:17}. This model has bar pattern speed $\Omega = 40\kms\kpc^{-1}$, mass-to-clump ratio 1000/\msun and nuclear stellar mass $2\times10^9\msun$.} & $217$ & $0.0092$ & $1.00$ \\
$h_{R,\star}=2.15\,{\rm kpc}$\footnote{Dynamical disk scale length measured by \citet{Bovy:13}. Has $\Sigma_\star(R_0)=32\msun/{\rm pc}^2$ to keep disk continuity at 5\kpc.} & $217$ & $0.0096$ & $0.98$ \\
$h_{R,\star}=2.68\,{\rm kpc}$\footnote{Dynamical disk scale length measured by \citet{Piffl:14}. Has $\Sigma_\star(R_0)=44\msun/{\rm pc}^2$ to keep disk continuity at 5\kpc.} & $218$ & $0.0091$ & $1.03$ \\
$h_{R,{\rm ism}}=3\times2.4\,{\rm kpc}$\footnote{Has $\Sigma_{\rm ism}(R_0)=16\msun/{\rm pc}^2$ to keep disk continuity at 5\kpc.} & $216$ & $0.0090$ & $1.04$ \\
$h_{R,{\rm ism}}=1.5\times2.4\,{\rm kpc}$\footnote{Has $\Sigma_{\rm ism}(R_0)=10\msun/{\rm pc}^2$ to keep disk continuity at 5\kpc.} & $218$ & $0.0094$ & $0.98$ \\
\citetalias{Portail:17} Boundary Model 1\footnote{Uses bar pattern speed $\Omega = 37.5\kms\kpc^{-1}$, mass-to-clump ratio 900/\msun and nuclear stellar mass $2.5\times10^9\msun$.} & $217$ & $0.0094$ & $0.99$ \\
\citetalias{Portail:17} Boundary Model 2\footnote{Uses bar pattern speed $\Omega = 42.5\kms\kpc^{-1}$, mass-to-clump ratio 1100/\msun and nuclear stellar mass $1.5\times10^9\msun$.} & $218$ & $0.0093$ & $1.01$ \\
RR Lyrae 0.03\,mag brighter\footnote{Estimated systematic uncertainty by \citetalias{Sesar:17}} & $216$ & $0.0095$ & $0.99$ \\
RR Lyrae 0.03\,mag fainter & $219$ & $0.0118$ & $0.99$ \\
$R_0=8.0\,{\rm kpc}$ & $217$ & $0.0090$ & $1.04$ \\
$R_0=8.4\,{\rm kpc}$ & $216$ & $0.0090$ & $1.00$ \\
$v_\odot=(11.1,255,7.25)\,{\rm km/s}$ & $217$ & $0.0089$ & $1.02$ \\
$v_\odot=(11.1,245,7.25)\,{\rm km/s}$ & $218$ & $0.0094$ & $1.01$ \\
Fitting including Sgr Stream\footnote{We remove the Sagittarius Dwarf, but leave the tail of the stream in the sample.} & $222$ & $0.0083$ & $1.06$ \\

\hline
\end{tabular}
\end{minipage}
\end{table}

\subsection{The Effects of Non-Axisymmetries}
\label{sec:nonaxi}

\begin{figure}
	\includegraphics[width=\columnwidth]{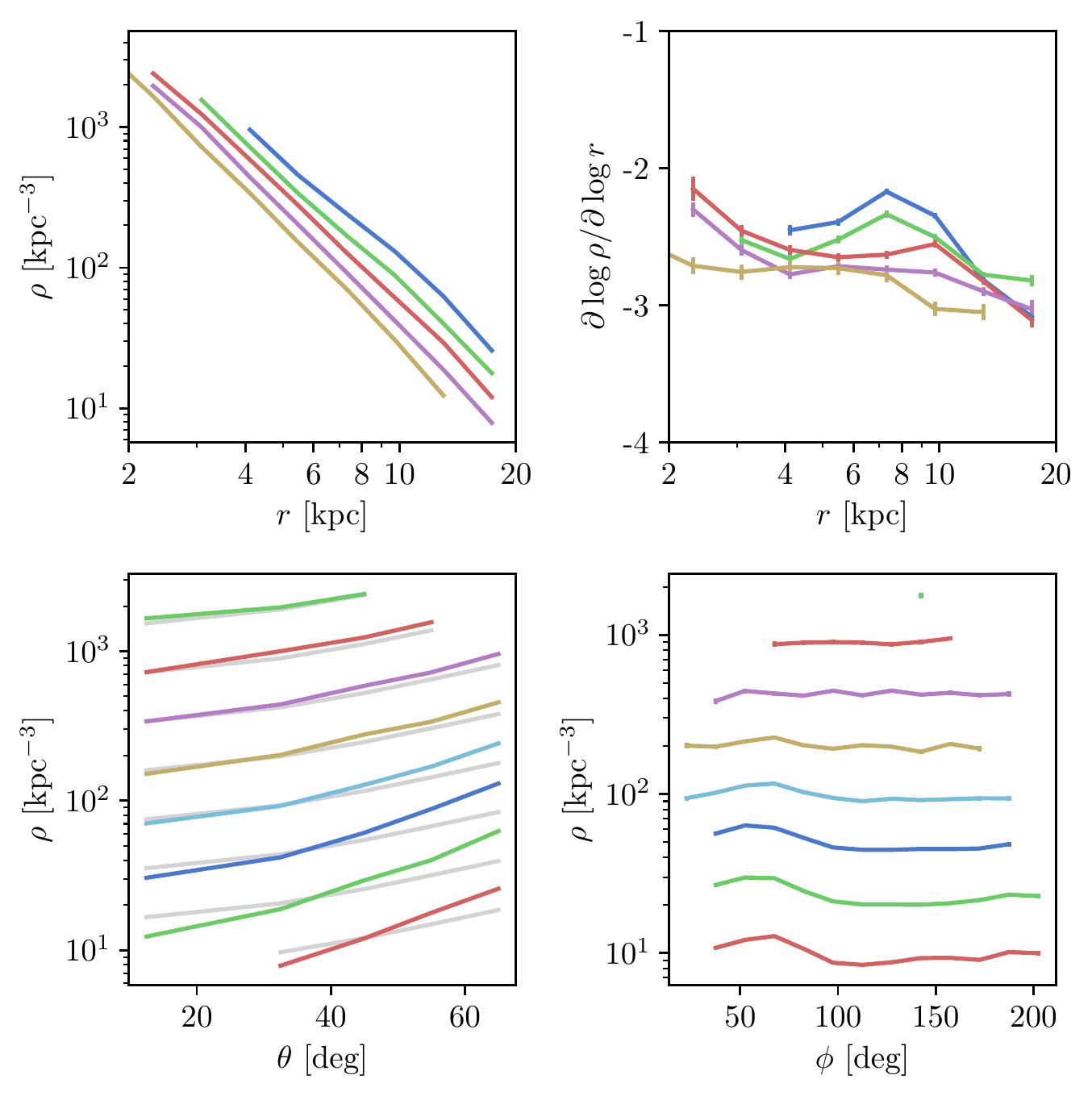}
    \caption{The density structure of our toy non-axisymmteric halo observed through the selection function and plotted similarly to \autoref{fig:density_structure}. By comparison to \autoref{fig:density_structure} we see that the toy halo has a similar density gradient gradient and a similar flattening. The variation with azimuth shown in the lower right plot is at a similar level, and oriented similarly. }
    \label{fig:sausage_density_structure}
\end{figure}

\begin{figure*}
	\includegraphics[width=\textwidth]{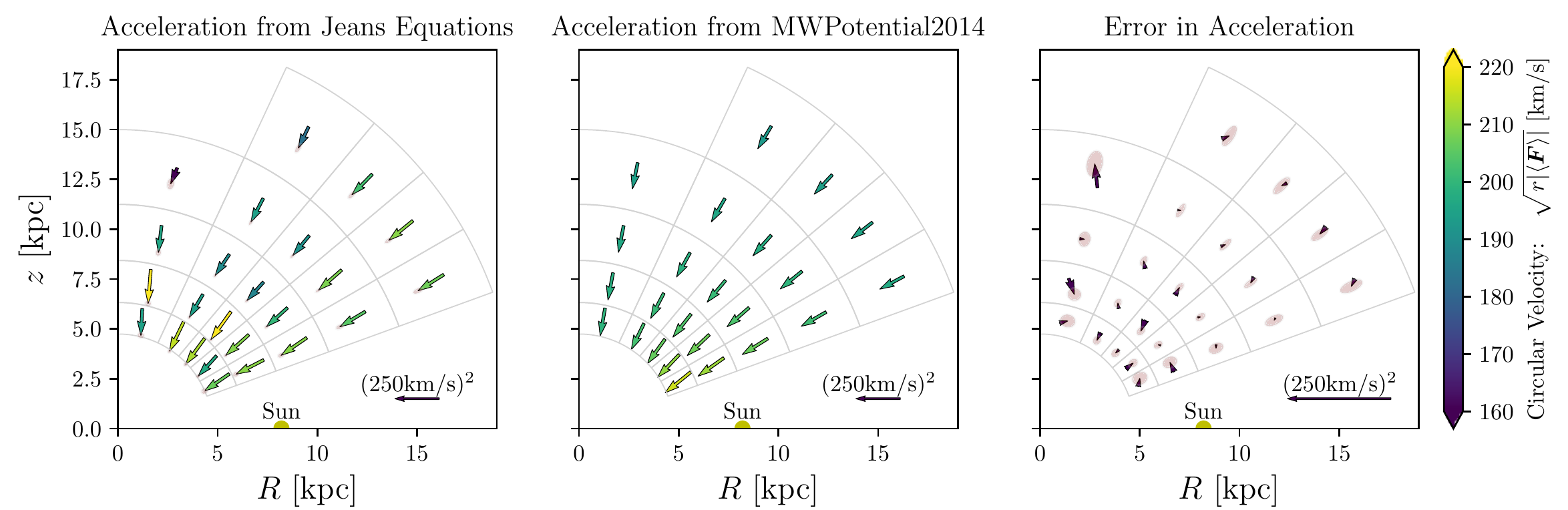}
    \caption{The reconstructed acceleration field of the non-axisymmetric toy halo described in \autoref{sec:nonaxi}. In the left plot, we show the azimuthally averaged gravitational acceleration field measured by applying the discretised Jeans equations (\autoref{eq:discrete_jeans}) to the kinematics of the toy halo measured from mock proper motions using the assumption that the kinematics do not vary with azimuth (\autoref{sec:kin}). In the middle panel we show the accelerations of the true background potential in which the N-body particles move. In the right panel we show the residual forces by subtracting these. Note that the scale here is enlarged. To keep the arrows of similar size in each panel we plot $r\azmean{\boldsymbol{F}}$ in $({\rm km}/{\rm s})^2$, which is analogous to the square circular velocity used in-plane, arrows are coloured by the square root of this quantity.}
    \label{fig:sausage_force_field}
\end{figure*}

%{\bf
We have reconstructed the azimuthally averaged acceleration field in the Galactic halo using Jeans equations that do not assume that the forces or the halo tracer population are axisymmetric (\hbox{\autoref{sec:jeanseqn}}). In these equations the second velocity moment terms that enter should be the tracer density weighted azimuthal average. However, the absence of radial velocities for our tracer population of RR Lyrae forced us to assume that the kinematics were independent of azimuth in order to evaluate these second velocity moments (\autoref{sec:kin}).

In the time since Gaia DR2 has been released it has become increasingly clear that a large fraction of the inner Halo was deposited in one accretion event, named Gaia-Enceladus or the Gaia Sausage \citep{Belokurov:18final,Helmi:18final}. It also appears that the merger debris as traced by RR Lyrae is not axisymmetric \citep{Iorio:18}. This is also visible in the density variation with azimuth present at large radii in the lower right panel of \autoref{fig:density_structure}. To investigate the possible effects of these non-axisymmetries on our results we have constructed a toy non-axisymmetric mock halo.

This halo was constructed by placing a Hernquist sphere of stars of mass $2\times10^9\msun$ and scale radius 0.5\kpc on a nearly radial orbit with apocenter $\sim 20\kpc$ in the fixed background potential of \verb!MWPotential2014! taken from GalPy \citep{Bovy:galpy}. This setup was integrated forwards for 6Gyr using the GyrFalcon integrator \citep{Dehnen:falcon}. This integration time was chosen so that our toy halo is likely to have a similar level of phase-mixing as the Milky Way's halo. The resulting distribution of particles is non-axisymmetric and composed of highly radial orbits with $\beta\sim0.9$ at solar galactocentric radii, but is less concentrated and more flattened than the Milky Way's halo. We therefore added to this an initially spherical halo with a $\rho \propto r^{-3}$ profile, which was relaxed in the same background potential for 6Gyr. This smooth halo was composed of $5\times10^5$ N-body particles, while the non-axisymmetric component contained $10^6$. When added they contribute roughly this same 2:1 proportion of particles within our volume. We emphasise that this is not a simulation of the formation of the halo, but rather a construction process through which we can produce a non-axisymmetric toy halo which is likely to have similar levels of phase mixing as the inner Galactic halo.

This toy halo was oriented similarly to the Milky Way's inner halo \citep{Iorio:18} and the N-body particles observed though our selection function as if they were RR Lyrae. The resultant density distribution seen in \autoref{fig:sausage_density_structure} shows that the variation and orientation of the density in the outer bins is comparable to the Milky Way's RR Lyrae (\autoref{fig:density_structure}. The densities themselves in are an order of magnitude higher that the actual RR Lyrae sample in order to better assess the size of the systematics. We have also analysed smaller toy halos which gave similar results but with correspondingly larger statistical errors.

We have analysed this mock halo using the same tools as to analyse the real data. The resultant acceleration field is shown in \autoref{fig:sausage_force_field} compared to the forces from the actual background potential of \verb!MWPotential2014!. The residual forces are reassuringly small. Subtracting the baryonic forces and fitting the dark matter halo we find that the fitted flattening of the dark matter density is $q_\rho=1.01\pm 0.03$, while the actual background potential had a spherical dark matter halo.

This test shows that by using the assumption of azimuthally invariant kinematics to fill the unobserved radial velocity with our mock non-axisymmetric halo we recover the forces and halo flattening to within the statistical errors. This accuracy is likely to be because proper motions provide 2 of the 3 kinematic components, while the halos major axis is at an intermediate angle i.e. neither face on or end on which would be more likely to provide a bias.  In the future surveys such as WEAVE and 4MOST will provide radial velocities of large samples of halo stars. The additional kinematic information should allow the velocity moments measured directly from the data, without the assumption that the kinematics are independent of azimuth made in \autoref{sec:kin}. %}

\section{Discussion}
\label{sec:discussion}

The Jeans modelling performed here assumes that the Galaxy, and our stellar sample, is in dynamical equilibrium. Because the Galactic halo is growing from the accretion of satellites, this assumption is broken in detail. Objects accreted into the inner halo are expected to phase mix relatively quickly, while retaining information in their integrals of motion or actions, making the distribution function less smooth. However, Jeans modelling does not require that the distribution function be featureless, only that the inner halo be well phase mixed \ie that the distribution function can be taken as time independent. Features that have not had enough time to well phase mix could however present a problem we investigated one such feature, Gaia Enceladus or the Gaia Sausage, in \autoref{sec:nonaxi}: the Sagittarius stream is another. There are likely to be other streams to be found in Gaia data \citep{Malhan:18} and some are already known in the volume that we have studied (\eg \citealt{Ibata:18}, and see also \citealt{Mateu:18}). However, because the fraction of RR Lyrae in these unmixed stellar streams in the halo inside 20\kpc is small, the effect on our results is expected to be similarly small and we have not attempted to excise all streams. This is supported by \autoref{sec:systematics}, in which we repeated our analysis without removing the Sagittarius  stream. Despite being the most prominent stream in our sample, the effects of not excising it are relatively small. 

The Jeans analysis presented here is attractive because it is non-parametric. However, as a result, its formal statistical power is lower than that of other methods. Non-parametric models have increased flexibility over parametric models to fit the underlying data, and as a result has increased formal statistical errors over parametric modelling, such as distribution function modelling. Moreover parametric modelling relies on the fitted functional form correctly representing the underlying stellar halo, something which is difficult to assess with complicated systems and high-dimensional data. In particular it can be difficult to assess possible model degeneracies: while fitting a parametric model to data and assessing that the fitted model reproduces the data is straightforward, evaluating whether a subtly different model could also reproduce the data, but with significantly different results requires a careful analysis. The non-parametric acceleration measurements here circumvent that problem.

The non-parametric method presented in this work has the further advantage that it is highly transparent: we can derive the acceleration field in a clear manner from the kinematic measurements, and fit models directly to this. For example, during the initial analysis, the Sagittarius stream was not completely removed by our selection cuts. This, however, was immediately clear when the first acceleration field was constructed because the bin from which the stream had not been excised was a clear outlier.

When deriving the dark matter distribution in \autoref{sec:dmparametricfits} we used  parametric fits. This is because deriving the dark matter density from the acceleration field  requires an additional derivative which, with the sample size analysed in this work, would result in density errors in each individual grid cell too large to be useful. In the future ground-based surveys such as WEAVE and 4MOST will provide larger samples of stars with full 6D phase space which will allow improved measurement of the Milky Way's dark matter distribution, and should even allow non-parametric measurement of its density using techniques similar to those presented here. For the time being we have used parametric fits which, by connecting the measurements, and fitting for a handful of numbers, provides smaller errors. In order to assess the possible biases introduced by this parametric approach we have fit for a range of dark matter profiles.  The reader concerned by this should concentrate on the less parametric approach of fitting an ellipsoidal potential at each radius which was taken in \autoref{sec:dmellipfits}.

\begin{figure}
	\includegraphics[width=\columnwidth]{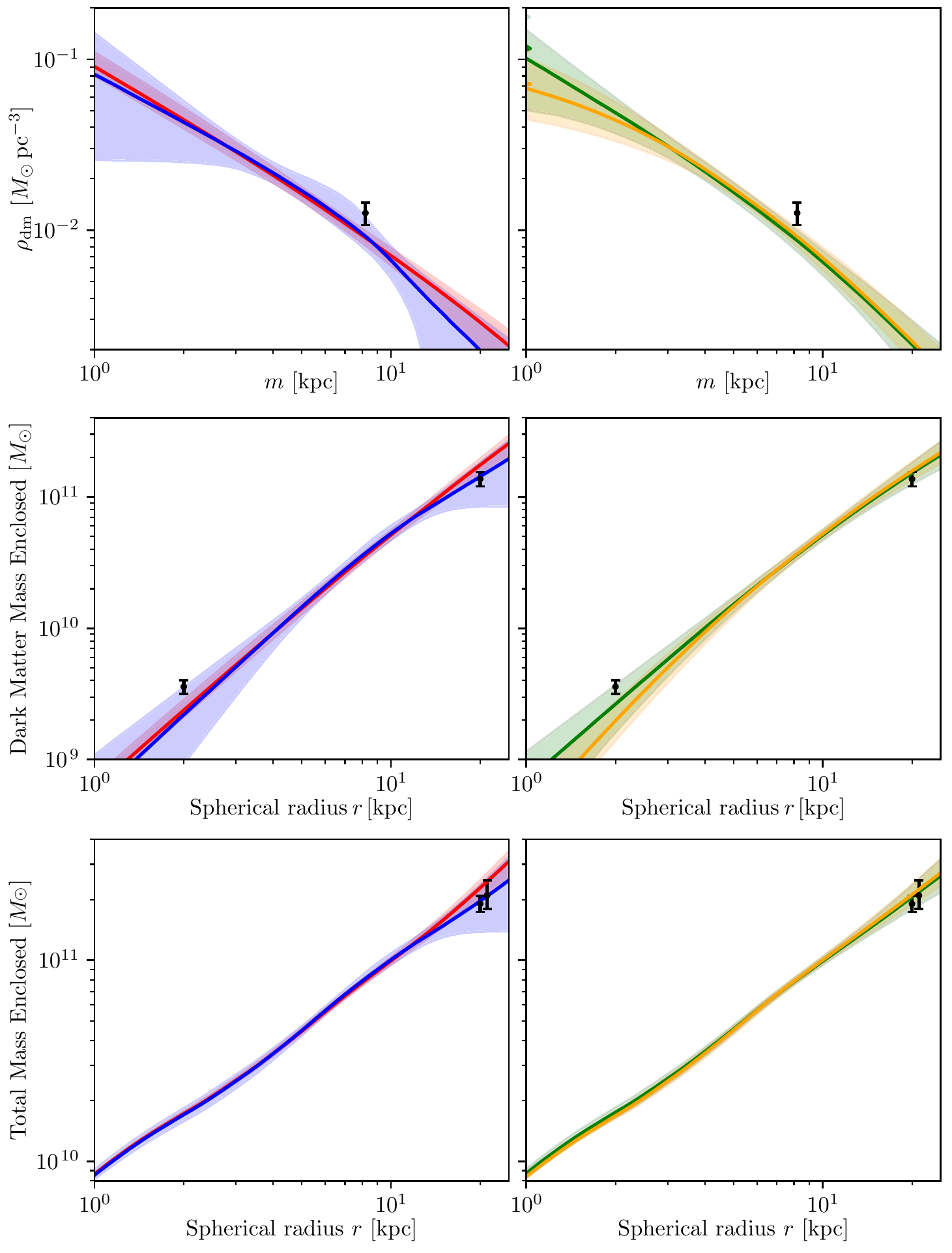}
    \caption{In the upper panels, we show dark matter density distributions that are consistent with the measured forces. We show the Einasto (blue), NFW (red), gNFW (green) and pseudo-Isothermal (orange) profiles. We also show the measurement made using the velocities of RAVE disk stars by \citet[][using our measured value of \qrho]{Piffl:14} as the data point. In the middle panels, we show the cumulative mass profile in spherical radii. The outer data point is the recent measurement using Gaia DR2 data by \citet{Posti:19}, the inner data point is the measurement in the bulge by \citetalias{Portail:17} converted to a spherical volume. In the lower panel, we show the total mass enclosed in spherical radii, also including the baryonic model. We also plot the measurements near the edge of our volume at 20\kpc by \citet{Posti:19}, and the measurement inside 21.1\kpc by \citet{Watkins:18}.
    \label{fig:dmprofiles}}
\end{figure}

Our measurement of the dark matter flattening of $\qrho=\qrhomeas$ agrees with several recent measurements, but with somewhat smaller error. Measurements from streams also point towards a near spherical halo: in particular \citet{Bovy:16} found $q_\rho=1.05\pm0.14$ towards the edge of our volume by combining measurements from the Pal-5 and GD-1 streams. In addition, very recently, \citet{Posti:19} performed action based modelling on 91 globular clusters with full 6D phase space information, finding $q_\rho = 1.30\pm0.25$. We expect these constraints to rapidly improve as the community begins to exploit Gaia DR2 in combination with other datasets.

If this emerging picture that the dark matter profile is nearly spherical holds into the Gaia era, then it appears in tension with the shapes expected from current cosmological simulations. Dissipation in baryonic simulations can make the highly triaxial halos seen in dark matter only simulations more spherical \citep{Dubinski:94,Kazantzidis:04,Debattista:08,Kazantzidis:10,Abadi:10}. However, a completely spherical, or even mildly prolate halo, would be in tension with these simulations which typically predict increases in axis ratio of $\Delta \qrho = 0.1-0.3$. It is possible that this tension could point to the physics of dark matter, one of the most studied examples being that halos are more spherical in self interacting dark matter models \citep{Spergel:00,Yoshida:00,Peter:13}. Careful assessment of these results is however needed: for example, \citet{Dai:18} reassessed the measurements of a near spherical halo by \citet{Bovy:16}, and concluded that the data could also be reproduced in the mildly oblate potential of the Eris simulation.
 
 We plot in \autoref{fig:dmprofiles} the dark matter densities of our fitted models. The density that we find near the Sun is consistent with, or slightly lower than, several recent measurements using the velocities of nearby disk stars. \citet{Piffl:14} finds a dark matter density of $0.0126 q_\rho^{-0.89} \msun/{\rm pc}^3$ with systematic errors estimated at 15\%, and we include this value in our plot. Likewise \citet{Bienayme:14} finds $(0.0143\pm0.0011)\msun/{\rm pc}^3$ which is slightly higher than our inferred value. We recall that our value was not measured near the Sun but is instead inferred from the in-plane extrapolation of the accelerations measured away from the Galactic plane. In particular our value of $\rho_{\rm dm}(R_0)$ is the dark matter density of our fitted ellipsoidal density profiles at the solar position. If, for example, there were a significant disk of dark matter (termed a `dark disk') this would not be included in our value of $\rho_{\rm dm}(R_0)$. The closeness of both kinds of measurements therefore points towards consistent picture of a near spherical dark matter halo at Solar galactocentric radii \citep{Read:14}.
 	
Our inferred value of the circular velocity at the Sun of $V_c(R_0)=(217 \pm 6) \kms$ is lower than some other recent measurements. In particular, it is lower than the $(238 \pm 9) \kms$ measured by \citet{Schonrich:12} and $(240 \pm 8) \kms$ measured by \citet{Reid:14}. This difference could be statistical, but non-axisymmetric motions could also play an role \citep{Ortwin:araa}, despite the care taken by \citet{Schonrich:12} and \citet{Reid:14}. Many non-axisymmetric motions are obvious in the new data from Gaia \citep[\eg][]{Katz:19,Antoja:18,Hunt:18spiral}, and we expect the value of $V_c(R_0)$ to soon be clarified with this data. Indeed our inferred value of the circular velocity at the Sun is quite close to the value of $V_c(R_0)=229\kms \pm 2\%$ very recently measured using the combination of Gaia DR2 and APOGEE data by \citet{Eilers:19}.

We also plot in \autoref{fig:dmprofiles} the spherical cumulative mass profiles of the dark matter and the total mass including baryonic matter. We see that our mass enclosed inside 20\kpc is consistent with the measurements using Gaia DR2 data of the mass inside the same volume by \citet{Posti:19} and \citet{Watkins:18}.

As expected, the profiles have quite similar dark matter densities at radii between $5\kpc$ and $20\kpc$, where our method provides its most accurate measurements. Interestingly they are quite different inside 5 kpc. The dynamical models of \citetalias{Portail:17} required fairly low dark matter fractions inside the bulge region of $17\%$, corresponding to a mass of $(3.2\pm 0.5)\times 10^{9} \msun$. When combined with $V_c(R_0)=(238 \pm 9) \kms$ and the stellar surface density, this required that the dark matter have a core or shallow cusp (with power-law slope $\gamma<0.7$). However, the dark matter mass inside the bulge would be consistent with all our profiles at $\approx 1\sigma$. The reason that the NFW profile (which has central power-law slope $\gamma=1$) is still consistent with the \citetalias{Portail:17} bulge mass measurement is most likely that the circular velocity found here is slightly smaller than the $V_c(R_0)=238\kms$ used in \citetalias{Portail:17}. This demonstrates the need for Gaia era dynamical modelling which connects data across the Galaxy in order to clarify whether the dark matter profile is shallow, as found by \citetalias{Portail:17}, or more steeply cusped as found in recent cosmological simulations \citep[\eg][]{Chan:15,Grand:17a}.

\section{Summary and Conclusions}
\label{sec:conc}

Gaia has produced a truly transformational dataset with which, over the coming years, we should learn much about how the Milky Way and similar galaxies formed and evolved. The sample we have analysed is remarkable in having accurate transverse kinematic measurements across the entire inner halo away from the Galactic plane. This is provided by combining the proper motions of Gaia with the accurate distances of the RR Lyrae sample of \citetalias{Sesar:17}. This allowed us to investigate the kinematics of the stellar halo between $1.5\kpc$ and $20\kpc$.

Statistically reconstructing the full 3d-kinematics from the proper motions in the absence of radial velocities, we found that, outside the central 5\kpc, halo RR Lyrae are highly radially anisotropic with $\beta \approx 0.8$ and have a nearly spherically aligned velocity ellipsoid. Between $1.5\kpc$ and $5\kpc$, that anisotropy drops, but even there it remains above $\beta \approx 0.25$. Inside 10\kpc, our sample of Halo RR Lyrae  rotates with a profile that rises to $50\kms$ in our innermost radial bin. This may reflect the early formation and accretion history of the halo, although there is significant transfer of angular momentum with the Galactic bar at these radii. Reaching firm conclusions regarding the origin of this rotation will require further modelling and data, including for example metallicity information.

By applying discretised versions of the Jeans equations to these kinematics, we subsequently measured the acceleration field inside $20\kpc$. The acceleration field is largely radial, particularly in the outer regions. While the in-plane acceleration field of the Milky Way is relatively well constrained by the rotation curve, away from the Galactic plane, our measurements give much-needed new constraints on the shape of the Galaxy's mass distribution. The most accurate previous measurements were limited to a small number of stellar streams in the halo.

 By subtracting baryonic models, we inferred the gravitational acceleration field produced by the Milky Way's dark matter halo. Because these accelerations from the dark matter halo are consistent with being directed in the radial direction, the resultant potential is consistent with spherical. We measured the profile of the shape of the dark matter potential between 5\kpc and 20\kpc and have found that these measurements are consistent with a single value of $q_\Phi=\qphimeas$.

We have also fit parametric dark matter profiles to the forces. These results are summarised in  \autoref{tab:dmparameters} while the impact of systematic changes to the data and baryonic model are summarised in \autoref{tab:systematics}. We found that the ellipsoidal flattening of these density profiles does not depend significantly on the profile and can be combined into a single constraint: $q_\rho=\qrhomeas$. This is consistent with a spherical profile, as expected given the radial nature of forces, and the measured sphericity of the potential. Our fits indicate that density profiles as flattened as $q_\rho=0.8$ are ruled out at higher than 99\% significance.

The fitted dark profiles are also interesting, however, using the present data alone, we cannot determine the steepness of the inner profile and further dynamical modelling and data are required. \citetalias{Portail:17} found dark matter density profiles that flattened on kpc scales to be less centrally steep than $\rho_{\rm dm} \propto r^{-0.7}$. The same physical processes which determine the dark matter profile in the inner regions also determine its shape. Therefore, if the less concentrated than expected dark matter halo found by \citetalias{Portail:17} persists with models of Gaia-era data, this may be related to the near spherical halo on larger scales derived in this work.

%In conclusion: Gaia is awesome. Our results are awesome. Everything is awesome.\footnote{\url{https://www.youtube.com/watch?v=StTqXEQ2l-Y}}

\section*{Acknowledgements}

We gratefully acknowledge useful discussions with Isabella S\"oldner-Rembold regarding the velocity ellipsoid, and numerous helpful discussions with Matias Bla\~na.

This project has received funding from the European Union's Horizon 2020 research and innovation programme under the Marie Sk\l{}odowska-Curie grant agreement No 798384.

This work has made use of data from the European Space Agency (ESA) mission  \href{https://www.cosmos.esa.int/gaia}{\it Gaia}, processed by the {\it Gaia} Data Processing and Analysis Consortium (\href{https://www.cosmos.esa.int/web/gaia/dpac/consortium}{DPAC}). Funding for the DPAC has been provided by national institutions, in particular the institutions participating in the {\it Gaia} Multilateral Agreement.
The Pan-STARRS1 Surveys (PS1) and the PS1 public science archive have been made possible through contributions by the Institute for Astronomy, the University of Hawaii, the Pan-STARRS Project Office, the Max-Planck Society and its participating institutes, the Max Planck Institute for Astronomy, Heidelberg and the Max Planck Institute for Extraterrestrial Physics, Garching, The Johns Hopkins University, Durham University, the University of Edinburgh, the Queen's University Belfast, the Harvard-Smithsonian Center for Astrophysics, the Las Cumbres Observatory Global Telescope Network Incorporated, the National Central University of Taiwan, the Space Telescope Science Institute, the National Aeronautics and Space Administration under Grant No. NNX08AR22G issued through the Planetary Science Division of the NASA Science Mission Directorate, the National Science Foundation Grant No. AST-1238877, the University of Maryland, Eotvos Lorand University (ELTE), the Los Alamos National Laboratory, and the Gordon and Betty Moore Foundation.

%%%%%%%%%%%%%%%%%%%%%%%%%%%%%%%%%%%%%%%%%%%%%%%%%%


%\bibliographystyle{mnras}
%\bibliography{mn-jour.bib,fullbib.bib}

\begin{thebibliography}{}
\makeatletter
\relax
\def\mn@urlcharsother{\let\do\@makeother \do\$\do\&\do\#\do\^\do\_\do\%\do\~}
\def\mn@doi{\begingroup\mn@urlcharsother \@ifnextchar [ {\mn@doi@}
  {\mn@doi@[]}}
\def\mn@doi@[#1]#2{\def\@tempa{#1}\ifx\@tempa\@empty \href
  {http://dx.doi.org/#2} {doi:#2}\else \href {http://dx.doi.org/#2} {#1}\fi
  \endgroup}
\def\mn@eprint#1#2{\mn@eprint@#1:#2::\@nil}
\def\mn@eprint@arXiv#1{\href {http://arxiv.org/abs/#1} {{\tt arXiv:#1}}}
\def\mn@eprint@dblp#1{\href {http://dblp.uni-trier.de/rec/bibtex/#1.xml}
  {dblp:#1}}
\def\mn@eprint@#1:#2:#3:#4\@nil{\def\@tempa {#1}\def\@tempb {#2}\def\@tempc
  {#3}\ifx \@tempc \@empty \let \@tempc \@tempb \let \@tempb \@tempa \fi \ifx
  \@tempb \@empty \def\@tempb {arXiv}\fi \@ifundefined
  {mn@eprint@\@tempb}{\@tempb:\@tempc}{\expandafter \expandafter \csname
  mn@eprint@\@tempb\endcsname \expandafter{\@tempc}}}

\bibitem[\protect\citeauthoryear{Abadi, Navarro, Fardal, Babul  \&
  Steinmetz}{Abadi et~al.}{2010}]{Abadi:10}
Abadi M.~G.,  Navarro J.~F.,  Fardal M.,  Babul A.,   Steinmetz M.,  2010,
  \mn@doi [MNRAS] {10.1111/j.1365-2966.2010.16912.x}, 407, 435

\bibitem[\protect\citeauthoryear{Allgood, Flores, Primack, Kravtsov, Wechsler,
  Faltenbacher  \& Bullock}{Allgood et~al.}{2006}]{Allgood:06}
Allgood B.,  Flores R.~A.,  Primack J.~R.,  Kravtsov A.~V.,  Wechsler R.~H.,
  Faltenbacher A.,   Bullock J.~S.,  2006, \mn@doi [MNRAS] {10.1086/428898},
  367, 1781

\bibitem[\protect\citeauthoryear{An \& Evans}{An \& Evans}{2016}]{An:16}
An J.,  Evans N.~W.,  2016, \mn@doi [ApJ] {10.3847/0004-637X/816/1/35}, 816, 35

\bibitem[\protect\citeauthoryear{Aniyan, Freeman, Gerhard, Arnaboldi  \&
  Flynn}{Aniyan et~al.}{2015}]{Aniyan:16}
Aniyan S.,  Freeman K.~C.,  Gerhard O.~E.,  Arnaboldi M.,   Flynn C.,  2015,
  \mn@doi [MNRAS] {10.1093/mnras/stv2730}, 456, 1484

\bibitem[\protect\citeauthoryear{Antoja et~al.,}{Antoja
  et~al.}{2018}]{Antoja:18}
Antoja T.,  et~al., 2018, \mn@doi [Nat] {10.1038/s41586-018-0510-7}, 561, 360

\bibitem[\protect\citeauthoryear{Athanassoula}{Athanassoula}{2003}]{Athanassoula:03}
Athanassoula E.,  2003, \mn@doi [MNRAS] {10.1046/j.1365-8711.2003.06473.x},
  341, 1179

\bibitem[\protect\citeauthoryear{Beers et~al.,}{Beers et~al.}{2012}]{Beers:12}
Beers T.~C.,  et~al., 2012, \mn@doi [ApJ] {10.1088/0004-637X/746/1/34}, 746, 34

\bibitem[\protect\citeauthoryear{Belokurov, Erkal, Evans, Koposov  \&
  Deason}{Belokurov et~al.}{2018}]{Belokurov:18final}
Belokurov V.,  Erkal D.,  Evans N.~W.,  Koposov S.~E.,   Deason A.~J.,  2018,
  \mn@doi [MNRAS] {10.1093/mnras/sty982}, 478, 611

\bibitem[\protect\citeauthoryear{Bienaym{\'e} et~al.,}{Bienaym{\'e}
  et~al.}{2014}]{Bienayme:14}
Bienaym{\'e} O.,  et~al., 2014, \mn@doi [A\&A] {10.1051/0004-6361/201424478},
  571, A92

\bibitem[\protect\citeauthoryear{Bird, Xue, Liu, Shen, Flynn  \& Yang}{Bird
  et~al.}{2018}]{Bird:18}
Bird S.~A.,  Xue X.-X.,  Liu C.,  Shen J.,  Flynn C.,   Yang C.,  2018, arXiv,
  1805.04503

\bibitem[\protect\citeauthoryear{Bland-Hawthorn \& Gerhard}{Bland-Hawthorn \&
  Gerhard}{2016}]{Ortwin:araa}
Bland-Hawthorn J.,  Gerhard O.,  2016, Annual Reviews of Astronomy {\&}
  Astrophysics, 54

\bibitem[\protect\citeauthoryear{Bond et~al.,}{Bond et~al.}{2010}]{Bond:10}
Bond N.~A.,  et~al., 2010, \mn@doi [ApJ] {10.1088/0004-637X/716/1/1}, 716, 1

\bibitem[\protect\citeauthoryear{Bovy}{Bovy}{2015}]{Bovy:galpy}
Bovy J.,  2015, \mn@doi [ApJS] {10.1088/0067-0049/216/2/29}, 216, 29

\bibitem[\protect\citeauthoryear{Bovy \& Rix}{Bovy \& Rix}{2013}]{Bovy:13}
Bovy J.,  Rix H.-W.,  2013, \mn@doi [ApJ] {10.1088/0004-637X/779/2/115}, 779,
  115

\bibitem[\protect\citeauthoryear{Bovy, Bahmanyar, Fritz  \& Kallivayalil}{Bovy
  et~al.}{2016}]{Bovy:16}
Bovy J.,  Bahmanyar A.,  Fritz T.~K.,   Kallivayalil N.,  2016, \mn@doi [ApJ]
  {10.3847/1538-4357/833/1/31}, 833, 31

\bibitem[\protect\citeauthoryear{Bowden, Belokurov  \& Evans}{Bowden
  et~al.}{2015}]{Bowden:15}
Bowden A.,  Belokurov V.,   Evans N.~W.,  2015, \mn@doi [MNRAS]
  {10.1093/mnras/stu892}, 449, 1391

\bibitem[\protect\citeauthoryear{Bowden, Evans  \& Williams}{Bowden
  et~al.}{2016}]{Bowden:16}
Bowden A.,  Evans N.~W.,   Williams A.~A.,  2016, \mn@doi [MNRAS]
  {10.1111/j.1365-2966.2009.15242.x}, 460, 329

\bibitem[\protect\citeauthoryear{Chan, Kere{\v s}, Onorbe, Hopkins, Muratov,
  Faucher-Gigu{\`e}re  \& Quataert}{Chan et~al.}{2015}]{Chan:15}
Chan T.~K.,  Kere{\v s} D.,  Onorbe J.,  Hopkins P.~F.,  Muratov A.~L.,
  Faucher-Gigu{\`e}re C.~A.,   Quataert E.,  2015, \mn@doi [MNRAS]
  {10.1093/mnras/stv2165}, 454, 2981

\bibitem[\protect\citeauthoryear{Dai, Robertson  \& Madau}{Dai
  et~al.}{2018}]{Dai:18}
Dai B.,  Robertson B.~E.,   Madau P.,  2018, \mn@doi [ApJ]
  {10.3847/1538-4357/aabb06}, 858, 73

\bibitem[\protect\citeauthoryear{Das \& Binney}{Das \& Binney}{2016}]{Das:16}
Das P.,  Binney J.,  2016, \mn@doi [MNRAS] {10.1086/163249}, 460, 1725

\bibitem[\protect\citeauthoryear{De~Lorenzi, Debattista, Gerhard  \&
  Sambhus}{De~Lorenzi et~al.}{2007}]{DeLorenzi:07}
De~Lorenzi F.,  Debattista V.,  Gerhard O.,   Sambhus N.,  2007, \mn@doi
  [MNRAS] {doi:10.1111/j.1365-2966.2007.11434.x}, 376, 71

\bibitem[\protect\citeauthoryear{Deason, Belokurov, Koposov, G{\'o}mez, Grand,
  Marinacci  \& Pakmor}{Deason et~al.}{2017}]{Deason:17}
Deason A.~J.,  Belokurov V.,  Koposov S.~E.,  G{\'o}mez F.~A.,  Grand R.~J.,
  Marinacci F.,   Pakmor R.,  2017, \mn@doi [MNRAS]
  {10.1088/0004-637X/702/2/1058}, 470, 1259

\bibitem[\protect\citeauthoryear{Deason, Belokurov, Koposov  \&
  Lancaster}{Deason et~al.}{2018}]{Deason:18}
Deason A.~J.,  Belokurov V.,  Koposov S.~E.,   Lancaster L.,  2018, \mn@doi
  [ApJ] {10.3847/2041-8213/aad0ee}, 862, L1

\bibitem[\protect\citeauthoryear{Debattista, Moore, Quinn, Kazantzidis, Maas,
  Mayer, Read  \& Stadel}{Debattista et~al.}{2008}]{Debattista:08}
Debattista V.~P.,  Moore B.,  Quinn T.,  Kazantzidis S.,  Maas R.,  Mayer L.,
  Read J.,   Stadel J.,  2008, \mn@doi [ApJ] {10.1086/587977}, 681, 1076

\bibitem[\protect\citeauthoryear{Debattista, Roskar, Valluri, Quinn, Moore  \&
  Wadsley}{Debattista et~al.}{2013}]{Debattista:13}
Debattista V.~P.,  Roskar R.,  Valluri M.,  Quinn T.,  Moore B.,   Wadsley J.,
  2013, \mn@doi [MNRAS] {10.1093/mnras/stt1217}, 434, 2971

\bibitem[\protect\citeauthoryear{Dehnen}{Dehnen}{2000}]{Dehnen:falcon}
Dehnen W.,  2000, \mn@doi [ApJ] {10.1086/312724}, 536, L39

\bibitem[\protect\citeauthoryear{Dehnen \& Binney}{Dehnen \&
  Binney}{1998}]{Dehnen:1998}
Dehnen W.,  Binney J.,  1998, \mn@doi [MNRAS]
  {10.1046/j.1365-8711.1998.01600.x}, 298, 387

\bibitem[\protect\citeauthoryear{Dehnen \& Gerhard}{Dehnen \&
  Gerhard}{1993}]{Dehnen:93}
Dehnen W.,  Gerhard O.~E.,  1993, \mn@doi [MNRAS] {10.1093/mnras/261.2.311},
  261, 311

\bibitem[\protect\citeauthoryear{Dubinski}{Dubinski}{1994}]{Dubinski:94}
Dubinski J.,  1994, \mn@doi [ApJ] {10.1086/174512}, 431, 617

\bibitem[\protect\citeauthoryear{Dubinski \& Carlberg}{Dubinski \&
  Carlberg}{1991}]{Dubinski:91}
Dubinski J.,  Carlberg R.~G.,  1991, \mn@doi [ApJ] {10.1086/170451}, 378, 496

\bibitem[\protect\citeauthoryear{Eilers, Hogg, Rix  \& Ness}{Eilers
  et~al.}{2019}]{Eilers:19}
Eilers A.-C.,  Hogg D.~W.,  Rix H.-W.,   Ness M.~K.,  2019, \mn@doi [ApJ]
  {10.3847/1538-4357/aaf648}, 871, 120

\bibitem[\protect\citeauthoryear{Einasto}{Einasto}{1965}]{Einasto:65}
Einasto J.,  1965, Trudy Astrofizicheskogo Instituta Alma-Ata, 5, 87

\bibitem[\protect\citeauthoryear{Evans, Sanders, Williams, An, Lynden-Bell  \&
  Dehnen}{Evans et~al.}{2016}]{Evans:16}
Evans N.~W.,  Sanders J.~L.,  Williams A.~A.,  An J.,  Lynden-Bell D.,   Dehnen
  W.,  2016, \mn@doi [MNRAS] {10.1093/mnras/stv2729}, 456, 4506

\bibitem[\protect\citeauthoryear{Fermani \& Sch{\"o}nrich}{Fermani \&
  Sch{\"o}nrich}{2013}]{Fermani:13}
Fermani F.,  Sch{\"o}nrich R.,  2013, \mn@doi [MNRAS] {10.1086/309386}, 432,
  2402

\bibitem[\protect\citeauthoryear{Foreman-Mackey, Hogg, Lang  \&
  Goodman}{Foreman-Mackey et~al.}{2013}]{ForemanMackey:13}
Foreman-Mackey D.,  Hogg D.~W.,  Lang D.,   Goodman J.,  2013, \mn@doi [PASP]
  {10.1086/670067}, 125, 306

\bibitem[\protect\citeauthoryear{Grand et~al.,}{Grand et~al.}{2017}]{Grand:17a}
Grand R. J.~J.,  et~al., 2017, \mn@doi [MNRAS] {10.1093/mnras/stx071}, 467,
  stx071

\bibitem[\protect\citeauthoryear{Grillmair \& Dionatos}{Grillmair \&
  Dionatos}{2006}]{Grillmair:gd1}
Grillmair C.~J.,  Dionatos O.,  2006, \mn@doi [ApJ] {10.1086/505111}, 643, L17

\bibitem[\protect\citeauthoryear{Harris}{Harris}{1996}]{Harris:96}
Harris W.~E.,  1996, \mn@doi [AJ] {10.1086/118116}, 112, 1487

\bibitem[\protect\citeauthoryear{Haywood, Di~Matteo, Lehnert, Snaith,
  Khoperskov  \& G{\'o}mez}{Haywood et~al.}{2018}]{Haywood:18}
Haywood M.,  Di~Matteo P.,  Lehnert M.~D.,  Snaith O.,  Khoperskov S.,
  G{\'o}mez A.,  2018, \mn@doi [ApJ] {10.3847/1538-4357/aad235}, 863, 113

\bibitem[\protect\citeauthoryear{Helmi}{Helmi}{2008}]{Helmi:08}
Helmi A.,  2008, \mn@doi [Astron Astrophys Rev] {10.1093/mnras/278.3.727}, 15,
  145

\bibitem[\protect\citeauthoryear{Helmi, Babusiaux, Koppelman, Massari,
  Veljanoski  \& Brown}{Helmi et~al.}{2018}]{Helmi:18final}
Helmi A.,  Babusiaux C.,  Koppelman H.~H.,  Massari D.,  Veljanoski J.,   Brown
  A. G.~A.,  2018, \mn@doi [Nat] {10.1088/0004-6256/141/4/130}, 563, 85

\bibitem[\protect\citeauthoryear{Hernitschek et~al.,}{Hernitschek
  et~al.}{2017}]{Hernitschek:17}
Hernitschek N.,  et~al., 2017, \mn@doi [ApJ] {10.3847/1538-4357/aa960c}, 850,
  96

\bibitem[\protect\citeauthoryear{Hernitschek et~al.,}{Hernitschek
  et~al.}{2018}]{Hernitschek:18}
Hernitschek N.,  et~al., 2018, \mn@doi [ApJ] {10.3847/1538-4357/aabfbb}, 859,
  31

\bibitem[\protect\citeauthoryear{Hunt, Hong, Bovy, Kawata  \& Grand}{Hunt
  et~al.}{2018}]{Hunt:18spiral}
Hunt J. A.~S.,  Hong J.,  Bovy J.,  Kawata D.,   Grand R. J.~J.,  2018, \mn@doi
  [MNRAS] {10.1093/mnras/sty2532}, 481, 3794

\bibitem[\protect\citeauthoryear{Ibata, Lewis, Irwin, Totten  \& Quinn}{Ibata
  et~al.}{2001}]{Ibata:01}
Ibata R.,  Lewis G.~F.,  Irwin M.,  Totten E.,   Quinn T.,  2001, \mn@doi [ApJ]
  {10.1086/320060}, 551, 294

\bibitem[\protect\citeauthoryear{Ibata, Malhan, Martin  \& Starkenburg}{Ibata
  et~al.}{2018}]{Ibata:18}
Ibata R.~A.,  Malhan K.,  Martin N.~F.,   Starkenburg E.,  2018, \mn@doi [ApJ]
  {10.3847/1538-4357/aadba3}, 865, 85

\bibitem[\protect\citeauthoryear{Iorio \& Belokurov}{Iorio \&
  Belokurov}{2018}]{Iorio:18}
Iorio G.,  Belokurov V.,  2018, \mn@doi [MNRAS] {10.1093/mnras/sty2806}

\bibitem[\protect\citeauthoryear{Jing \& Suto}{Jing \& Suto}{2002}]{Jing:02}
Jing Y.~P.,  Suto Y.,  2002, \mn@doi [ApJ] {10.1086/341065}, 574, 538

\bibitem[\protect\citeauthoryear{Kafle, Sharma, Lewis  \& Bland-Hawthorn}{Kafle
  et~al.}{2013}]{Kafle:13}
Kafle P.~R.,  Sharma S.,  Lewis G.~F.,   Bland-Hawthorn J.,  2013, \mn@doi
  [MNRAS] {10.1088/0004-637X/702/2/1058}, 430, 2973

\bibitem[\protect\citeauthoryear{Kafle, Sharma, Robotham, Pradhan, Guglielmo,
  Davies  \& Driver}{Kafle et~al.}{2017}]{Kafle:17}
Kafle P.~R.,  Sharma S.,  Robotham A. S.~G.,  Pradhan R.~K.,  Guglielmo M.,
  Davies L. J.~M.,   Driver S.~P.,  2017, \mn@doi [MNRAS]
  {10.3847/1538-4357/aa70e6}, 470, 2959

\bibitem[\protect\citeauthoryear{Katz et~al.,}{Katz et~al.}{2019}]{Katz:19}
Katz D.,  et~al., 2019, \mn@doi [A\&A] {10.1093/mnras/sty2293}, 622, A205

\bibitem[\protect\citeauthoryear{Kazantzidis, Kravtsov, Zentner, Allgood, Nagai
   \& Moore}{Kazantzidis et~al.}{2004}]{Kazantzidis:04}
Kazantzidis S.,  Kravtsov A.~V.,  Zentner A.~R.,  Allgood B.,  Nagai D.,
  Moore B.,  2004, \mn@doi [ApJ] {10.1086/423992}, 611, L73

\bibitem[\protect\citeauthoryear{Kazantzidis, Abadi  \& Navarro}{Kazantzidis
  et~al.}{2010}]{Kazantzidis:10}
Kazantzidis S.,  Abadi M.~G.,   Navarro J.~F.,  2010, \mn@doi [ApJ]
  {10.1088/2041-8205/720/1/L62}, 720, L62

\bibitem[\protect\citeauthoryear{Koposov, Rix  \& Hogg}{Koposov
  et~al.}{2010}]{Koposov:10}
Koposov S.~E.,  Rix H.-W.,   Hogg D.~W.,  2010, \mn@doi [ApJ]
  {10.1088/0004-637X/712/1/260}, 712, 260

\bibitem[\protect\citeauthoryear{Koppelman, Helmi  \& Veljanoski}{Koppelman
  et~al.}{2018}]{Koppelman:18}
Koppelman H.,  Helmi A.,   Veljanoski J.,  2018, \mn@doi [ApJ]
  {10.3847/2041-8213/aac882}, 860, L11

\bibitem[\protect\citeauthoryear{Kunder et~al.,}{Kunder
  et~al.}{2012}]{Kunder:12}
Kunder A.,  et~al., 2012, \mn@doi [AJ] {10.1088/0004-6256/143/3/57}, 143, 57

\bibitem[\protect\citeauthoryear{K{\"u}pper, Balbinot, Bonaca, Johnston, Hogg,
  Kroupa  \& Santiago}{K{\"u}pper et~al.}{2015}]{Kupper:15}
K{\"u}pper A. H.~W.,  Balbinot E.,  Bonaca A.,  Johnston K.~V.,  Hogg D.~W.,
  Kroupa P.,   Santiago B.~X.,  2015, \mn@doi [ApJ]
  {10.1088/0004-637X/803/2/80}, 803, 80

\bibitem[\protect\citeauthoryear{Law \& Majewski}{Law \&
  Majewski}{2010}]{Law:10}
Law D.~R.,  Majewski S.~R.,  2010, \mn@doi [ApJ] {10.1088/0004-637X/714/1/229},
  714, 229

\bibitem[\protect\citeauthoryear{Lindegren et~al.,}{Lindegren
  et~al.}{2018}]{Lindegren:18}
Lindegren L.,  et~al., 2018, \mn@doi [A\&A] {10.1051/aas:2000332}, 616, A2

\bibitem[\protect\citeauthoryear{Loebman et~al.,}{Loebman
  et~al.}{2014}]{Loebman:14}
Loebman S.~R.,  et~al., 2014, \mn@doi [ApJ] {10.1088/0004-637X/794/2/151}, 794,
  151

\bibitem[\protect\citeauthoryear{Majewski, Skrutskie, Weinberg  \&
  Ostheimer}{Majewski et~al.}{2003}]{Majewski:03}
Majewski S.~R.,  Skrutskie M.~F.,  Weinberg M.~D.,   Ostheimer J.~C.,  2003,
  \mn@doi [ApJ] {10.1086/379504}, 599, 1082

\bibitem[\protect\citeauthoryear{Malhan \& Ibata}{Malhan \&
  Ibata}{2018}]{Malhan:18}
Malhan K.,  Ibata R.~A.,  2018, \mn@doi [MNRAS] {10.1093/mnras/sty912}, 477,
  4063

\bibitem[\protect\citeauthoryear{Martinsson, Verheijen, Westfall, Bershady,
  Andersen  \& Swaters}{Martinsson et~al.}{2013}]{Martinsson:13}
Martinsson T. P.~K.,  Verheijen M. A.~W.,  Westfall K.~B.,  Bershady M.~A.,
  Andersen D.~R.,   Swaters R.~A.,  2013, \mn@doi [A\&A]
  {10.1051/0004-6361/201321390}, 557, A131

\bibitem[\protect\citeauthoryear{Mateu, Read  \& Kawata}{Mateu
  et~al.}{2017}]{Mateu:18}
Mateu C.,  Read J.~I.,   Kawata D.,  2017, \mn@doi [MNRAS]
  {10.1093/mnras/183.3.341}, 474, 4112

\bibitem[\protect\citeauthoryear{McMillan \& Binney}{McMillan \&
  Binney}{2009}]{McMillan:09}
McMillan P.~J.,  Binney J.~J.,  2009, \mn@doi [MNRAS]
  {10.1093/mnras/291.4.683}, 400, L103

\bibitem[\protect\citeauthoryear{Navarro, Eke  \& Frenk}{Navarro
  et~al.}{1996a}]{Navarro:96}
Navarro J.~F.,  Eke V.~R.,   Frenk C.~S.,  1996a, MNRAS, 283, L72

\bibitem[\protect\citeauthoryear{Navarro, Frenk  \& White}{Navarro
  et~al.}{1996b}]{nfw}
Navarro J.~F.,  Frenk C.~S.,   White S. D.~M.,  1996b, \mn@doi [ApJ]
  {10.1086/177173}, 462, 563

\bibitem[\protect\citeauthoryear{Ness et~al.,}{Ness et~al.}{2013}]{Ness:13IV}
Ness M.,  et~al., 2013, \mn@doi [MNRAS] {10.1093/mnras/stt533}, 432, 2092

\bibitem[\protect\citeauthoryear{Odenkirchen et~al.,}{Odenkirchen
  et~al.}{2001}]{Odenkirchen:01}
Odenkirchen M.,  et~al., 2001, \mn@doi [ApJ] {10.1086/319095}, 548, L165

\bibitem[\protect\citeauthoryear{P{\'e}rez-Villegas, Portail  \&
  Gerhard}{P{\'e}rez-Villegas et~al.}{2016}]{PerezVillegas:16}
P{\'e}rez-Villegas A.,  Portail M.,   Gerhard O.,  2016, \mn@doi [MNRAS]
  {10.1093/mnrasl/slw189}, 464, L80

\bibitem[\protect\citeauthoryear{Peter, Rocha, Bullock  \& Kaplinghat}{Peter
  et~al.}{2013}]{Peter:13}
Peter A. H.~G.,  Rocha M.,  Bullock J.~S.,   Kaplinghat M.,  2013, \mn@doi
  [MNRAS] {10.1111/j.1365-2966.2009.16188.x}, 430, 105

\bibitem[\protect\citeauthoryear{Piffl et~al.,}{Piffl et~al.}{2014}]{Piffl:14}
Piffl T.,  et~al., 2014, \mn@doi [MNRAS] {10.1093/mnras/stu1948}, 445, 3133

\bibitem[\protect\citeauthoryear{Portail, Gerhard, Wegg  \& Ness}{Portail
  et~al.}{2017}]{Portail:17}
Portail M.,  Gerhard O.,  Wegg C.,   Ness M.,  2017, \mn@doi [MNRAS]
  {10.1093/mnras/stw2819}, 465, 1621

\bibitem[\protect\citeauthoryear{Posti \& Helmi}{Posti \&
  Helmi}{2019}]{Posti:19}
Posti L.,  Helmi A.,  2019, \mn@doi [A\&A] {10.1086/163249}, 621, A56

\bibitem[\protect\citeauthoryear{Posti, Helmi, Veljanoski  \& Breddels}{Posti
  et~al.}{2018}]{Posti:18ellipsoid}
Posti L.,  Helmi A.,  Veljanoski J.,   Breddels M.~A.,  2018, \mn@doi [A\&A]
  {10.1051/0004-6361/201732277}

\bibitem[\protect\citeauthoryear{Ratnatunga, Bahcall  \& Casertano}{Ratnatunga
  et~al.}{1989}]{Ratnatunga:1989}
Ratnatunga K.~U.,  Bahcall J.~N.,   Casertano S.,  1989, \mn@doi [ApJ]
  {10.1086/167281}, 339, 106

\bibitem[\protect\citeauthoryear{Read}{Read}{2014}]{Read:14}
Read J.~I.,  2014, \mn@doi [J. Phys. G: Nucl. Part. Phys.]
  {10.1088/0954-3899/41/6/063101}, 41, 063101

\bibitem[\protect\citeauthoryear{Reid et~al.,}{Reid et~al.}{2014}]{Reid:14}
Reid M.~J.,  et~al., 2014, \mn@doi [ApJ] {10.1088/0004-637X/783/2/130}, 783,
  130

\bibitem[\protect\citeauthoryear{Sackett, Rix, Jarvis  \& Freeman}{Sackett
  et~al.}{1994}]{Sackett:94}
Sackett P.~D.,  Rix H.-W.,  Jarvis B.~J.,   Freeman K.~C.,  1994, \mn@doi [ApJ]
  {10.1086/174938}, 436, 629

\bibitem[\protect\citeauthoryear{Schneider, Frenk  \& Cole}{Schneider
  et~al.}{2012}]{Schneider:12}
Schneider M.~D.,  Frenk C.~S.,   Cole S.,  2012, \mn@doi [J. Cosmol. Astropart.
  Phys.] {10.1088/1475-7516/2012/05/030}, 2012, 030

\bibitem[\protect\citeauthoryear{Sch{\"o}nrich}{Sch{\"o}nrich}{2012}]{Schonrich:12}
Sch{\"o}nrich R.,  2012, \mn@doi [MNRAS] {10.1111/j.1365-2966.2012.21631.x},
  427, 274

\bibitem[\protect\citeauthoryear{Sch{\"o}nrich \& Dehnen}{Sch{\"o}nrich \&
  Dehnen}{2018}]{Schonrich:18}
Sch{\"o}nrich R.,  Dehnen W.,  2018, \mn@doi [MNRAS] {10.1093/mnras/sty1256},
  478, 3809

\bibitem[\protect\citeauthoryear{Sch{\"o}nrich, Binney  \&
  Asplund}{Sch{\"o}nrich et~al.}{2011}]{Schonrich:11}
Sch{\"o}nrich R.,  Binney J.,   Asplund M.,  2011, \mn@doi [MNRAS]
  {10.1051/0004-6361/201014922}, 420, 1281

\bibitem[\protect\citeauthoryear{Sesar et~al.,}{Sesar et~al.}{2017}]{Sesar:17}
Sesar B.,  et~al., 2017, \mn@doi [AJ] {10.3847/1538-3881/aa661b}, 153, 204

\bibitem[\protect\citeauthoryear{Smith et~al.,}{Smith et~al.}{2009a}]{Smith:09}
Smith M.~C.,  et~al., 2009a, \mn@doi [MNRAS] {10.1093/mnras/184.2.311}, 399,
  1223

\bibitem[\protect\citeauthoryear{Smith, Wyn~Evans  \& An}{Smith
  et~al.}{2009b}]{Smith:09tilt}
Smith M.~C.,  Wyn~Evans N.,   An J.~H.,  2009b, \mn@doi [ApJ]
  {10.1088/0004-637X/698/2/1110}, 698, 1110

\bibitem[\protect\citeauthoryear{Spergel \& Steinhardt}{Spergel \&
  Steinhardt}{2000}]{Spergel:00}
Spergel D.~N.,  Steinhardt P.~J.,  2000, \mn@doi [Physical Review Letters]
  {10.1016/S1384-1076(96)00018-8}, 84, 3760

\bibitem[\protect\citeauthoryear{Syer \& Tremaine}{Syer \&
  Tremaine}{1996}]{Syer:96}
Syer D.,  Tremaine S.,  1996, MNRAS, 282, 223

\bibitem[\protect\citeauthoryear{Tian, Liu, Xu  \& Xue}{Tian
  et~al.}{2019}]{Tian:19}
Tian H.,  Liu C.,  Xu Y.,   Xue X.,  2019, \mn@doi [ApJ]
  {10.3847/1538-4357/aaf6e8}, 871, 184

\bibitem[\protect\citeauthoryear{Vera-Ciro \& Helmi}{Vera-Ciro \&
  Helmi}{2013}]{VeraCiro:13}
Vera-Ciro C.,  Helmi A.,  2013, \mn@doi [ApJ] {10.1088/2041-8205/773/1/L4},
  773, L4

\bibitem[\protect\citeauthoryear{Watkins, van~der Marel, Sohn  \&
  Evans}{Watkins et~al.}{2018}]{Watkins:18}
Watkins L.~L.,  van~der Marel R.~P.,  Sohn S.~T.,   Evans N.~W.,  2018, arXiv,
  1804.11348

\bibitem[\protect\citeauthoryear{Wegg \& Gerhard}{Wegg \&
  Gerhard}{2013}]{Wegg:13}
Wegg C.,  Gerhard O.,  2013, \mn@doi [MNRAS] {10.1093/mnras/stt1376}, 435, 1874

\bibitem[\protect\citeauthoryear{Wegg, Gerhard  \& Portail}{Wegg
  et~al.}{2015}]{Wegg:15}
Wegg C.,  Gerhard O.,   Portail M.,  2015, \mn@doi [MNRAS]
  {10.1093/mnras/stv745}, 450, 4050

\bibitem[\protect\citeauthoryear{Wegg, Gerhard  \& Portail}{Wegg
  et~al.}{2016}]{Wegg:16}
Wegg C.,  Gerhard O.,   Portail M.,  2016, \mn@doi [MNRAS]
  {10.1093/mnras/stw1954}, 463, 557

\bibitem[\protect\citeauthoryear{Yoshida, Springel, White  \& Tormen}{Yoshida
  et~al.}{2000}]{Yoshida:00}
Yoshida N.,  Springel V.,  White S. D.~M.,   Tormen G.,  2000, \mn@doi [ApJ]
  {10.1086/312707}, 535, L103

\bibitem[\protect\citeauthoryear{Zhao}{Zhao}{1996}]{Zhao:96gnfw}
Zhao H.,  1996, \mn@doi [MNRAS] {10.1093/mnras/278.2.488}, 278, 488

\bibitem[\protect\citeauthoryear{de Zeeuw, Evans  \& Schwarzschild}{de~Zeeuw
  et~al.}{1996}]{deZeeuw:96}
de Zeeuw P.~T.,  Evans N.~W.,   Schwarzschild M.,  1996, \mn@doi [MNRAS]
  {10.1093/mnras/280.3.903}, 280, 903

\bibitem[\protect\citeauthoryear{van Uitert, Hoekstra, Schrabback, Gilbank,
  Gladders  \& Yee}{van Uitert et~al.}{2012}]{vanUitert:12}
van Uitert E.,  Hoekstra H.,  Schrabback T.,  Gilbank D.~G.,  Gladders M.~D.,
  Yee H. K.~C.,  2012, \mn@doi [A\&A] {10.1086/301513}, 545, A71

\makeatother
\end{thebibliography}

%%%%%%%%%%%%%%%%%%%%%%%%%%%%%%%%%%%%%%%%%%%%%%%%%%

%%%%%%%%%%%%%%%%% APPENDICES %%%%%%%%%%%%%%%%%%%%%

\appendix

\section{Mock Halo Analysis}
\label{sec:mockhalos}

\subsection{Construction of Mock Samples}
\label{sec:mockconstruction}

We have constructed observations of mock halos in order to test that our analysis reliably recovers both the intrinsic kinematics of the sample, and that we can reliably use these measurements to infer the Galactic potential. These mock halos were constructed from the made-to-measure models of the inner Galaxy constructed by \citet{Portail:17}. The models do not include a stellar halo and so, instead, a sample of dark matter halo particles were selected. To do so, each dark matter particle was treated as a test particle with a statistical weighting factor determined so that the resultant halo had the desired properties.

To construct stellar halos, this statistical weighting was determined on the basis of each particle's energy.  We used a piecewise exponential of the energy:
\begin{equation}
	\log f(E) = \alpha_i \frac{E-E_i}{E_{i+1}-Ei} + \beta_i\quad\mbox{when}~E_{i+1}>E>Ei~. \label{eq:logf}
\end{equation}
We found that using 7 bins equally spaced in energy was sufficient to construct accurate radial profiles. Enforcing continuity leaves $8$ free parameters. These were determined by fitting the density of particles within $20\degr$ of the minor axis between $1\kpc < |z| <15\kpc$ to the target profile $\nu \propto {-3}$. %Throughout we have used mock halos with a target density profile $\nu \propto {-3}$. %The free parameters of \autoref{eq:logf} 

The energy is not an integral of motion in the non-axisymmetric inner Galaxy\footnote{An alternative choice would be the Jacobi Energy, which is an integral of motion. However using this in place of the energy resulted in unrealistic halo shapes.}. Therefore, after choosing these statistical weights, we integrated the particles forwards in the models potential. After first integrating  for 500Myr (approximately the orbital timescale at 15\kpc) we selected particles every 100Myr to be part of our model halo proportional to the statistical weight given by \autoref{eq:logf}. 

\begin{figure*}
	\includegraphics[width=0.8\textwidth]{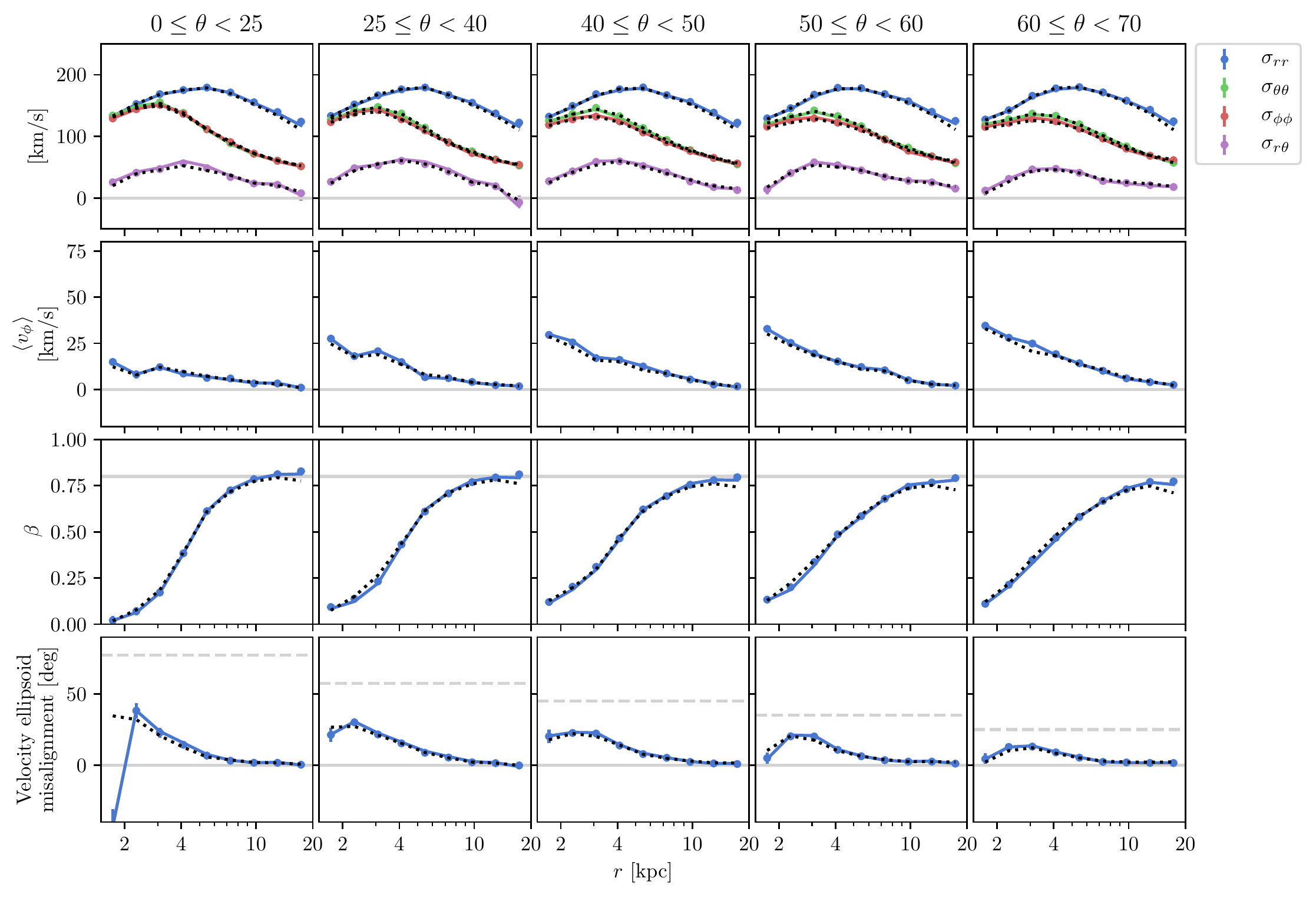}
    \caption{As \autoref{fig:mwkino} but instead showing the kinematics of our mock halo. The points correspond the reconstruction of the kinematics without radial velocity information assuming the the velocity distribution is Gaussian (\autoref{sec:gauss}). The coloured lines correspond to reconstruction using the method inspired by \citetalias{Dehnen:1998} (\autoref{sec:nongauss}). Note that both are indistinguishable across the entire volume probed, and that the points have error bars, but our mock halo has so many particles that they are mostly smaller than the points themselves. The dotted black lines are the intrinsic kinematics using full 6D phase space information without errors.}
    \label{fig:mockkino}
\end{figure*}

The resultant mock halo had a minor axis profile near to the observed $\nu \propto r^{-3}$ but the velocity distribution was nearly isotropic. This is because the dark halo particles in the \citetalias{Portail:17} model from which the stellar halo was constructed had a nearly isotropic velocity distribution. The Galaxy's stellar halo is however highly non-isotropic and therefore we constructed a family of mock anisotropic halos. To do so, we used an approximate third integral inspired by axisymmetric models of \citet{Dehnen:93}. We integrated each particle backwards in time for 10\Gyr and computed the normalised radial extent of each particles orbit in the equatorial plane: $D_r=(R_+-R_-)/R_+$ where $R_+$ and $R_-$ are the maximum and minimum radii reached at an equatorial crossing. We then adapted the statistical weight to be
\begin{equation}
	\log f(E,D_r) = \alpha_i \frac{E-E_i}{E_{i+1}-Ei} + \beta_i + h(D_r,E) \quad\mbox{when}~E_{i+1}>E>Ei~, \label{eq:logfd}
\end{equation}
where we chose $h(D_r,E)$ to provide a bias towards more radially extended orbits. Inspired by model 3 of \citet{Dehnen:93} we chose
\begin{equation}
h(D_r,E) = -(1-D_r)^2 q^2(E)/2y_0^2
\end{equation}
where $y_0$ is the parameter which determines the degree of radial anisotropy of our mock halo and $q(E)$ is a function that ensures the halo is centrally isotropic:
\begin{equation}
q(E) = \frac{R_c^2(E)}{R_a^2 + R_c^2(E)}
\end{equation}
where $R_c(E)$ is the radius of an in-plane circular orbit with the energy $E$ and $R_a$ is an anisotropy radius for which we used 0.5\kpc.

We note in passing that a more elegant solution to constructing the mock halos might be to use the actions of the particles to select the statistical weights of each particle. However, both the energy and orbital turning points would be integrals of motion in an axisymmetric potential, and would therefore be functions of the actions, making both methods not fundamentally different. In addition, our method has the advantage of being straightforward and computationally fast, while still resulting in mock halos with a minor axis profile and anisotropy very similar to the Milky Way's halo.

\subsection{Mock Halo Kinematics}
\label{sec:mockkinematics}

\begin{figure*}
	\includegraphics[width=0.8\textwidth]{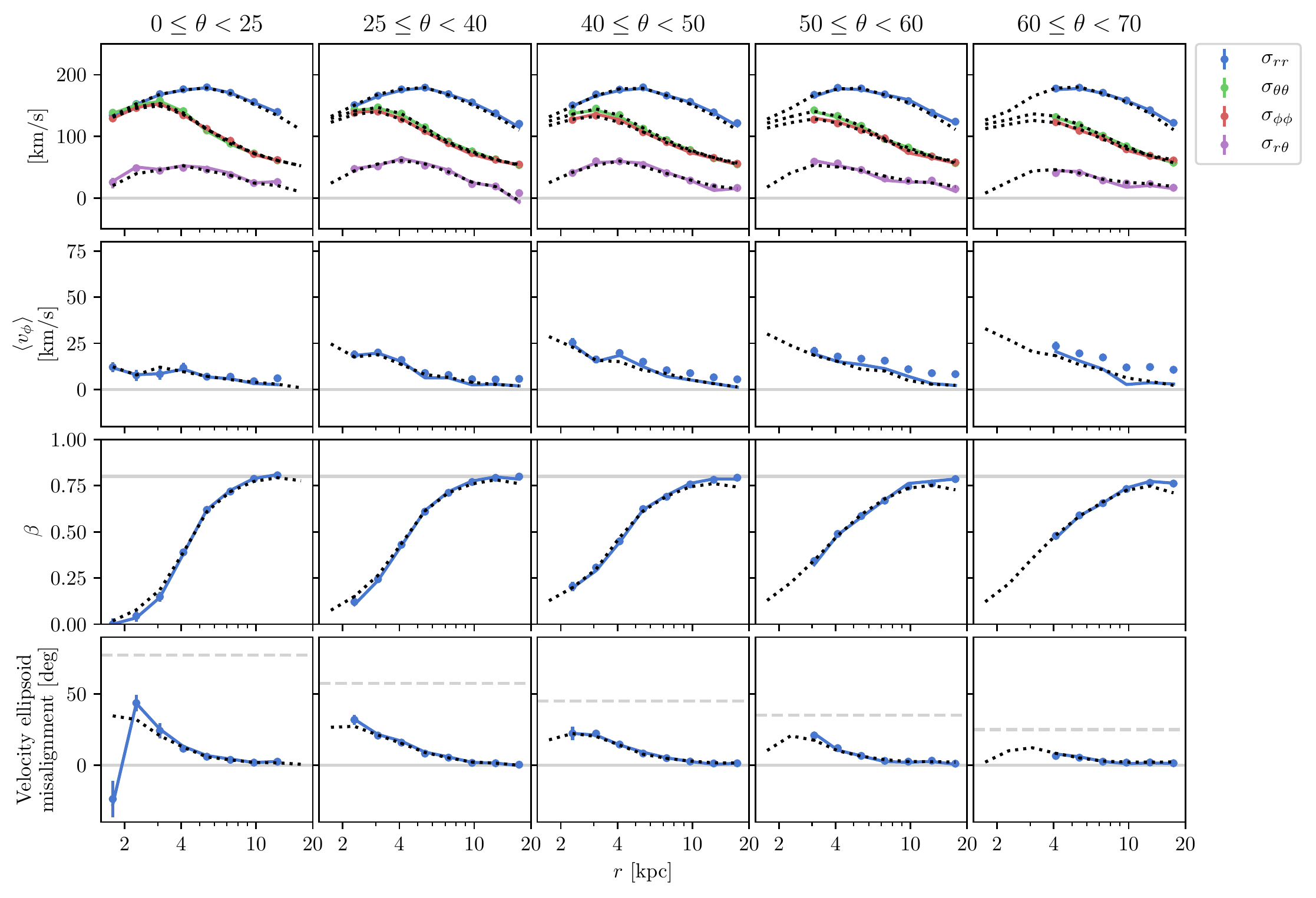}
    \caption{As \autoref{fig:mockkino} but instead showing the kinematics of our mock halo folded though the selection function described in \autoref{sec:sample}.}
    \label{fig:mockkinocut}
\end{figure*}

From the mock halos, we selected the halo with anisotropy parameter $y_0=0.15$ as being most similar to the Milky Way's halo and show the results of its analysis here. Comparing our mock halo kinematics in \autoref{fig:mockkino} to the measured Milky Way kinematics in \autoref{fig:mwkino}, we see that they are qualitatively very similar. We therefore proceed to test our analysis methods and code on this mock. We have also tested extensively with anisotropy $y_0=0.10$ and $y_0=0.20$ and obtained similar results. We use throughout exactly the code and methods developed for analysing the Milky Way halo, but here we have the advantage that we know both full 6D phase space information to test the kinematic reconstruction, and the potential in which the stars are moving in order to test the potential reconstruction.

\begin{figure*}
	\includegraphics[width=0.8\textwidth]{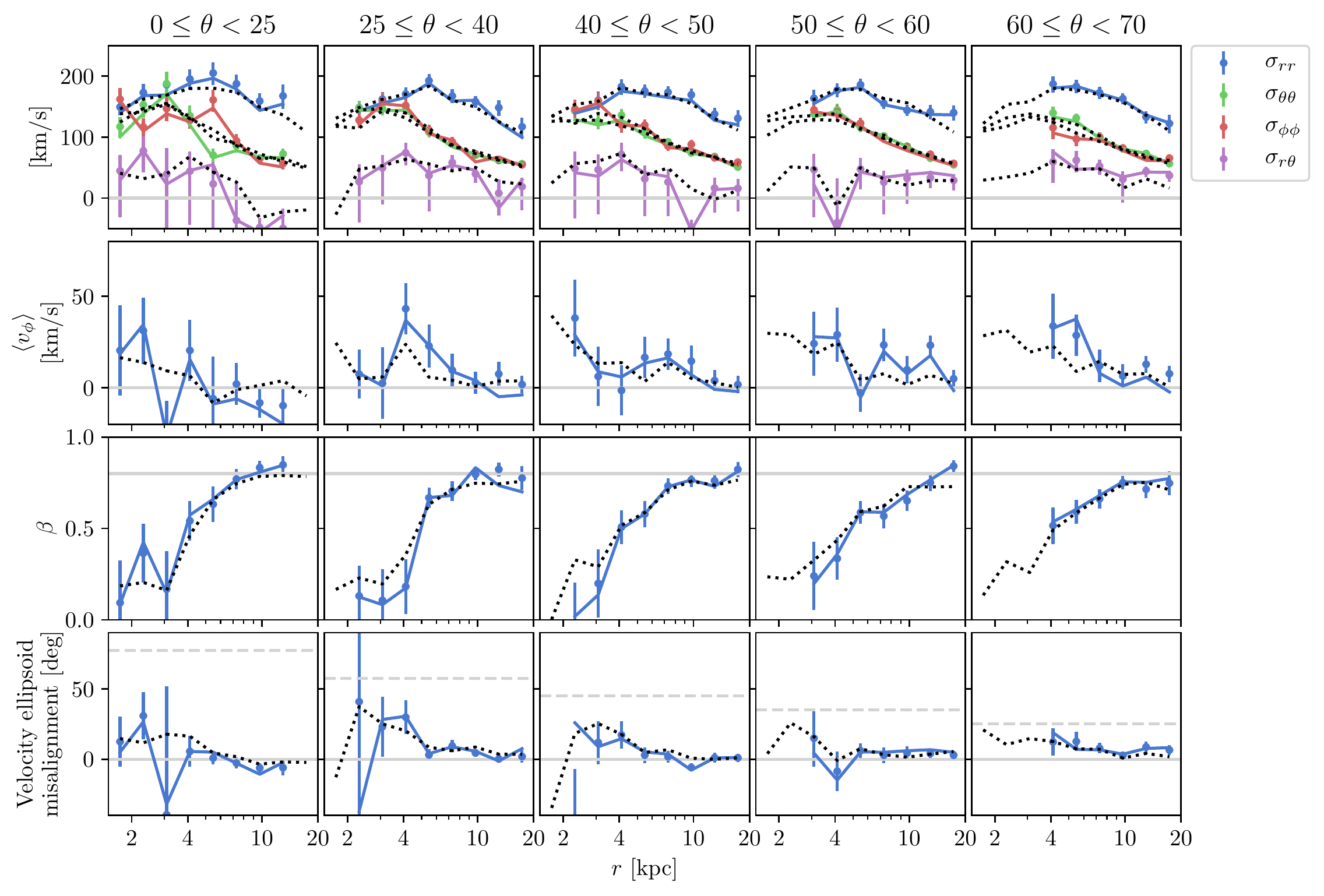}
    \caption{As \autoref{fig:mockkino} but instead showing the kinematics of our smaller mock halo folded though the selection function described in \autoref{sec:sample}. Because this halo has only 16,000 mock stars, similar to the actual sample this shows that the size of the error bars are appropriate.}
    \label{fig:mockkinocutsmall}
\end{figure*}

\begin{figure*}
\centering
	\includegraphics[width=\columnwidth]{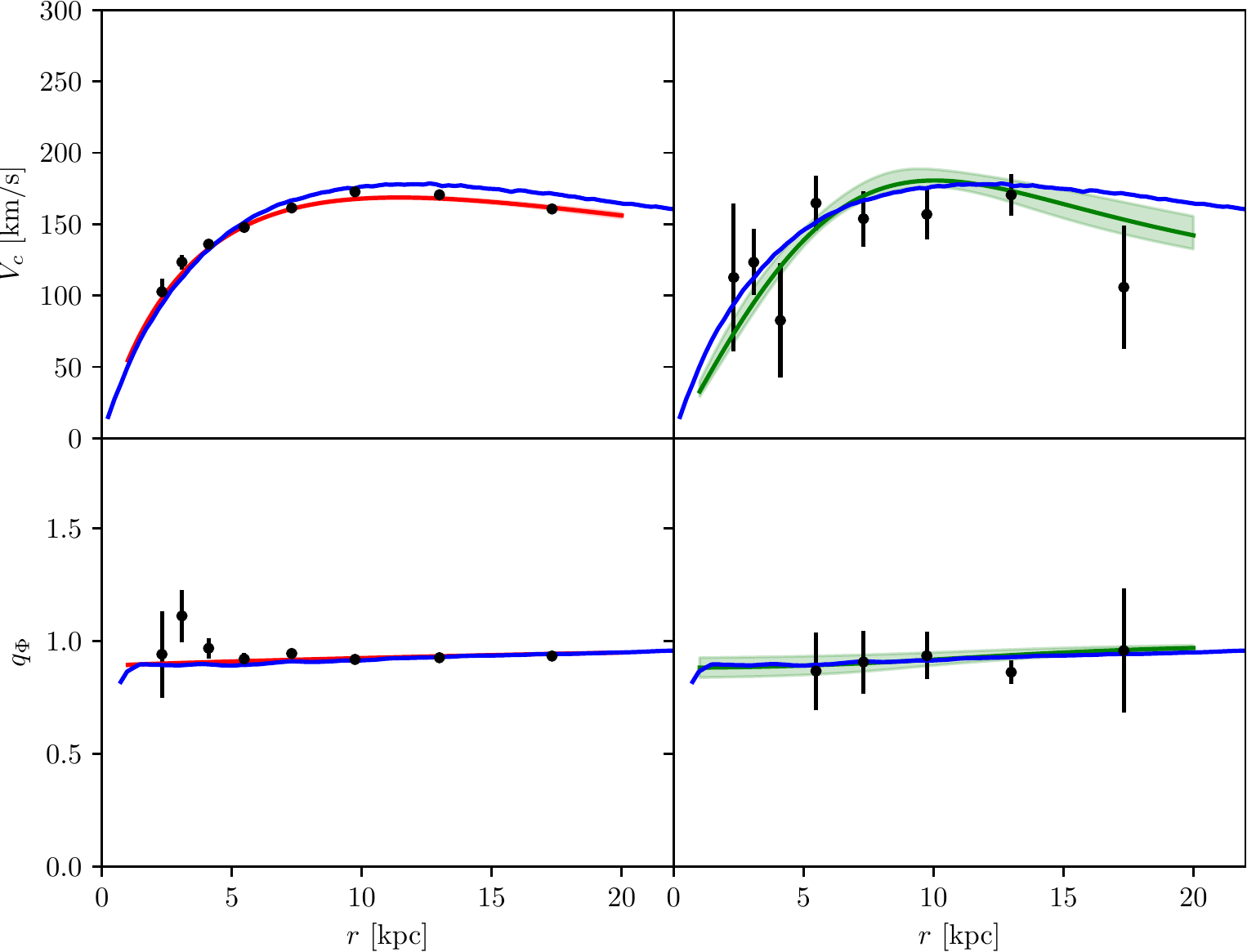}
    \caption{As \autoref{fig:ellipdmhalo} but instead showing the results of fitting an ellipsoidal shaped dark matter potential to the forces in the mock halo. On the left we show the results of fitting the mock halo with 870,000 stars, and on the right the realistic sized mock sample of 16,000 stars. The true underlying potential is shown in blue. We also plot the results of fitting an Einasto dark matter density profile to the forces in red on the left for the mock halo with 870,000 stars, and in green on the right for the mock halo with 16,000 stars.}
    \label{fig:mock_ellipdmhalo}
\end{figure*}

\begin{figure*}
\centering
	\includegraphics[width=\columnwidth]{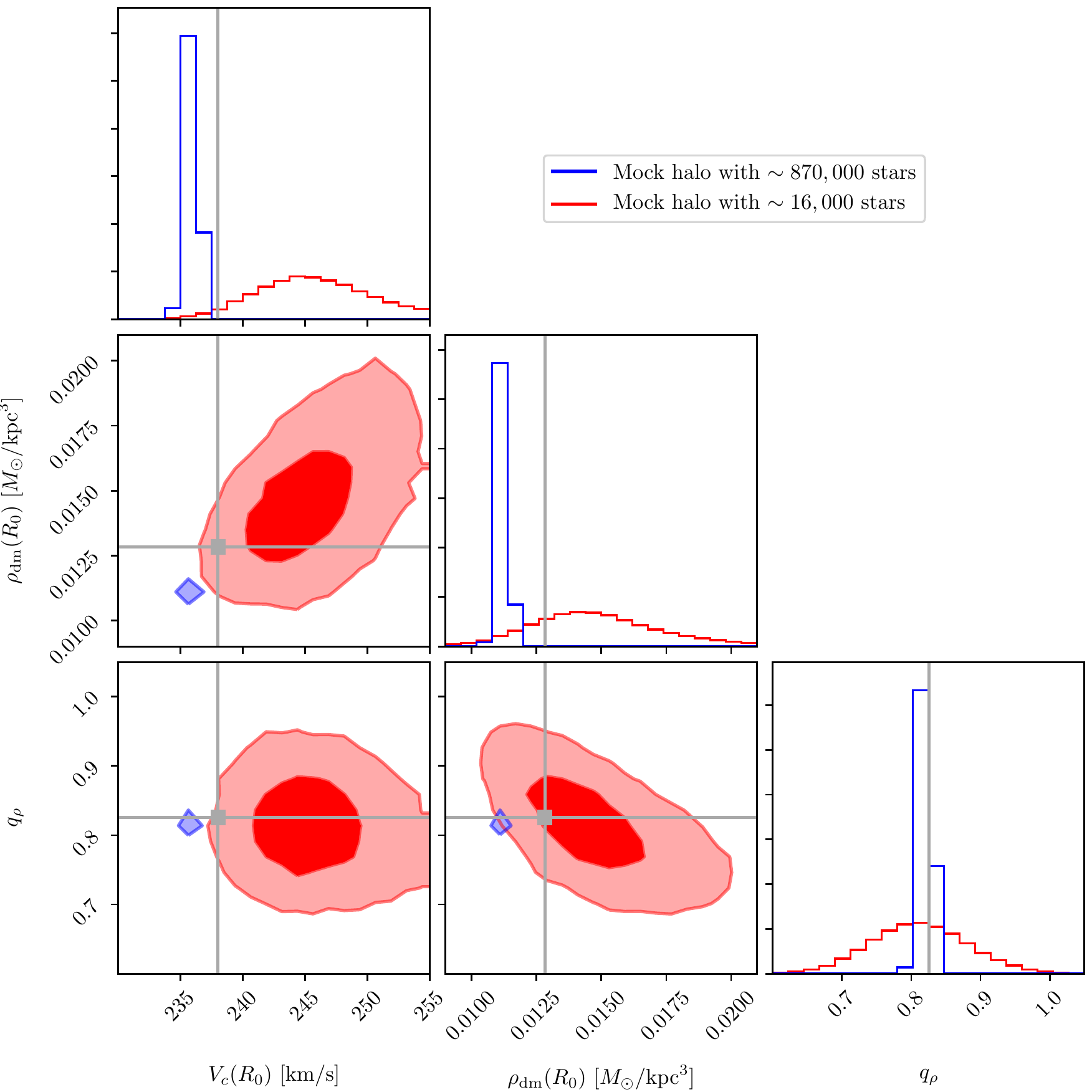}
    \caption{As \autoref{fig:dmproperties} but instead showing the parameters recovered after fitting an Einasto dark matter density profile to the forces. The results of fitting the large mock halo with 870,000 stars are shown in blue, and the results of fitting the mock halo with 16,000 stars are shown in red. The true parameters of the underlying potential are shown in dark gray. }
    \label{fig:mock_dmproperties}
\end{figure*}

We have constructed our mock halos to have a large number of stars in order to test that our methods asymptotically recover the kinematics, and the properties of the potential. In particular, we have \nbigmock stars in our mock sample. \Autoref{fig:mockkino,fig:mockkinocut} use this large sample to show that our analysis is asymptotically correct. In \Autoref{fig:mock_ellipdmhalo,fig:mock_dmproperties}, we compare the results obtained with the large sample and with a smaller subset of $\sim 16,000$ mock stars, similar to the number in the real sample.

In \autoref{fig:mockkino}, we compare the kinematics computed using the three dimensional galactocentric velocities of the particles (and therefore full 6D phase-space), to the kinematics reconstructed without radial velocity measurements (and therefore only 5D measurements). We use both methods described in \autoref{sec:kin}: the method assuming Gaussian velocities (\autoref{sec:gauss}) and method inspired by \citetalias{Dehnen:1998} (\autoref{sec:nongauss}). Both reconstruct the intrinsic kinematics accurately, being almost indistinguishable from the intrinsic kinematics.

\autoref{fig:mockkino} was constructed assuming a complete sample over the entire inner Halo. In \autoref{fig:mockkinocut}, we instead apply the selection function described in \autoref{sec:sample} to the mock halo before observing it. Again, both methods accurately reconstruct the velocity dispersion tensor. There are slight differences in the reconstruction of rotation, $\azmean{v_\phi}$, although at a level $<10\kms$. In general, the method inspired by \citetalias{Dehnen:1998} performs slightly better. For this reason and because it explicitly does not depend on the assumption of a Gaussian velocity distribution, we use this reconstruction method as our fiducial, testing only that we find equivalent  results with both in \autoref{sec:systematics}. 

In \autoref{fig:mockkinocutsmall} we make the same plot as \autoref{fig:mockkinocut}, but for the same smaller mock halo which has roughly the same number of particles as the sample of RR Lyrae analysed. This plot demonstrates that our statistical errors are of an appropriate size.

Because of the large numbers of stars in the mock sample, the errors in the forces are extremely small. In \autoref{fig:mock_ellipdmhalo}, we show the result of fitting an ellipsoidal  potential to these forces. In the left hand side, we show the result for the large mock. We accurately recover the flattening, but slightly underestimate the circular velocity due to the dark matter. However, this effect is much smaller than the errors in the real sample. The analysis of a mock sample of 16,000 stars is shown on the right of this figure.

Finally, in \autoref{fig:mock_dmproperties}, we show the results of fitting an Einasto profile to the force measurements. The mock halos were constructed in the potential of the \citetalias{Portail:17} model. The dark matter in this model had a profile chosen to be close to an Einasto profile with circular velocity at the sun $V_c(R_0)=238\kms$, dark matter density at the Sun $V_c(R_0)=0.013\msun/{\rm pc^3}$, and has average flattening over the range 5\kpc to 20\kpc of $q_\rho \approx 0.83$. With the large mock halo, we recover the parameters to outside the formal statistical errors, but this bias is well within the size of statistical errors of the sample of real RR Lyrae analysed in the main text. We also recover the parameters to within the errors when we use the mock sample of $\sim 16,000$ halo stars.

The small discrepancies when fitting the large mock halo were not reduced by increasing the number of bins to reduce the discretisation errors. The reason for the difference may result from the mock dark matter halo being an N-body system which is not strictly an ellipsoid Einasto profile. For example in the mock the halo flattening increases from $q_\rho \approx 0.85$ at 5\kpc to $\approx 0.8$ at 20\kpc, and we plot the average value 0.83. In any case because of the smaller sample sizes, the bias is smaller than the errors of both the smaller mock and the data.

%%%%%%%%%%%%%%%%%%%%%%%%%%%%%%%%%%%%%%%%%%%%%%%%%%

% Don't change these lines
\bsp	% typesetting comment
\label{lastpage}
\end{document}